\documentclass[letterpaper,onecolumn]{IEEEtran}
\usepackage{amsmath,amssymb,amsthm,mathtools,bm,mathrsfs}
\usepackage{booktabs,array,cite,graphicx}
\usepackage{needspace,placeins}
\usepackage[hidelinks]{hyperref}
\usepackage{color}
\newcommand{\F}{\mathbb F}
\newcommand{\C}{\mathbb C}
\newcommand{\HH}{\mathcal H}
\newcommand{\BB}{\mathcal B}
\newcommand{\id}{\operatorname{id}}
\newcommand{\Tr}{\operatorname{Tr}}
\newcommand{\rank}{\operatorname{rank}}
\newcommand{\supp}{\operatorname{supp}}
\newcommand{\swt}{\operatorname{swt}}

\newcommand{\rad}{\operatorname{rad}}

\newcommand{\wt}{\operatorname{wt}}
\newcommand{\perpt}{\perp_{\mathrm t}}
\newcommand{\perps}{\perp_{\mathrm s}}
\newcommand{\cV}{\mathcal V}
\newtheoremstyle{titstatement}{6pt}{6pt}{\itshape}{}{\bfseries\boldmath}{.}{0.5em}
 {\thmname{#1}\thmnumber{ #2}\thmnote{ (#3)}}
\newtheoremstyle{titdefinition}{6pt}{6pt}{\normalfont}{}{\bfseries\boldmath}{.}{0.5em}
 {\thmname{#1}\thmnumber{ #2}\thmnote{ (#3)}}
\theoremstyle{titstatement}
\newtheorem{theorem}{Theorem}[section]
\newtheorem{lemma}[theorem]{Lemma}
\newtheorem{proposition}[theorem]{Proposition}
\newtheorem{corollary}[theorem]{Corollary}
\theoremstyle{titdefinition}
\newtheorem{definition}[theorem]{Definition}
\newtheorem{example}[theorem]{Example}
\newtheorem{remark}[theorem]{Remark}
\theoremstyle{titstatement}
\hypersetup{pdftitle={Subsystem Stabilizer Codes with (r,delta)-Locality},pdfauthor={Yaozong Zhang and Sihem Mesnager}}
\allowdisplaybreaks
\begin{document}

\title{Subsystem Stabilizer $(r,\delta)$-Locally Recoverable Codes:
Characterization, Bounds and Constructions}

\author{Yaozong Zhang,~Sihem Mesnager,~\IEEEmembership{Member,~IEEE},~Xiaoqiang Wang~and Jian Gao{}

\thanks{Yaozong Zhang is with the School of Mathematical Sciences and LPMC, Nankai University, Tianjin 300071, China (e-mail: zyzsdutms@163.com).}
\thanks{Sihem Mesnager is with the Department of Mathematics, University of Paris VIII, 93526 Saint-Denis, France, with the Laboratory Analysis, Geometry and Applications, LAGA, University Sorbonne Paris Nord, CNRS, UMR7539, 93430 Villetaneuse, France, and also with the Telecom Paris, Polytechnic Institute of Paris, 91120 Palaiseau, France~(e-mail: smesnager@univ-paris8.fr).}
\thanks{Xiaoqiang Wang is with the Hubei Key Laboratory of Applied Mathematics, Faculty of Mathematics and Statistics, Hubei University, Wuhan 430062, China (e-mail: waxiqq@163.com).}
\thanks{Jian Gao is with the School of Mathematics and Statistics, Shandong University of Technology, Zibo 255000, China (e-mail: dezhougaojian@163.com).}

\thanks{{\it Corresponding author: Jian Gao.}}
}
\markboth{}%
{}
\maketitle

\begin{abstract}
Local erasure recovery limits the number of physical systems accessed to repair a distributed quantum memory.
For subsystem stabilizer codes (SSCs), a basic question is whether erasures can be corrected within a prescribed repair set.
In this paper, we study subsystem stabilizer $(r,\delta)$-locally recoverable codes ($(r,\delta)$-SSLRCs),
in which each coordinate belongs to a repair set of size between $\delta$ and $r+\delta-1$ that corrects any set of up to $\delta-1$ erasures while preserving the logical subsystem and allowing the gauge subsystem to change.
We first characterize local erasure recovery algebraically and apply this criterion to Calderbank-Shor-Steane (CSS) and Hermitian hull constructions for $(r,\delta)$-SSLRCs.
For $(r,\delta)$-SSLRCs, without assuming purity, we derive a redundancy bound for disjoint repair sets, a supported-subspace bound based on supported stabilizer and gauge subspaces, and a mixed counting bound for known erasures within selected repair sets and errors at unknown positions outside their union. For pure $(r,\delta)$-SSLRCs with overlapping repair sets, we establish the Singleton-type and alphabet-dependent bounds, together with explicit Griesmer, Hamming, and Plotkin bounds.
Moreover, we construct nine classes of $\F_q$-linear pure $(r,\delta)$-SSLRCs from maximum distance separable (MDS) stabilizer codes, Reed-Solomon (RS) codes, and linear codes with MDS Hermitian hulls, which are simultaneously optimal with respect to several bounds for $(r,\delta)$-SSLRCs.
\end{abstract}

\section{Introduction}

\IEEEPARstart{Q}{uantum} error-correcting codes protect quantum information against noise and the loss of physical systems~\cite{kribs,poulin}. An erasure is the loss of a physical system at a known location. In a distributed quantum memory, recovery may require access to surviving systems at different locations. Local erasure recovery limits the number
of systems accessed by confining recovery to a prescribed repair set~\cite{golowich}. Under $(r,\delta)$-locality, every coordinate belongs to a repair set of size between $\delta$ and $r+\delta-1$, within which any set of up to $\delta-1$ erasures can be corrected~\cite{galindo}. In contrast, the global minimum distance $d$ guarantees correction of any set of up to $d-1$ erasures when recovery can access all surviving systems. Thus, $\delta$ specifies a local erasure tolerance under an access restriction, whereas $d$ describes protection without this restriction. For codes with a nontrivial logical subsystem, $(r,\delta)$-locality implies $d\ge\delta$ (see Corollary~\ref{cor:globaldistance}).

Subsystem stabilizer codes (SSCs) decompose the encoded space into a logical subsystem and a gauge subsystem~\cite{kribs,poulin}. Recovery preserves the logical subsystem and its correlations with any reference system, while a completely positive trace-preserving (CPTP) map acts on the gauge subsystem. Pauli errors differing by a gauge operator are equivalent for logical recovery. We construct a syndrome decoder that distinguishes these equivalence classes using stabilizers supported in the repair set. Stabilizer and subsystem codes identify errors modulo $S$ and $G$, respectively; the additional freedom is described by the quotient space $G/S$.

In this paper, we study subsystem stabilizer $(r,\delta)$-locally recoverable codes ($(r,\delta)$-SSLRCs) and determine how local recovery requirements constrain the logical dimension, gauge dimension, and global minimum distance. More precisely, we address the following three questions:
\begin{itemize}
\item[(Q1)] \emph{Algebraic characterization:} what conditions on the stabilizer and gauge label spaces characterize erasure recovery within a prescribed repair set for arbitrary joint states of the logical and gauge subsystems?
\item[(Q2)] \emph{Parameter bounds:} what bounds follow from $(r,\delta)$-locality without a purity assumption, and what further bounds hold for pure codes?
\item[(Q3)] \emph{Optimal pure constructions:} can explicit pure $(r,\delta)$-SSLRCs simultaneously attain the resulting bounds?
\end{itemize}

\subsection{Related Work}

\textit{Classical locally recoverable codes (cLRCs).}
In classical coding theory, Gopalan et al.~\cite{gopalan} introduced the notion of symbol locality. A symbol has locality at most $r$ if it can be recovered from at most $r$ other symbols. They established the Singleton-type bound $d\le n-k-\lceil k/r\rceil+2$ for an $[n,k,d]_q$ linear code whose information symbols have locality at most $r$.
Prakash et al.~\cite{prakash} introduced $(r,\delta)$-locality to allow local correction of multiple erasures and obtained the bound
$d\le n-k+1-(\lceil k/r\rceil-1)(\delta-1)$, which reduces to the previous bound when $\delta=2$. Cadambe and Mazumdar~\cite{cadambe} derived alphabet-dependent bounds, while Grezet et al.~\cite{grezet} obtained such bounds for linear codes with $(r,\delta)$-locality by using residual codes. By accounting for the alphabet size, these bounds can improve on Singleton-type bounds over small alphabets.

\textit{Quantum locally recoverable codes (qLRCs).}
Golowich and Guruswami~\cite{golowich} studied quantum local recovery from single erasures. Luo et al.~\cite{luoCSS} derived parameter bounds and constructed pure qLRCs attaining them through the CSS construction. Galindo et al.~\cite{galindo} characterized $(r,\delta)$-locality for stabilizer codes in terms of puncturing and shortening. For stabilizer codes obtained from Euclidean or Hermitian dual-containing classical codes, their Singleton-type bound involves the classical minimum distance, which can be replaced by the quantum minimum distance under purity~\cite[Theorem~31 and Remark~34]{galindo}. Galindo et al.~\cite{impure} subsequently constructed impure qLRCs exceeding the corresponding bounds of pure qLRCs. For Hermitian constructions, Li et al.~\cite{lipure} established further bounds and constructions for pure qLRCs, while Li et al.~\cite{lihermitian} derived bounds for pure and impure qLRCs and used $t$-designs to determine their locality. Matsumoto~\cite{matsumoto} characterized stabilizer measurements sufficient for erasure correction and determined the minimum number of measurements within the corresponding decoding model. Entanglement-assisted qLRCs were studied in~\cite{galindoea,liea}. In these models, local recovery may also access the receiver's halves of preshared maximally entangled pairs; these halves are assumed to be unaffected by erasures. The $(r,\delta)$-SSLRCs considered in this paper use no entanglement assistance; recovery preserves the logical subsystem and allows the gauge subsystem to change.

\textit{Quantum bounds for qLRCs.}
Kumar and Bandi~\cite{kumar} obtained explicit Griesmer, Hamming, and Plotkin bounds for pure qLRCs from the Hermitian construction. Ball, Lavrauw, and Popatia~\cite{ball} established a Griesmer-type bound for additive codes without requiring linearity over the extension field. Since the associated classical codes need not be linear over $\F_{q^2}$, we use this additive bound. Stylianou and Boche~\cite{stylianou} used blockwise weight enumerators to derive bounds for qLRCs with disjoint repair sets under a blockwise purity condition. Their formulation does not require a stabilizer structure.

\textit{SSCs and gauge freedom.}
Operator quantum error correction~\cite{kribs} and the subsystem stabilizer formalism~\cite{poulin} allow recovery to preserve logical information while changing the gauge subsystem. Aly and Klappenecker~\cite{aly} developed additive constructions and propagation rules for trading logical and gauge dimensions with guarantees on minimum distance and purity. Further constructions arise from Hermitian hulls~\cite{luo,luomds}, while Bacon-Shor codes~\cite{bacon} provide impure examples with weight-two
gauge generators. Quantum code concatenation~\cite{knill}, including concatenation of subsystem codes~\cite[Lemma~20]{dai}, provides bounds on the minimum distance in terms of the constituent-code distances. The geometric locality in~\cite{dai} concerns interaction lengths, whereas the locality studied here concerns erasure recovery within prescribed coordinate sets. B\'eny et al.~\cite{beny} studied approximate recovery under locality and symmetry constraints. Cao, Cotardo, and Lackey~\cite{caoanticodes} studied supported symplectic subspaces and local invariants of stabilizer and subsystem codes through quantum anticodes.

\textit{Global and local constraints.}
Klappenecker and Sarvepalli~\cite{klappenecker} proved a Singleton bound for $\F_q$-linear subsystem codes without a purity assumption. This bound constrains the global code parameters without incorporating a prescribed family of repair sets. For $(r,\delta)$-SSLRCs, a further question is which errors can be distinguished modulo the gauge space using only stabilizers supported in a repair set. We give an algebraic characterization of this requirement and use it to derive parameter bounds.

\subsection{Recovery Model and Methods}
We require local recovery to return the state to the original code space and preserve the logical subsystem together with its correlations with an arbitrary reference system.
We impose this requirement on every joint state of the logical, gauge, and reference systems. The gauge state may change and need not be known.

Let $G$ and $S$ denote the gauge and stabilizer label spaces, respectively. For a coordinate set $T$, let $W_T$ be the space of Pauli labels supported in $T$, and let $L_T=L\cap W_T$. Theorem~\ref{thm:local} shows that a nonempty erased set $E\subseteq\Gamma$ is recoverable within $\Gamma$ if and only if $W_E\cap S_\Gamma^{\perpt}=G_E$, where $\perpt$ denotes trace-symplectic orthogonality. To prove sufficiency, we construct an explicit local recovery channel based on stabilizer syndrome measurements. Without assuming purity, we derive bounds by studying supported stabilizer and gauge subspaces and counting gauge-equivalence classes of errors. For pure codes with a nontrivial logical subsystem, the stabilizer eigenspace defines a stabilizer code with the same minimum distance~\cite[Lemma~10]{alylp}. We show that this reduction preserves the prescribed repair cover and obtain further bounds by shortening an associated additive classical code.

\subsection{Main Contributions}

In this paper, the parameters of an $(r,\delta)$-SSC $\mathcal Q$ are denoted by $((n,K,R,d))_q$, where $K$ and $R$ are the logical and gauge dimensions, and $d$ is the global minimum distance. When $K=q^k$ and $R=q^g$ for nonnegative integers $k$ and $g$, we say that $\mathcal Q$ is an $(r,\delta)$-SSC with parameters $[[n,k,g,d]]_q$.
The notation $\dim_p$ denotes dimension over $\F_p$. Our main results for $(r,\delta)$-SSLRCs are summarized as follows. Item 1) answers (Q1), items 2) and 3) answer (Q2), and item 4) answers (Q3).

\begin{enumerate}
\item \emph{Algebraic characterization.}
For an $(r,\delta)$-SSC, we establish necessary and sufficient algebraic conditions for local erasure recovery within a prescribed repair set without a purity assumption (see Theorem~\ref{thm:local}). We construct a recovery channel for arbitrary joint logical-gauge states, including correlations with a reference system.
We establish algebraic criteria for the CSS and Hermitian hull constructions of $\mathcal Q$ (see Propositions~\ref{CSS} and~\ref{prop:hermitian}). If $R=1$, the criterion reduces to the stabilizer local recovery condition in~\cite{galindo}.
\item \emph{Bounds without a purity assumption.}
We derive a redundancy bound for $(r,\delta)$-SSLRCs with disjoint repair sets (Proposition~\ref{prop:redundancy}).
When two disjoint erased sets of an $(r,\delta)$-SSLRC are globally correctable, their supported gauge spaces and the stabilizers on the remaining coordinates bound $K$ and $R$ (see Theorem~\ref{thm:profilebound}).
The mixed counting bound in Theorem~\ref{thm:mixedpacking} constrains $KR$ for fewer than $\delta$ known erasures in each of several pairwise disjoint repair sets and at most $t$ errors at unknown positions outside their union, where $2t<d$.
\item \emph{Bounds for pure codes.}
We establish a rank-accumulation lemma for additive $(r,\delta)$-cLRCs based on the arguments in \cite[Lemma~1]{cadambe} and \cite[proof of Corollary~2]{grezet} (see Lemma~\ref{lem:rankaccumulation}). Using this result to shorten the comparison codes, we derive Singleton-type and alphabet-dependent bounds for pure $(r,\delta)$-SSLRCs (see Theorems~\ref{thm:singleton} and~\ref{thm:alphabet}). Moreover, by applying the corresponding classical bounds to these shortened codes, we obtain explicit Griesmer, Hamming, and Plotkin bounds (see Theorems~\ref{cor:griesmer} and~\ref{cor:hammingplotkin}).
\item \emph{Optimal constructions of pure codes.}
We determine the parameters of tensor products of SSCs and trade logical and gauge dimensions while preserving the stabilizer label space, minimum distance, and repair cover (see Propositions~\ref{prop:blocks} and~\ref{prop:local-dimension-trading}).
We apply these results to construct nine classes of pure $(r,\delta)$-SSLRCs from MDS stabilizer codes, Reed-Solomon codes, and MDS Hermitian hulls (see Table~\ref{tab:constructions}). These pure codes simultaneously attain the Singleton-type bound~\eqref{eq:lrcbound}, the alphabet-dependent bound~\eqref{eq:alphabetbound}, the $\F_{q^2}$-linear Griesmer bound~\eqref{eq:lineargriesmer}, and the redundancy bound~\eqref{eq:pureredundancy}.
\end{enumerate}

\subsection{Organization of the Paper}

The rest of this paper is organized as follows. Section~\ref{sec:prelim} reviews the notation and preliminary results for $(r,\delta)$-SSLRCs. Section~\ref{sec:recovery} establishes the algebraic characterization of local recovery and presents the CSS and Hermitian hull constructions for $(r,\delta)$-SSLRCs.
Section~\ref{sec:bounds} derives bounds without assuming purity, while Section~\ref{sec:pure} establishes further bounds for pure $(r,\delta)$-SSLRCs.
Section~\ref{sec:construction} constructs nine classes of pure $(r,\delta)$-SSLRCs that simultaneously attain the stated bounds from Sections~\ref{sec:bounds} and~\ref{sec:pure}.
Section~\ref{con-out} concludes the paper and discusses further questions.

\section{Preliminaries}\label{sec:prelim}
In this section, we recall basic definitions and results on classical $(r,\delta)$-cLRCs and SSCs, and state the global erasure correction criterion needed for $(r,\delta)$-SSLRCs.

\subsection{Classical Codes in Coding Theory}\label{sec:classicalcodes}
Let $q=p^m$, where $p$ is prime and $m\ge1$ is an integer, and let $\F_q$ denote the finite field of order $q$.
Let $\mathbb{F}_q^*$ denote the multiplicative group of $\mathbb{F}_q$, and $\mathbb{F}_q^n$ the $n$-dimensional vector space over $\mathbb{F}_q$.
A linear code $C$ with parameters $[n,k,d]_q$ is a $k$-dimensional linear subspace of $\mathbb{F}_q^n$ with minimum Hamming distance $d=d(C)=\min\{\wt(\bm{c}):\bm{c}\in C\setminus\{\bm{0}\}\},$ where $\wt(\bm{c})$ is the number of nonzero coordinates of $\bm{c}$.
The support of $\bm{c}=(c_1,\ldots,c_n)$ is the set $\supp(\bm{c})=\{i:c_i\ne0,1\leq i\leq n\}$.
Thus $\wt(\bm c)=|\supp(\bm c)|$.
A generator matrix for an $[n,k,d]_q$ code $C$ is a $k\times n$ matrix $M$ whose rows form a basis for $C$.
Let $\bm{u}=(u_1,\ldots,u_n),\bm{v}=(v_1,\ldots,v_n)\in\F_q^n$.
The Euclidean inner product of $\bm{u}$ and $\bm{v}$ is defined by $\langle\bm{u},\bm{v}\rangle=\sum_{i=1}^nu_iv_i$, and the Euclidean dual code of $C$ is defined by
$$C^\perp = \left\{\bm{v} \in \mathbb{F}_q^n : \langle \bm{u}, \bm{v} \rangle = 0 \text{ for all } \bm{u} \in C\right\}.$$
For $\bm u,\bm v\in\F_{q^2}^n$, define the Hermitian inner product by $\langle\bm u,\bm v\rangle_H=\sum_{i=1}^n u_iv_i^q$.
For an $\F_{q^2}$-linear code $C\le\F_{q^2}^n$, define
\[
 C^{\perp_H}=\{\bm v\in\F_{q^2}^n:\langle\bm u,\bm v\rangle_H=0\text{ for all }\bm u\in C\}.
\]
Moreover, the Hermitian hull of $C$ is defined as
$$\operatorname{Hull}_H(C)=C\cap C^{\perp_H}.$$

The parameters of an $[n,k,d]_q$ linear code $C$ satisfy the classical Singleton bound $d\leq n-k+1$ \cite{macwilliams}.
We call $C$ a maximum distance separable (MDS) code if $d=n-k+1$.
Generalized Reed-Solomon (GRS) codes are MDS.
\begin{definition}\cite{Huffman2003}\label{GRS}
Let $n$ be any integer with $1\leq n\leq q$. Let $\bm{a}=(a_1,\ldots,a_n)$ be an $n$-tuple of distinct elements of $\F_q$, and let $\bm{v}=(v_1,\ldots,v_n)$ be an $n$-tuple of nonzero, but not necessarily distinct, elements of $\F_q$. Let $1\leq k\leq n$. Then the code
$$\operatorname{GRS}_k(\bm{a},\bm{v})=\left\{(v_1f(a_1),\ldots,v_nf(a_n)):f\in\F_q[x],\ \deg f<k\right\}$$
is called a GRS code. For $\bm{v}=\bm{1}$, it is called a Reed-Solomon (RS) code.
\end{definition}
We recall the following standard results on GRS codes.
\begin{lemma}\cite{Huffman2003}\label{le:GRS}
With the above notation, the following statements hold.
\begin{itemize}
  \item [(i)] $\operatorname{GRS}_k(\bm{a},\bm{v})$ is an $[n,k,n-k+1]_q$ MDS code,
  \item [(ii)] If $k<n$, its Euclidean dual code $\operatorname{GRS}_k(\bm{a},\bm{v})^\perp$ is an $[n,n-k,k+1]_q$ MDS code,
  \item [(iii)] For a GRS code over $\F_{q^2}$ with $1\le k<n\le q^2$, its Hermitian dual has parameters $[n,n-k,k+1]_{q^2}$,
  \item [(iv)] A generator matrix of $\operatorname{GRS}_k(\bm{a},\bm{v})$ is
  $$
  \left(
    \begin{array}{cccc}
      v_1 & v_2 & \cdots & v_n \\
      v_1a_1 & v_2a_2 & \cdots & v_na_n \\
      v_1a_1^2 & v_2a_2^2 & \cdots & v_na_n^2 \\
      \vdots & \vdots & \ddots & \vdots \\
      v_1a_1^{k-1} & v_2a_2^{k-1} & \cdots & v_na_n^{k-1} \\
    \end{array}
  \right).
$$
\end{itemize}
\end{lemma}

We next recall basic definitions and results on additive codes.
\begin{definition}\label{def:additive-code}\cite{Bartoli2025,Delsarte1973}
Let $Q=p^e$, where $p$ is prime and $e\ge1$.
An additive code of length $n$ over $\F_Q$ is an additive subgroup $C$ of $\F_Q^n$, or equivalently, an $\F_p$-linear subspace of $\F_Q^n$.
The normalized dimension of $C$ is
\[
 \kappa=\log_Q |C|=\frac{\dim_p C}{e}.
\]
For $C\ne\{\bm{0}\}$, its minimum Hamming distance is
\[
 d=d(C)=\min_{\bm{c}\in C\setminus\{\bm{0}\}}\wt(\bm{c}).
\]
We denote the parameters of $C$ by $(n,M,d)_Q$, where $M=|C|$.
If $C$ is $\F_Q$-linear, then $\kappa=\dim_Q C$ and its parameters are written as $[n,\kappa,d]_Q$.
\end{definition}

Let $C\subseteq\F_Q^n$ be an additive code.
Let $n\geq1$ be an integer, and let $[n]\triangleq\{1,2,\ldots,n\}$.
For $T\subseteq[n]$, let $\pi_T: \F_Q^n\rightarrow\F_Q^{|T|}$ denote the coordinate projection onto $T$.
The punctured code obtained by deleting the coordinates in $T^c=[n]\backslash T$ is $C|_T=\pi_T(C)$.
Let
\[
 C_T=\{\bm{c}\in C:\supp(\bm{c})\subseteq T\}
\]
be the subcode of $C$ supported in $T$.
The shortened code obtained by deleting the coordinates in $T^c$ from the codewords of $C_T$ is denoted by $\operatorname{sh}_T(C)=\pi_T(C_T)$~(see \cite[Definition~20]{ball}).

Let $\F$ be a finite field. For a linear map $\varphi:U\to V$ between finite-dimensional vector spaces over $\mathbb F$, the rank-nullity formula states that
\begin{equation}\label{eq:rank-nullity}
 \dim_{\mathbb F}U=\dim_{\mathbb F}\ker\varphi+\rank_{\mathbb F}\varphi.
\end{equation}
In particular, the restriction $\pi_{T^c}|_C$ has kernel $C_T$ and image $\pi_{T^c}(C)$.
Since $\pi_T$ maps $C_T$ isomorphically onto $\operatorname{sh}_T(C)$, it follows from \eqref{eq:rank-nullity} that
\[
 \dim_p\operatorname{sh}_T(C)=\dim_p C_T=\dim_p C-\dim_p\pi_{T^c}(C).
\]

\subsection{Classical and Quantum $(r,\delta)$-LRCs}
We begin by defining classical codes with $(r,\delta)$-locality.
\begin{definition}\cite{galindo,prakash}
An $[n,k,d]_q$ linear code $C\subseteq\F_q^n$ is an $(r,\delta)$-LRC (or an LRC with locality $(r,\delta)$) if for each index $i\in [n]$, there exists a set of indices $\Gamma_i\subseteq[n]$ that contains $i$ and satisfies the following two conditions:
\begin{itemize}
  \item $\delta\leq|\Gamma_i|\leq r+\delta-1$,
  \item the minimum distance of the code $C|_{\Gamma_i}$ obtained by deleting code symbols $c_i$, $i\in [n]\backslash \Gamma_i$, is at least $\delta$.
\end{itemize}
\end{definition}
\begin{remark}
The sets $\Gamma_i$ form a repair cover.
If $C|_{\Gamma_i}$ is nonzero, then $\delta\le d(C|_{\Gamma_i})\le|\Gamma_i|$.
\end{remark}
For $(r,\delta)$-cLRCs, we recall the following Singleton-like bound from \cite{prakash}.
\begin{lemma}\label{bound:cLRC}
Let $C$ be an $(r,\delta)$-cLRC with parameters $[n,k,d]_q$. Then
\[
 d\le n-k+1-\left(\left\lceil\frac{k}{r}\right\rceil-1\right)(\delta-1).
\]
\end{lemma}

For standard definitions and notation concerning quantum stabilizer codes, we refer the reader to \cite{Ketkar2006,galindo}. Following \cite{galindo}, we recall the definitions and basic results for $(r,\delta)$-qLRCs.
\begin{definition}\label{de:qLRCs1}\cite[Definition~9]{galindo}
Given two sets of indices $\emptyset\neq I\varsubsetneq J\subseteq [n]$, a quantum code $\mathcal Q\subseteq\C^{q^n}$ is said to be an $(I,J)$-locally recoverable code if there exists a trace-preserving quantum operation $\mathcal R_{\mathcal Q,I}^J$, which acts only on the qudits corresponding to $J$ and leaves the remaining qudits unchanged, such that
$$\mathcal R_{\mathcal Q,I}^J\circ\Gamma^{I}(|\varphi\rangle\langle\varphi|)=|\varphi\rangle\langle\varphi|$$
for all $|\varphi\rangle\in \mathcal Q$, where $\Gamma^I$ is defined in \cite[p.~9]{galindo}.
\end{definition}
Using this definition, we introduce $(r,\delta)$-qLRCs as follows.
\begin{definition}\label{de:qLRCs2}\cite[Definition~10]{galindo}
A quantum code $\mathcal Q$ is said to be an $(r,\delta)$-qLRC if for each index $i\in[n]$, there exists a set $J\subseteq[n]$ containing $i$ with $|J|\leq r+\delta-1$ such that, for every subset $I\subseteq J$ of cardinality $\delta-1$, $\mathcal Q$ is $(I,J)$-locally recoverable.
\end{definition}
For $(r,\delta)$-qLRCs, we recall the Singleton-type bound in \cite[Theorem~31]{galindo}.
\begin{lemma}\cite[Theorem~31]{galindo}
An $(r,\delta)$-qLRC with parameters $[[n,k,\geq d(C)]]_q$ satisfies
\begin{equation}\label{qLRC:Singleton}
k+2d(C)+2\left(\left\lceil\frac{n+k}{2r}\right\rceil-1\right)(\delta-1)\leq n+2,
\end{equation}
where $C$ is a Euclidean or Hermitian dual-containing $(r,\delta)$-cLRC with parameters $[n,\frac{n+k}{2},d(C)]$.
\end{lemma}

\subsection{Pauli Label Space and Additive Symplectic Geometry}\label{sec:label-geometry}
Let $q=p^m$, where $p$ is prime and $m\ge1$ is an integer.
For a quantum register $X$, let $\HH_X$ denote its finite-dimensional complex Hilbert space.
The space of linear operators on $\HH_X$ is
\[
 \BB(\HH_X)=\{M\colon\HH_X\to\HH_X \mid M \text{ is } \C\text{-linear}\}.
\]
Let $I_X$ denote the identity operator on $\HH_X$.
For $M\in\BB(\HH_X)$, let $M^\dagger$ denote its adjoint and $\Tr(M)$ its trace.
A state of $X$ is represented by a density operator $\rho\in\BB(\HH_X)$ satisfying $\rho\succeq0$ and $\Tr(\rho)=1$.

Each physical coordinate $i\in[n]$ corresponds to a qudit with Hilbert space $\HH_i=\C^q$.
For $T\subseteq[n]$, write $\HH_T=\bigotimes_{i\in T}\HH_i$, with $\HH_\varnothing=\C$.
The joint Hilbert space of the $n$ qudits is $\HH_{[n]}=\bigotimes_{i=1}^n\HH_i=(\C^q)^{\otimes n}$.
For a state $\rho$ on $\HH_{[n]}$, its reduced state on $T^c=[n]\setminus T$ is $\rho_{T^c}=\Tr_T(\rho)$, where $\Tr_T$ denotes the partial trace over the qudits in $T$.

Let $\zeta_p=e^{2\pi\mathrm{i}/p}$, and let $\Tr_{\F_q/\F_p}$ denote the field trace from $\F_q$ to $\F_p$.
For $a,b\in\F_q$, define the single-qudit Pauli operators $X(a),Z(b)\colon\C^q\to\C^q$ by
\[
 X(a)|u\rangle=|u+a\rangle,~
 Z(b)|u\rangle=\zeta_p^{\Tr_{\F_q/\F_p}(bu)}|u\rangle,
\]
where $\{|u\rangle:u\in\F_q\}$ is the computational basis of $\C^q$.

\begin{definition}\cite{Ketkar2006}\label{def:pauli-labels}
Let $W=\F_q^n\times\F_q^n$.
For $\bm{w}=(\bm{x},\bm{z})\in W$, with $\bm{x}=(x_1,\ldots,x_n)$ and $\bm{z}=(z_1,\ldots,z_n)$, define the $n$-qudit Pauli operator
\[
 P_{\bm{w}}=\bigotimes_{i=1}^n X(x_i)Z(z_i)
 \colon \HH_{[n]}\longrightarrow\HH_{[n]}.
\]
For every $n$-qudit Pauli operator $U$, there is a unique $\bm{w}\in W$ such that $U=\xi P_{\bm{w}}$ for some $\xi\in\C$ with $|\xi|=1$.
We call $\bm{w}$ the \emph{Pauli label} of $U$ and $W$ the \emph{Pauli label space}.
\end{definition}

For $\bm{w}=(\bm{x},\bm{z})$ and $\bm{w}'=(\bm{x}',\bm{z}')$ in $W$, define the symplectic and trace-symplectic forms, respectively, by
\[
 \begin{aligned}
 [\bm{w},\bm{w}']_{\mathrm s}
 =\bm{x}\cdot\bm{z}'-\bm{z}\cdot\bm{x}',~
 [\bm{w},\bm{w}']_{\mathrm t}
 =\Tr_{\F_q/\F_p}\bigl([\bm{w},\bm{w}']_{\mathrm s}\bigr),
 \end{aligned}
\]
where $\bm{x}\cdot\bm{z}'=\sum_{i=1}^n x_i z_i'$.
The support and symplectic weight of $\bm{w}$ are
\[
 \supp(\bm{w})=\{i\in[n]:(x_i,z_i)\ne(0,0)\},~
 \swt(\bm{w})=|\supp(\bm{w})|.
\]

\begin{lemma}\cite{Ketkar2006}\label{lem:pauli-identities}
For $\bm{w}=(\bm{x},\bm{z})$ and $\bm{w}'=(\bm{x}',\bm{z}')$ in $W$, the following statements hold.
\begin{itemize}
  \item [(i)] The multiplication and commutation relations are
\begin{equation}\label{eq:pauli-commutation}
 \begin{aligned}
 P_{\bm{w}}P_{\bm{w}'}
 =\zeta_p^{\Tr_{\F_q/\F_p}(\bm{z}\cdot\bm{x}')}
   P_{\bm{w}+\bm{w}'},~
 P_{\bm{w}}P_{\bm{w}'}
 =\zeta_p^{-[\bm{w},\bm{w}']_{\mathrm t}}
   P_{\bm{w}'}P_{\bm{w}}.
 \end{aligned}
\end{equation}
  \item [(ii)] The set $\mathcal E_n=\{P_{\bm{w}}:\bm{w}\in W\}$ is a nice error basis on $\HH_{[n]}$,
  \item [(iii)] The operators in $\mathcal E_n$ form an orthogonal basis of $\BB(\HH_{[n]})$ and satisfy
   \begin{equation}\label{eq:pauli-trace}
   \Tr(P_{\bm{w}}^\dagger P_{\bm{w}'})=q^n\delta_{\bm{w},\bm{w}'}.
   \end{equation}
\end{itemize}
\end{lemma}

By~\eqref{eq:pauli-commutation}, two Pauli operators commute if and only if their labels are trace-symplectically orthogonal.
We now consider orthogonality for additive subspaces of $W$.

\begin{definition}\label{def:label-subspaces}
An \emph{additive subspace} of $W$ is an $\F_p$-linear subspace of $W$.
For an additive subspace $L\le W$, define its trace-symplectic and symplectic orthogonal complements, respectively, by
\[
 \begin{aligned}
 L^{\perpt}
 &=\{\bm{w}\in W:[\bm{w},\bm{v}]_{\mathrm t}=0
       \text{ for all }\bm{v}\in L\},\\
 L^{\perps}
 &=\{\bm{w}\in W:[\bm{w},\bm{v}]_{\mathrm s}=0
       \text{ for all }\bm{v}\in L\}.
 \end{aligned}
\]
We call $L$ \emph{isotropic} if $L\subseteq L^{\perpt}$.
The \emph{radical} of $L$ with respect to the trace-symplectic form is $\rad L=L\cap L^{\perpt}$.
\end{definition}

The trace-symplectic form on $W$ is alternating and nondegenerate over $\F_p$ (see~\cite[proof of Lemma~6]{alylp}).
Since $\dim_p W=2mn$, every additive subspace $L\le W$ satisfies
\begin{equation}\label{eq:orthogonal-complement}
 \dim_p L+\dim_p L^{\perpt}=2mn.
\end{equation}
If $L$ is $\F_q$-linear, then $L^{\perpt}=L^{\perps}$~\cite[proof of Theorem~2]{GalindoEA2019}.

For $T\subseteq[n]$, let $\pi_T\colon W\to\F_q^T\times\F_q^T$ denote the coordinate projection given by
\[
 \pi_T(\bm{x},\bm{z})=((x_i)_{i\in T},(z_i)_{i\in T}).
\]
Let $W_T=\{\bm{w}\in W:\supp(\bm{w})\subseteq T\}$ and $L_T=L\cap W_T$.
The puncturing and shortening of $L$ to $T$ are, respectively,
\[
 L|_T=\pi_T(L),~
 \operatorname{sh}_T(L)=\pi_T(L_T).
\]
Thus $\pi_T$ maps $L_T$ isomorphically onto $\operatorname{sh}_T(L)$; its inverse extends each label by zero outside $T$.

\begin{lemma}\label{lem:duality}
Let $L\le W$ be an additive subspace and let $T\subseteq[n]$.
By projection--shortening duality~\cite[Sec.~II-H]{ForneyGL2013} for the trace-symplectic form, we have
\[
 \operatorname{sh}_T(L^{\perpt})
 =\bigl(\pi_T(L)\bigr)^{\perpt},
\]
where the orthogonal complement on the right is taken in $\F_q^T\times\F_q^T$.
Consequently,
\[
 \dim_p(W_T\cap L^{\perpt})
 =2m|T|-\dim_p\pi_T(L).
\]
\end{lemma}

We next identify the Pauli label space $W$ with $\F_{q^2}^n$ to relate trace-symplectic orthogonality to Hermitian duality.
Fix $\alpha\in\F_{q^2}\setminus\F_q$.
Since $\{1,\alpha\}$ is an $\F_q$-basis of $\F_{q^2}$, the map
\begin{equation}\label{eq:psi}
 \psi\colon W\longrightarrow\F_{q^2}^n,~
 \psi(\bm{x},\bm{z})=\bm{x}+\alpha\bm{z},
\end{equation}
is an $\F_q$-linear isomorphism.
For each coordinate $i$, we have $x_i+\alpha z_i=0$ if and only if $(x_i,z_i)=(0,0)$.
Hence, for every $\bm{w}\in W$,
$$\supp(\psi(\bm{w}))=\supp(\bm{w}),~\wt(\psi(\bm{w}))=\swt(\bm{w}).$$
We also use $\psi$ for the corresponding map on coordinates in $T$, so that $\pi_T(\psi(\bm{w}))=\psi(\pi_T(\bm{w}))$.

\begin{lemma}\cite{Ketkar2006}\label{lem:hermitian-correspondence}
Let $D\le\F_{q^2}^n$ be an $\F_{q^2}$-linear code.
Under the map $\psi$ in~\eqref{eq:psi}, its Hermitian dual satisfies
\[
 D^{\perp_H}=\psi\bigl((\psi^{-1}(D))^{\perpt}\bigr).
\]
\end{lemma}

\subsection{Subsystem Stabilizer Codes}

A subsystem stabilizer code encodes information in a logical subsystem $A$.
Its stabilizer group determines the code space, on which gauge operators act only on the gauge subsystem $B$~\cite{poulin}.

Let $G$ be an additive subspace of the Pauli label space $W$, and let $\mathcal G$ be the group of all Pauli operators whose labels lie in $G$.
We call $\mathcal G$ the \emph{gauge group} and $G$ its \emph{gauge label space}.
Set
\begin{equation}\label{eq:radical}
 S=G\cap G^{\perpt}.
\end{equation}
By~\eqref{eq:pauli-commutation}, $S$ is the label space of the center $Z(\mathcal G)$.
Choose a subgroup $\mathcal S\le Z(\mathcal G)$ with label space $S$ whose only scalar operator is the identity.
We call $\mathcal S$ the \emph{stabilizer group} and denote its common $+1$ eigenspace by
\[
 \mathcal Q
 =\{|\psi\rangle\in\HH_{[n]}:
      U|\psi\rangle=|\psi\rangle
      \text{ for every }U\in\mathcal S\}.
\]

\begin{definition}\cite{poulin,alylp}\label{def:ssc}
A \emph{subsystem stabilizer code} (SSC) is the code space $\mathcal Q$ together with the decomposition $\mathcal Q\simeq\HH_A\otimes\HH_B$ induced by $\mathcal G$.
In this decomposition, each $U\in\mathcal G$ acts on $\mathcal Q$ as $I_A\otimes U_B$ for some unitary $U_B$ on $\HH_B$, and the operators $U_B$ act irreducibly on $\HH_B$.
We call $A$ the \emph{logical subsystem} and $B$ the \emph{gauge subsystem}.
\end{definition}

Let $\mathcal Q\simeq\HH_A\otimes\HH_B$ be an SSC with gauge label space $G$, stabilizer label space $S$, and stabilizer group $\mathcal S$.
Write $h=\dim_p G$, $s=\dim_p S$, $K=\dim\HH_A$, and $R=\dim\HH_B$.
Let $P$ be the orthogonal projector onto $\mathcal Q$, and
let $V\colon\HH_A\otimes\HH_B\to\HH_{[n]}$ be an encoding isometry for this decomposition.
Then $V^\dagger V=I_{AB}$ and $VV^\dagger=P$.
The corresponding encoding channel $\mathcal V\colon\BB(\HH_A\otimes\HH_B)\to\BB(\HH_{[n]})$ is defined by
$\mathcal V(M)=VMV^\dagger$, where $M\in\BB(\HH_A\otimes\HH_B)$.

\begin{lemma}\label{lem:ssc-representation}\cite{alylp,Ketkar2006,poulin}
With the above notation, the following statements hold.
\begin{itemize}
  \item [(i)] The logical and gauge dimensions satisfy
\begin{equation}\label{eq:dimensions}
 K=p^{mn-(h+s)/2},~R=p^{(h-s)/2}.
\end{equation}
  \item [(ii)] The projector $P$ satisfies
\begin{equation}\label{eq:stabilizer-projector}
 P=p^{-s}\sum_{U\in\mathcal S}U,~\Tr(P)=p^{mn-s}=KR.
\end{equation}
  \item [(iii)] For each $\bm{w}\in W$,
\begin{equation}\label{eq:encoded-pauli}
 \begin{aligned}
 V^\dagger P_{\bm{w}}V
 &=0
 &&\text{if }\bm{w}\notin S^{\perpt},\\
 V^\dagger P_{\bm{w}}V
 &=I_A\otimes U_{\bm{w},B}
 &&\text{if }\bm{w}\in G,\\
 V^\dagger P_{\bm{w}}V
 &\notin I_A\otimes\BB(\HH_B)
 &&\text{if }\bm{w}\in S^{\perpt}\setminus G,
 \end{aligned}
\end{equation}
where $U_{\bm{w},B}$ is unitary on $\HH_B$.
\end{itemize}
\end{lemma}
\begin{remark}
Since $S=\rad G$, the induced trace-symplectic form on $G/S$ is nondegenerate, and by~\eqref{eq:dimensions}, we have $\dim_p(G/S)=h-s=2\log_p R$.
Moreover, Lemma~\ref{lem:ssc-representation}(iii) states the Pauli error classification of~\cite{poulin} using the additive notation of~\cite[Theorem~1]{alylp}.
\end{remark}

The encoding channel is $\cV(M)=VMV^\dagger$ for $M\in\BB(\HH_A\otimes\HH_B)$.
The encoded subsystems $A$ and $B$ need not correspond to disjoint sets of physical coordinates.
By~\eqref{eq:encoded-pauli}, the labels in $S^{\perpt}\setminus G$ are precisely those of Pauli operators that preserve $\mathcal Q$ and act nontrivially on $A$.
Thus, for $K>1$, the global minimum distance is~\cite[Theorem~1]{alylp}
\begin{equation}\label{eq:distance}
 d=\min\{\swt(\bm{w}):\bm{w}\in S^{\perpt}\setminus G\}.
\end{equation}

We denote the parameters of $\mathcal Q$ by $((n,K,R,d))_q$.
When $K=q^k$ and $R=q^g$ for nonnegative integers $k$ and $g$, we write $[[n,k,g,d]]_q$.
In particular, if $G$ is $\F_q$-linear, then it follows from~\eqref{eq:dimensions} that
\begin{equation}\label{eq:qdimensions}
 k=n-\frac{\dim_qG+\dim_qS}{2},~
 g=\frac{\dim_qG-\dim_qS}{2}.
\end{equation}
If $R=1$, then $G=S$ and $\mathcal Q$ is an ordinary stabilizer code with parameters $((n,K,d))_q$, or $[[n,k,d]]_q$ when $K=q^k$.

\begin{definition}\cite{alylp}\label{def:purity}
An SSC with gauge label space $G$ and minimum distance $d$ is \emph{pure} if $\swt(\bm{w})\ge d$ for every nonzero $\bm{w}\in G$.
Otherwise, it is \emph{impure}.
\end{definition}

For a pure SSC with $K>1$, every nonzero label in $G$ has symplectic weight at least $d$, and by~\eqref{eq:distance}, the minimum weight in $S^{\perpt}$ is $d$.
Hence
\[
 d=\min\{\swt(\bm{w}):\bm{w}\in S^{\perpt}\setminus\{\bm{0}\}\}.
\]

\subsection{Global Erasure Recovery}

A quantum channel is a completely positive trace-preserving (CPTP) linear map between operator spaces~\cite[Definition~2.13]{watrous}.
We write $\id_X$ for the identity channel on $\BB(\HH_X)$.
For a known erased set $E\subseteq[n]$, the erasure channel is the partial trace $\Tr_E$.
In global recovery, the decoder has access to all surviving qudits in $E^c$ and returns a state on the full set of coordinates $[n]$.

We use the definition of subsystem correctability in~\cite[Definition~3.1]{kribs} to formulate recovery from the erasure of a known set of coordinates.

\begin{definition}\label{def:global-recovery}
Let $\cV$ be the encoding channel of an SSC $\mathcal Q$ with logical subsystem $A$ and gauge subsystem $B$.
A known erased set $E\subseteq[n]$ is \emph{globally correctable} if there exist CPTP maps
\[
 \mathcal R_E\colon\BB(\HH_{E^c})\longrightarrow\BB(\HH_{[n]}),~
 \Phi_E\colon\BB(\HH_B)\longrightarrow\BB(\HH_B)
\]
such that
\begin{equation}\label{eq:global-recovery-channel}
 \mathcal R_E\circ\Tr_E\circ\cV
 =\cV\circ(\id_A\otimes\Phi_E).
\end{equation}
\end{definition}

The recovery channel $\mathcal R_E$ has access to all surviving qudits in $E^c$.
For each fixed $E$, the same maps $\mathcal R_E$ and $\Phi_E$ satisfy~\eqref{eq:global-recovery-channel} for every joint input state of $A$ and $B$.
The channel identity also holds after tensoring both sides with $\id_F$ for any reference register $F$.
Thus, recovery returns the output to $\mathcal Q$ and preserves the joint state of $A$ and $F$, while allowing the gauge state to change.

By the subsystem Knill-Laflamme condition~\cite[Theorem~3.5]{kribs} and~\eqref{eq:encoded-pauli}, we obtain the following criterion for a known erasure.

\begin{lemma}\label{lem:global}
Let $\mathcal Q$ be an SSC with gauge label space $G$ and stabilizer label space $S$.
A known erased set $E\subseteq[n]$ is globally correctable if and only if
\begin{equation}\label{eq:global}
 W_E\cap S^{\perpt}=G_E.
\end{equation}
By Lemma~\ref{lem:duality}, this condition is equivalent to
\begin{equation}\label{eq:globalrank}
 \dim_p\pi_E(S)+\dim_p G_E=2m|E|.
\end{equation}
\end{lemma}

\begin{remark}
If $K>1$ and $|E|<d$, then each label in $W_E$ has symplectic weight less than $d$.
By~\eqref{eq:distance}, no such label belongs to $S^{\perpt}\setminus G$.
Hence $E$ is globally correctable by Lemma~\ref{lem:global}.
\end{remark}

\section{Local Recovery and Algebraic Characterization}\label{sec:recovery}
In this section, we characterize local erasure recovery for $(r,\delta)$-SSLRCs without a purity assumption.
We also specialize the criterion to CSS and Hermitian hull constructions.

\subsection{Local Recovery and $(r,\delta)$-Locality}
Let $\emptyset\ne E\subseteq\Gamma\subseteq[n]$, and set $H=\Gamma\setminus E$ and $O=\Gamma^c$.
Recovery within $\Gamma$ has the same correction requirement as global recovery in Definition~\ref{def:global-recovery}, but the decoder may access only the surviving qudits in $H$.

\begin{definition}\label{def:local}
Let $\cV$ be the encoding channel of an SSC $\mathcal Q$ with logical subsystem $A$ and gauge subsystem $B$.
The erasure of $E$ is \emph{recoverable within $\Gamma$} if there exist CPTP maps
\[
 \mathcal R_{\Gamma,E}\colon\BB(\HH_H)\longrightarrow\BB(\HH_\Gamma),~
 \Phi_{\Gamma,E}\colon\BB(\HH_B)\longrightarrow\BB(\HH_B)
\]
such that
\begin{equation}\label{eq:recovery}
 (\mathcal R_{\Gamma,E}\otimes\id_O)\circ\Tr_E\circ\cV
 =\cV\circ(\id_A\otimes\Phi_{\Gamma,E}).
\end{equation}
The maps may depend on $\Gamma$ and $E$, but the same pair must apply to every joint input state of $A$ and $B$.
\end{definition}

\begin{remark}
For every joint input state of the logical subsystem $A$, the gauge subsystem $B$, and a reference register $F$, equation~\eqref{eq:recovery} ensures that recovery within $\Gamma$ preserves the reduced state on $AF$.
When $R=1$, the gauge subsystem is one-dimensional, so recovery within $\Gamma$ must restore the entire encoded state, as in the stabilizer recovery model of~\cite{galindo}.
\end{remark}

We now define $(r,\delta)$-SSLRCs in terms of local erasure recovery.

\begin{definition}\label{def:lrc}
Let $r\ge1$ and $\delta\ge2$ be integers.
An SSC $\mathcal Q$ of length $n$ is a \emph{subsystem stabilizer $(r,\delta)$-locally recoverable code} (\emph{$(r,\delta)$-SSLRC}) if, for each $i\in[n]$, there exists a repair set $\Gamma_i\subseteq[n]$ containing $i$ such that
\begin{equation}\label{eq:groupsize}
 \delta\le|\Gamma_i|\le r+\delta-1
\end{equation}
and every nonempty erased set $E\subseteq\Gamma_i$ with $|E|<\delta$ is recoverable within $\Gamma_i$ in the sense of Definition~\ref{def:local}.
The family $\{\Gamma_i:i\in[n]\}$ is called a \emph{repair cover}.
Repair sets for different coordinates may overlap or coincide.
\end{definition}

The erased set $E$ need not contain $i$.
For $e=|E|<\delta$, the number of surviving qudits in $\Gamma_i$ is
\[
 |\Gamma_i\setminus E|=|\Gamma_i|-e\le r+\delta-1-e.
\]
In particular, when $e=\delta-1$, at most $r$ qudits in $\Gamma_i$ are available for recovery.

\subsection{An Algebraic Characterization for Local Recovery}

In this subsection, we characterize local erasure recovery in SSCs without assuming purity (Theorem~\ref{thm:local}).
We first establish two lemmas on the correction operators and the local stabilizer measurement needed for Proposition~\ref{lem:local-syndrome}.

For the next two lemmas, let $\mathcal Q$ be an SSC with gauge label space $G$, stabilizer label space $S$, and stabilizer group $\mathcal S$.
Let $\varnothing\ne E\subseteq\Gamma\subseteq[n]$, and set $H=\Gamma\setminus E$ and $O=\Gamma^c$.
Let $V\colon\HH_A\otimes\HH_B\to\HH_{[n]}$ be an encoding isometry for $\mathcal Q$, with associated encoding channel $\cV$.
Let $\nu=\dim_p S_\Gamma$, and choose an $\F_p$-basis $\bm{s}_1,\ldots,\bm{s}_\nu$ of $S_\Gamma$.
For each $j$, let $U_j\in\mathcal S$ be the stabilizer operator with label $\bm{s}_j$.
Define the syndrome map by
\[
 \chi\colon W_E\longrightarrow\F_p^\nu,~
 \chi(\bm{a})
 =\bigl([\bm{a},\bm{s}_1]_{\mathrm t},\ldots,
        [\bm{a},\bm{s}_\nu]_{\mathrm t}\bigr).
\]

The decoder selects a correction operator according to the measured syndrome. The following lemma provides a sufficient condition under which this operator restores the logical subsystem for all Pauli errors with the same syndrome.

\begin{lemma}\label{lem:syndrome-correction}

With the above notation, suppose that
$
 W_E\cap S_\Gamma^{\perpt}=G_E.
$
Then $\chi$ is $\F_p$-linear and $\ker\chi=G_E$.
For each $\bm{y}\in\chi(W_E)$, choose a representative label $\bm{a}_{\bm{y}}\in W_E$ with $\chi(\bm{a}_{\bm{y}})=\bm{y}$.
For every $\bm{a}\in W_E$, the residual operator
$
 D_{\bm{a}}
 :=P_{\bm{a}_{\chi(\bm{a})}}^\dagger P_{\bm{a}}
$
satisfies
$$
 D_{\bm{a}}V=V(I_A\otimes U_{\bm{a},B})
$$
for some unitary $U_{\bm{a},B}$ on $\HH_B$.
\end{lemma}

\begin{proof}
Let $\bm{s}_j=(\bm{x}_j,\bm{z}_j)$.
Then
\[
 U_j=\xi_j\bigotimes_{i=1}^n X(x_{j,i})Z(z_{j,i})
\]
for some scalar phase $\xi_j$.
Since $\supp(\bm{s}_j)\subseteq\Gamma$, we have $(x_{j,i},z_{j,i})=(0,0)$ for every $i\in O=\Gamma^c$.
Thus each factor on $O$ is $X(0)Z(0)=I_i$, and $U_j=\widehat U_j\otimes I_O$, where
\[
 \widehat U_j
 =\xi_j\bigotimes_{i\in\Gamma}X(x_{j,i})Z(z_{j,i}),~
 I_O=\bigotimes_{i\in O}I_i.
\]
For every $|\varphi\rangle\in\HH_A\otimes\HH_B$, we have $V|\varphi\rangle\in\mathcal Q$.
Since $U_j\in\mathcal S$, we have $U_jV|\varphi\rangle=V|\varphi\rangle$ for every $|\varphi\rangle$, and hence $U_jV=V$.
For every $\bm{a}\in W_E$, it follows from~\eqref{eq:pauli-commutation} that
\begin{equation}\label{eq:(18)}
 \begin{aligned}
 U_jP_{\bm{a}}V
 &=\xi_jP_{\bm{s}_j}P_{\bm{a}}V\\
 &=\zeta_p^{-[\bm{s}_j,\bm{a}]_{\mathrm t}}
   P_{\bm{a}}\xi_jP_{\bm{s}_j}V\\
 &=\zeta_p^{-[\bm{s}_j,\bm{a}]_{\mathrm t}}
   P_{\bm{a}}U_jV\\
 &=\zeta_p^{[\bm{a},\bm{s}_j]_{\mathrm t}}
   P_{\bm{a}}V,
 \end{aligned}
\end{equation}
where the last equality follows from $U_jV=V$ and $-[\bm{s}_j,\bm{a}]_{\mathrm t}=[\bm{a},\bm{s}_j]_{\mathrm t}$.
Let $|\psi\rangle=V|\varphi\rangle$ be any unit vector in $\mathcal Q$.
After the error $P_{\bm{a}}$, the state becomes $P_{\bm{a}}|\psi\rangle$. By~\eqref{eq:(18)},
\[
 U_j\bigl(P_{\bm{a}}|\psi\rangle\bigr)
 =\zeta_p^{[\bm{a},\bm{s}_j]_{\mathrm t}}
  \bigl(P_{\bm{a}}|\psi\rangle\bigr).
\]
Thus $P_{\bm{a}}|\psi\rangle$ is an eigenvector of $U_j$ with eigenvalue $\zeta_p^{[\bm{a},\bm{s}_j]_{\mathrm t}}$.
For fixed $j$, the measurement outcome depends only on the error label $\bm{a}$ and is the same for every encoded state.

The map $\chi$ is $\F_p$-linear.
Since $\bm{s}_1,\ldots,\bm{s}_\nu$ span $S_\Gamma$, the hypothesis implies
$
 \ker\chi=W_E\cap S_\Gamma^{\perpt}=G_E.
$
Consequently, for $\bm{a},\bm{a}'\in W_E$,
\[
 \chi(\bm{a})=\chi(\bm{a}')
 \Longleftrightarrow
 \bm{a}-\bm{a}'\in G_E.
\]
Thus the local stabilizer measurements distinguish errors with different syndromes, whereas errors with the same syndrome differ only by a gauge operator.

For each measured syndrome $\bm{y}$, we next specify a correction operator and verify that it restores the logical subsystem for every Pauli error with that syndrome.
For each $\bm{y}\in\chi(W_E)$, choose a representative label $\bm{a}_{\bm{y}}\in W_E$ with $\chi(\bm{a}_{\bm{y}})=\bm{y}$.
For the measured syndrome $\bm{y}$, the decoder applies $P_{\bm{a}_{\bm{y}}}^\dagger$, where $P_{\bm{a}_{\bm{y}}}^\dagger=P_{\bm{a}_{\bm{y}}}^{-1}$.
For $\bm{y}\in\F_p^\nu\setminus\chi(W_E)$, set $\bm{a}_{\bm{y}}=\bm{0}$; the corresponding correction operator is then $P_{\bm{0}}^\dagger=I$.
Now suppose that the actual error is $P_{\bm{a}}$, where $\bm{a}\in W_E$, and let $\bm{y}=\chi(\bm{a})$.
After the decoder applies the correction operator, the residual operator is
\[
 D_{\bm{a}}
 :=P_{\bm{a}_{\bm{y}}}^\dagger P_{\bm{a}}.
\]
If $\bm{a}=\bm{a}_{\bm{y}}$, then $D_{\bm{a}}=I$.
If $\bm{a}\ne\bm{a}_{\bm{y}}$, write $\bm{a}=(\bm{x},\bm{z})$ and $\bm{a}_{\bm{y}}=(\bm{x}_{\bm{y}},\bm{z}_{\bm{y}})$.
By~\eqref{eq:pauli-commutation},
\[
 \begin{aligned}
 P_{\bm{a}_{\bm{y}}}P_{\bm{a}-\bm{a}_{\bm{y}}}
 &=\zeta_p^{\Tr_{\F_q/\F_p}
   (\bm{z}_{\bm{y}}\cdot(\bm{x}-\bm{x}_{\bm{y}}))}
   P_{\bm{a}_{\bm{y}}+(\bm{a}-\bm{a}_{\bm{y}})}\\
 &=\zeta_p^{\Tr_{\F_q/\F_p}
   (\bm{z}_{\bm{y}}\cdot(\bm{x}-\bm{x}_{\bm{y}}))}
   P_{\bm{a}}.
 \end{aligned}
\]
Multiplying both sides on the left by $P_{\bm{a}_{\bm{y}}}^\dagger$ and using $P_{\bm{a}_{\bm{y}}}^\dagger P_{\bm{a}_{\bm{y}}}=I$, we obtain
\[
 D_{\bm{a}}
 =P_{\bm{a}_{\bm{y}}}^\dagger P_{\bm{a}}
 =\zeta_p^{-\Tr_{\F_q/\F_p}
   (\bm{z}_{\bm{y}}\cdot(\bm{x}-\bm{x}_{\bm{y}}))}
   P_{\bm{a}-\bm{a}_{\bm{y}}}.
\]
Thus $D_{\bm{a}}$ has label $\bm{a}-\bm{a}_{\bm{y}}$.
By definition, $\bm{y}=\chi(\bm{a})$, and the choice of $\bm{a}_{\bm{y}}$ ensures that $\chi(\bm{a}_{\bm{y}})=\bm{y}$.
By the linearity of $\chi$,
\[
 \chi(\bm{a}-\bm{a}_{\bm{y}})
 =\chi(\bm{a})-\chi(\bm{a}_{\bm{y}})
 =\bm{y}-\bm{y}
 =\bm{0}.
\]
Hence $\bm{a}-\bm{a}_{\bm{y}}\in\ker\chi=G_E$.
Since $\bm{a}-\bm{a}_{\bm{y}}$ is the label of $D_{\bm{a}}$, it follows that $D_{\bm{a}}$ is a gauge operator supported on $E$.

Let $U\in\mathcal S$ have label $\bm{s}$.
Since $\bm{s}\in S\subseteq G^{\perpt}$ and $\bm{a}-\bm{a}_{\bm{y}}\in G$, we have $[\bm{s},\bm{a}-\bm{a}_{\bm{y}}]_{\mathrm t}=0$.
Hence $UD_{\bm{a}}=D_{\bm{a}}U$ by~\eqref{eq:pauli-commutation}.
For every $|\psi\rangle\in\mathcal Q$, we have
\[
 UD_{\bm{a}}|\psi\rangle
 =D_{\bm{a}}U|\psi\rangle
 =D_{\bm{a}}|\psi\rangle,
\]
which implies that $D_{\bm{a}}|\psi\rangle\in\mathcal Q$.
For every $|\varphi\rangle\in\HH_A\otimes\HH_B$, we have $V|\varphi\rangle=|\psi\rangle\in\mathcal Q$.
Then $D_{\bm{a}}V|\varphi\rangle\in\mathcal Q$.
Hence $PD_{\bm{a}}V=D_{\bm{a}}V$, where $P=VV^\dagger$ is the orthogonal projector onto $\mathcal Q$.
Since $\bm{a}-\bm{a}_{\bm{y}}\in G_E\subseteq G$, it follows from~\eqref{eq:encoded-pauli} that
\[
 V^\dagger D_{\bm{a}}V=I_A\otimes U_{\bm{a},B}
\]
for some unitary $U_{\bm{a},B}$ on $\HH_B$.
Therefore,
\begin{equation}\label{eq:local-residual-gauge}
 \begin{aligned}
 D_{\bm{a}}V
 =PD_{\bm{a}}V
 =V(V^\dagger D_{\bm{a}}V)
 =V(I_A\otimes U_{\bm{a},B}).
 \end{aligned}
\end{equation}
Hence the correction operator $P_{\bm{a}_{\bm{y}}}^\dagger$ returns every encoded state affected by an error of syndrome $\bm{y}$ to $\mathcal Q$ and leaves the logical subsystem unchanged.
\end{proof}

Lemma~\ref{lem:syndrome-correction} specifies a correction operator for each syndrome under the stated hypothesis.
{We next construct a projective measurement on $\Gamma$ that determines the syndrome.
This construction does not require the condition $W_E\cap S_\Gamma^{\perpt}=G_E$.

\begin{lemma}
\label{lem:syndrome-measurement}
For the SSC and coordinate sets fixed above, there is a projective measurement $\{\Pi_{\bm{y}}:\bm{y}\in\F_p^\nu\}$ on $\HH_\Gamma$ such that
\[
 (\Pi_{\bm{y}}\otimes I_O)P_{\bm{a}}V
 =\delta_{\bm{y},\chi(\bm{a})}P_{\bm{a}}V,
\]
where $\bm{a}\in W_E$ and $\bm{y}\in\F_p^\nu$. For every encoded state affected by $P_{\bm{a}}$, this measurement returns the syndrome $\chi(\bm{a})$ with probability $1$.
\end{lemma}

\begin{proof}
Let $U_j=\widehat U_j\otimes I_O$, where $\widehat U_j$ acts on $\HH_\Gamma$.
For any $j,k\in\{1,\ldots,\nu\}$, the stabilizer operators $U_j$ and $U_k$ commute.
Since $U_j=\widehat U_j\otimes I_O$ and $U_k=\widehat U_k\otimes I_O$, we have
\[
 (\widehat U_j\widehat U_k)\otimes I_O
 =U_jU_k
 =U_kU_j
 =(\widehat U_k\widehat U_j)\otimes I_O.
\]
Hence $\widehat U_j\widehat U_k=\widehat U_k\widehat U_j$.
By~\eqref{eq:pauli-commutation}, the label of $U_j^p$ is $p\bm{s}_j=\bm{0}$.
Thus $U_j^p$ is a scalar operator in $\mathcal S$.
Since the identity is the only scalar operator in $\mathcal S$, we obtain
\[
 \widehat U_j^{\,p}\otimes I_O
 =(\widehat U_j\otimes I_O)^{\,p}
 =U_j^p
 =I_\Gamma\otimes I_O.
\]
Therefore $\widehat U_j^{\,p}=I_\Gamma$, and every eigenvalue of $\widehat U_j$ has the form $\zeta_p^c$ for some $c\in\F_p$.

For each $j\in\{1,\ldots,\nu\}$ and $y_j\in\F_p$, we construct the orthogonal projector onto the eigenspace with eigenvalue $\zeta_p^{y_j}$.
Define
\[
 Q_{j,y_j}
 =\frac1p\sum_{b=0}^{p-1}
   (\zeta_p^{-y_j}\widehat U_j)^b
 =\frac1p\sum_{b=0}^{p-1}
   \zeta_p^{-by_j}\widehat U_j^{\,b}.
\]
Let $|v\rangle$ be an eigenvector satisfying $\widehat U_j|v\rangle=\zeta_p^c|v\rangle$, where $c\in\F_p$.
Then
\[
 \begin{aligned}
 Q_{j,y_j}|v\rangle
 &=\frac1p\sum_{b=0}^{p-1}
   \zeta_p^{-by_j}\widehat U_j^{\,b}|v\rangle\\
 &=\left(\frac1p\sum_{b=0}^{p-1}
   \zeta_p^{b(c-y_j)}\right)|v\rangle.
 \end{aligned}
\]
If $c=y_j$, the coefficient in parentheses equals $1$.
If $c\ne y_j$, then $\zeta_p^{c-y_j}\ne1$, and
\[
 \frac1p\sum_{b=0}^{p-1}\zeta_p^{b(c-y_j)}
 =\frac{1-\zeta_p^{p(c-y_j)}}{p(1-\zeta_p^{c-y_j})}
 =0.
\]
Consequently,
\[
 Q_{j,y_j}|v\rangle
 =\begin{cases}
   |v\rangle,&c=y_j,\\
   0,&c\ne y_j.
  \end{cases}
\]
Since $\widehat U_j$ is unitary, the spectral theorem~\cite[Theorem~1.3]{watrous} implies that its eigenspaces form an orthogonal decomposition of $\HH_\Gamma$.
Hence $Q_{j,y_j}$ is the required orthogonal projector.
For each fixed $j$, we therefore have
\[
 Q_{j,c}^\dagger=Q_{j,c},~
 Q_{j,c}Q_{j,c'}=\delta_{c,c'}Q_{j,c},~
 \sum_{c\in\F_p}Q_{j,c}=I_\Gamma.
\]
Each $Q_{j,c}$ is a polynomial in $\widehat U_j$.
Since $\widehat U_1,\ldots,\widehat U_\nu$ commute, we have $Q_{j,c}Q_{k,c'}=Q_{k,c'}Q_{j,c}$ for all $j,k\in\{1,\ldots,\nu\}$ and $c,c'\in\F_p$.

For $\bm{y}=(y_1,\ldots,y_\nu)\in\F_p^\nu$, define
\[
 \Pi_{\bm{y}}
 =\prod_{j=1}^{\nu}Q_{j,y_j}
 =\prod_{j=1}^{\nu}
   \left(\frac1p\sum_{b=0}^{p-1}
   \zeta_p^{-by_j}\widehat U_j^{\,b}\right).
\]
Since the factors commute and are self-adjoint, we have $\Pi_{\bm{y}}^\dagger=\Pi_{\bm{y}}$.
Moreover,
\[
 \begin{aligned}
 \Pi_{\bm{y}}\Pi_{\bm{y}'}
 &=\prod_{j=1}^{\nu}
   \bigl(Q_{j,y_j}Q_{j,y_j'}\bigr)\\
 &=\left(\prod_{j=1}^{\nu}\delta_{y_j,y_j'}\right)
   \prod_{j=1}^{\nu}Q_{j,y_j}\\
 &=\delta_{\bm{y},\bm{y}'}\Pi_{\bm{y}},
 \end{aligned}
\]
and
\[
 \begin{aligned}
 \sum_{\bm{y}\in\F_p^\nu}\Pi_{\bm{y}}
 &=\sum_{y_1\in\F_p}\sum_{y_\nu\in\F_p}
   Q_{1,y_1}\cdots Q_{\nu,y_\nu}\\
 &=\prod_{j=1}^{\nu}
   \left(\sum_{y_j\in\F_p}Q_{j,y_j}\right)\\
 &=I_\Gamma.
 \end{aligned}
\]
Thus $\{\Pi_{\bm{y}}:\bm{y}\in\F_p^\nu\}$ defines a projective measurement on $\Gamma$.
Since the factors are commuting orthogonal projectors, we also have
\[
 \begin{aligned}
 \operatorname{im}\Pi_{\bm{y}}
 &=\bigcap_{j=1}^{\nu}\operatorname{im}Q_{j,y_j}\\
 &=\left\{
 |v\rangle\in\HH_\Gamma:
 \widehat U_j|v\rangle=\zeta_p^{y_j}|v\rangle
 \text{ for all }j=1,\ldots,\nu
 \right\}.
 \end{aligned}
\]
We label the joint measurement outcome $(\zeta_p^{y_1},\ldots,\zeta_p^{y_\nu})$ by $\bm{y}$.
For $\nu=0$, we use the empty-product convention $\Pi_{\bm{0}}=I_\Gamma$, where $\bm{0}$ is the unique element of $\F_p^0$.

It remains to verify that this measurement returns $\chi(\bm{a})$ for every $\bm{a}\in W_E$.
Since $U_jV=V$, it follows from~\eqref{eq:pauli-commutation} that
\[
 (\widehat U_j\otimes I_O)P_{\bm{a}}V
 =\zeta_p^{[\bm{a},\bm{s}_j]_{\mathrm t}}P_{\bm{a}}V.
\]
By induction on $b$ with $0\le b<p$, we obtain
\[
 (\widehat U_j^{\,b}\otimes I_O)P_{\bm{a}}V
 =\zeta_p^{b[\bm{a},\bm{s}_j]_{\mathrm t}}P_{\bm{a}}V.
\]
By the definition of $Q_{j,y_j}$, we have
\[
 \begin{aligned}
 (Q_{j,y_j}\otimes I_O)P_{\bm{a}}V
 &=\frac1p\sum_{b=0}^{p-1}
   \zeta_p^{-by_j}
   (\widehat U_j^{\,b}\otimes I_O)P_{\bm{a}}V\\
 &=\left(\frac1p\sum_{b=0}^{p-1}
   \zeta_p^{b([\bm{a},\bm{s}_j]_{\mathrm t}-y_j)}\right)
   P_{\bm{a}}V\\
 &=\delta_{y_j,[\bm{a},\bm{s}_j]_{\mathrm t}}P_{\bm{a}}V,
 \end{aligned}
\]
where the last equality follows from the geometric sum computed above.
By the definitions of $\Pi_{\bm{y}}$ and $\chi$, we obtain
\begin{equation}\label{eq:syndrome-projection}
 \begin{aligned}
 (\Pi_{\bm{y}}\otimes I_O)P_{\bm{a}}V
 &=\left(\prod_{j=1}^{\nu}
   \delta_{y_j,[\bm{a},\bm{s}_j]_{\mathrm t}}\right)
   P_{\bm{a}}V\\
 &=\delta_{\bm{y},\chi(\bm{a})}P_{\bm{a}}V.
 \end{aligned}
\end{equation}
For any unit vector $|\varphi\rangle\in\HH_A\otimes\HH_B$, the probability of outcome $\bm{y}$ on the state $P_{\bm{a}}V|\varphi\rangle$ is
\[
 \begin{aligned}
 \Pr(\bm{y})
 &=\bigl\|(\Pi_{\bm{y}}\otimes I_O)
   P_{\bm{a}}V|\varphi\rangle\bigr\|^2\\
 &=\delta_{\bm{y},\chi(\bm{a})}
   \bigl\|P_{\bm{a}}V|\varphi\rangle\bigr\|^2\\
 &=\delta_{\bm{y},\chi(\bm{a})},
 \end{aligned}
\]
since $V$ is an isometry and $P_{\bm{a}}$ is unitary.
By linearity in the density operator, the same probability formula holds for mixed encoded states.
Thus, for any encoded state after the error $P_{\bm{a}}$, the projective measurement $\{\Pi_{\bm{y}}:\bm{y}\in\F_p^\nu\}$ on $\Gamma$ returns the syndrome $\bm{y}=\chi(\bm{a})$ with probability $1$.
\end{proof}

By Lemmas~\ref{lem:syndrome-correction} and~\ref{lem:syndrome-measurement}, the measurement outcome $\chi(\bm{a})$ determines the correction operator $P_{\bm{a}_{\chi(\bm{a})}}^\dagger$ whenever $W_E\cap S_\Gamma^{\perpt}=G_E$.
We now use this measurement and the corresponding correction operators to construct a recovery channel for the erasure of $E$.

\begin{proposition}\label{lem:local-syndrome}
Let $\mathcal Q$ be an SSC with gauge label space $G$ and stabilizer label space $S$.
Let $\emptyset\ne E\subseteq\Gamma\subseteq[n]$.
If
$
 W_E\cap S_\Gamma^{\perpt}=G_E,
$
then the erasure of $E$ is recoverable within $\Gamma$.
\end{proposition}

\begin{proof}
Let $H=\Gamma\setminus E$ and $O=\Gamma^c$.
Let $V\colon\HH_A\otimes\HH_B\to\HH_{[n]}$ be an encoding isometry for $\mathcal Q$, with associated encoding channel $\cV$.
We construct a CPTP recovery map $\mathcal R_{\Gamma,E}\colon\BB(\HH_H)\to\BB(\HH_\Gamma)$ and a CPTP gauge map $\Phi_{\Gamma,E}\colon\BB(\HH_B)\to\BB(\HH_B)$ such that~\eqref{eq:recovery} holds.

Choose the basis of $S_\Gamma$ and the syndrome map $\chi$ as above.
By Lemma~\ref{lem:syndrome-correction}, choose the representative labels $\bm{a}_{\bm{y}}$ for $\bm{y}\in\chi(W_E)$ and the unitaries $U_{\bm{a},B}$ satisfying~\eqref{eq:local-residual-gauge}.
Set $\bm{a}_{\bm{y}}=\bm{0}$ for $\bm{y}\in\F_p^\nu\setminus\chi(W_E)$, and let $\{\Pi_{\bm{y}}:\bm{y}\in\F_p^\nu\}$ be the measurement from Lemma~\ref{lem:syndrome-measurement}.

To perform the measurement on $\Gamma$, the decoder first prepares replacement qudits at $E$ in the maximally mixed state.
Let $d_E=q^{|E|}$ and define the replacement map
\[
 \mathcal J_E\colon\BB(\HH_H)\longrightarrow\BB(\HH_\Gamma),~
 \mathcal J_E(\tau)=\frac{I_E}{d_E}\otimes\tau.
\]
For every operator $\rho\in\BB(\HH_{[n]})$, we have
\begin{equation}\label{eq:twirl}
 (\mathcal J_E\otimes\id_O)(\Tr_E\rho)
 =\frac{I_E}{d_E}\otimes\Tr_E\rho
 =\frac{1}{d_E^2}\sum_{\bm{a}\in W_E}
   P_{\bm{a}}\rho P_{\bm{a}}^\dagger.
\end{equation}
To prove the second equality, expand $\rho$ in the Pauli basis on $E$ as
\[
 \rho=\sum_{\bm{b}\in W_E}
 P_{\bm{b},E}\otimes M_{\bm{b}},
\]
where $P_{\bm{b}}=P_{\bm{b},E}\otimes I_{E^c}$ and $M_{\bm{b}}\in\BB(\HH_{E^c})$.
By~\eqref{eq:pauli-trace},
\[
 \Tr_E\rho
 =\sum_{\bm{b}\in W_E}
   \Tr(P_{\bm{b},E})M_{\bm{b}}
 =d_E M_{\bm{0}}.
\]
By~\eqref{eq:pauli-commutation}, conjugation by $P_{\bm{a}}$ multiplies the term indexed by $\bm{b}$ by $\zeta_p^{-[\bm{a},\bm{b}]_{\mathrm t}}$.
For $\bm{b}\ne\bm{0}$, the nondegeneracy of the trace-symplectic form on $W_E$ implies that $\bm{a}\mapsto[\bm{a},\bm{b}]_{\mathrm t}$ is a nonzero $\F_p$-linear functional.
Each value in $\F_p$ therefore occurs $d_E^2/p$ times.
By the geometric sum formula, we obtain
\[
 \frac{1}{d_E^2}
 \sum_{\bm{a}\in W_E}
 \zeta_p^{-[\bm{a},\bm{b}]_{\mathrm t}}
 =\delta_{\bm{b},\bm{0}},
\]
where all summands equal $1$ when $\bm{b}=\bm{0}$.
Consequently,
\[
 \frac{1}{d_E^2}
 \sum_{\bm{a}\in W_E}
 P_{\bm{a}}\rho P_{\bm{a}}^\dagger
 =I_E\otimes M_{\bm{0}}
 =\frac{I_E}{d_E}\otimes\Tr_E\rho,
\]
which proves~\eqref{eq:twirl}.

Since $\bm{a}_{\bm{y}}\in W_E$ and $E\subseteq\Gamma$, write $P_{\bm{a}_{\bm{y}}}=C_{\bm{y}}\otimes I_O$, where $C_{\bm{y}}$ is unitary on $\HH_\Gamma$.
After the replacement, the decoder performs the projective measurement $\{\Pi_{\bm{y}}:\bm{y}\in\F_p^\nu\}$ and applies $C_{\bm{y}}^\dagger$ upon obtaining outcome $\bm{y}$.
The resulting recovery map is
\begin{equation}\label{eq:explicitlocalchannel}
 \mathcal R_{\Gamma,E}(\tau)
 =\sum_{\bm{y}\in\F_p^\nu}
   C_{\bm{y}}^\dagger\Pi_{\bm{y}}
   \mathcal J_E(\tau)
   \Pi_{\bm{y}}C_{\bm{y}}.
\end{equation}
The map $\mathcal J_E$ is CPTP because it appends the fixed state $I_E/d_E$.
The subsequent measurement and correction define a channel with Kraus operators $K_{\bm{y}}=C_{\bm{y}}^\dagger\Pi_{\bm{y}}$, since
\[
 \begin{aligned}
 \sum_{\bm{y}\in\F_p^\nu}
 K_{\bm{y}}^\dagger K_{\bm{y}}
 &=\sum_{\bm{y}\in\F_p^\nu}
   \Pi_{\bm{y}}C_{\bm{y}}C_{\bm{y}}^\dagger\Pi_{\bm{y}}\\
 &=\sum_{\bm{y}\in\F_p^\nu}\Pi_{\bm{y}}\\
 &=I_\Gamma.
 \end{aligned}
\]
Hence $\mathcal R_{\Gamma,E}$ is CPTP, with input on $H$ and output on $\Gamma$.

It remains to verify~\eqref{eq:recovery}.
Let $M\in\BB(\HH_A\otimes\HH_B)$ be arbitrary.
By~\eqref{eq:twirl} and linearity, we may apply the measurement and correction to each Pauli term separately.
For the term indexed by $\bm{a}$, it follows from~\eqref{eq:syndrome-projection} that only the outcome $\bm{y}=\chi(\bm{a})$ contributes.
Thus
\[
 \begin{aligned}
 &(\mathcal R_{\Gamma,E}\otimes\id_O)
   \bigl(\Tr_E\cV(M)\bigr)\\
 &\quad=\frac{1}{d_E^2}\sum_{\bm{a}\in W_E}
   D_{\bm{a}}VMV^\dagger D_{\bm{a}}^\dagger\\
 &\quad=\frac{1}{d_E^2}\sum_{\bm{a}\in W_E}
   V(I_A\otimes U_{\bm{a},B})M
   (I_A\otimes U_{\bm{a},B}^\dagger)V^\dagger,
 \end{aligned}
\]
where the last equality follows from~\eqref{eq:local-residual-gauge}.
Define
\[
 \Phi_{\Gamma,E}(\sigma)
 =\frac{1}{d_E^2}\sum_{\bm{a}\in W_E}
   U_{\bm{a},B}\sigma U_{\bm{a},B}^\dagger,~
 \sigma\in\BB(\HH_B).
\]
Since $|W_E|=d_E^2$, the map $\Phi_{\Gamma,E}$ is a uniform average of unitary channels and is therefore CPTP.
With this definition, we can write the preceding identity as
\[
 (\mathcal R_{\Gamma,E}\otimes\id_O)
 \bigl(\Tr_E\cV(M)\bigr)
 =\cV\bigl((\id_A\otimes\Phi_{\Gamma,E})(M)\bigr).
\]
Since this equality holds for every $M\in\BB(\HH_A\otimes\HH_B)$, we obtain
\[
 (\mathcal R_{\Gamma,E}\otimes\id_O)\circ\Tr_E\circ\cV
 =\cV\circ(\id_A\otimes\Phi_{\Gamma,E}).
\]
By Definition~\ref{def:local}, the erasure of $E$ is recoverable within $\Gamma$.
\end{proof}

We next prove a necessary condition for local erasure recovery in terms of the stabilizer label space.

\begin{lemma}\label{le:1)to4)}
Let $\mathcal Q$ be an SSC with stabilizer label space $S$,
and let $\emptyset\ne E\subseteq\Gamma\subseteq[n]$.
If the erasure of $E$ is recoverable within $\Gamma$
in the sense of Definition~\ref{def:local}, then
$
 \pi_E(S_\Gamma)=\pi_E(S).
$
\end{lemma}

\begin{proof}
Set $H=\Gamma\setminus E$ and $O=\Gamma^c$.
Since $S_\Gamma\subseteq S$, we have
$\pi_E(S_\Gamma)\subseteq\pi_E(S)$.
It remains to prove the reverse inclusion.
We first show that
$
 \pi_O(S)=\pi_O(S_{E^c}).
$
Let $\mathcal S$ be the stabilizer group of $\mathcal Q$,
and let $P$ be the orthogonal projector onto $\mathcal Q$.
We consider the maximally mixed state on $\mathcal Q$, whose Pauli expansion contains only stabilizer terms.
By~\eqref{eq:stabilizer-projector}, $\Tr P=KR$, so
$
 \omega=\frac{P}{KR}
$
is a normalized state.
Since $\cV(M)=VMV^\dagger$ and $VV^\dagger=P$, we also have
\[
 \begin{aligned}
 \cV\left(\frac{I_A}{K}\otimes\frac{I_B}{R}\right)
 &=\frac{V(I_A\otimes I_B)V^\dagger}{KR}\\
 &=\frac{P}{KR}
 =\omega.
 \end{aligned}
\]
For each $\bm{s}\in S$, let $U_{\bm{s}}\in\mathcal S$
be its stabilizer representative.
By~\eqref{eq:stabilizer-projector},
$KR|\mathcal S|=q^n$, and hence
\[
 \omega
 =\frac{1}{KR|\mathcal S|}
   \sum_{\bm{s}\in S}U_{\bm{s}}
 =q^{-n}\sum_{\bm{s}\in S}U_{\bm{s}}.
\]
By Definition~\ref{def:local}, there exist CPTP maps
\[
 \mathcal R_{\Gamma,E}:
 \BB(\HH_H)\longrightarrow\BB(\HH_\Gamma),~
 \Phi_{\Gamma,E}:
 \BB(\HH_B)\longrightarrow\BB(\HH_B)
\]
satisfying~\eqref{eq:recovery}.
Applying this identity to
$I_A/K\otimes I_B/R$, we obtain
\[
 \begin{aligned}
 \eta
 &:=(\mathcal R_{\Gamma,E}\otimes\id_O)(\Tr_E\omega)\\
 &=\cV\left(
     \frac{I_A}{K}\otimes
     \Phi_{\Gamma,E}\left(\frac{I_B}{R}\right)
   \right).
 \end{aligned}
\]
Since the channels are trace-preserving and
$\omega$ is normalized, $\Tr(\eta)=1$.
The last expression represents an encoded state, and hence $P\eta P=\eta$.
For every $\bm{s}\in S$, the operator
$U_{\bm{s}}^\dagger$ acts as the identity on $\mathcal Q$,
and therefore $U_{\bm{s}}^\dagger P=P$.
It follows that
\begin{equation}\label{eq:stabilizer-expectation}
 \begin{aligned}
 \Tr(U_{\bm{s}}^\dagger\eta)
 &=\Tr(U_{\bm{s}}^\dagger P\eta P)\\
 &=\Tr(P\eta P)\\
 &=\Tr(\eta)
 =1.
 \end{aligned}
\end{equation}

We next compute the Pauli expansion of $\eta$.
Let
\[
 U_{\bm{s}}
 =\xi_{\bm{s}}\,
   P_{\pi_\Gamma(\bm{s})}
   \otimes P_{\pi_O(\bm{s})},~|\xi_{\bm{s}}|=1,
\]
where each projected Pauli operator acts on the indicated
coordinate set.
After grouping the tensor factors according to
$E$, $H$, and $O$, we have
\[
 \Tr_E U_{\bm{s}}
 =\xi_{\bm{s}}\,
   \Tr\bigl(P_{\pi_E(\bm{s})}\bigr)\,
   P_{\pi_H(\bm{s})}
   \otimes P_{\pi_O(\bm{s})}.
\]
By~\eqref{eq:pauli-trace},
\[
 \Tr\bigl(P_{\pi_E(\bm{s})}\bigr)
 =
 \begin{cases}
 q^{|E|},&\pi_E(\bm{s})=\bm{0},\\
 0,&\pi_E(\bm{s})\ne\bm{0}.
 \end{cases}
\]
Thus a term survives the partial trace over $E$
precisely when its label belongs to $S_{E^c}$.
Consequently,
\[
 \begin{aligned}
 \Tr_E\omega
 &=q^{-n}\sum_{\bm{s}\in S}\Tr_E U_{\bm{s}}\\
 &=q^{-(n-|E|)}
   \sum_{\bm{u}\in S_{E^c}}
   \xi_{\bm{u}}\,
   P_{\pi_H(\bm{u})}
   \otimes P_{\pi_O(\bm{u})}.
 \end{aligned}
\]
The recovery channel acts on $H$ and leaves $O$ unchanged.
By linearity,
\[
 \eta
 =q^{-(n-|E|)}
   \sum_{\bm{u}\in S_{E^c}}
   \xi_{\bm{u}}\,
   \mathcal R_{\Gamma,E}
      \bigl(P_{\pi_H(\bm{u})}\bigr)
   \otimes P_{\pi_O(\bm{u})}.
\]
In this expansion, every Pauli factor on $O$
has its label in $\pi_O(S_{E^c})$.
Suppose that there exists $\bm{s}\in S$ such that
$
 \pi_O(\bm{s})\notin\pi_O(S_{E^c}).
$
Then $\pi_O(\bm{s})\ne\pi_O(\bm{u})$ for every
$\bm{u}\in S_{E^c}$.
By the Pauli orthogonality relation
in~\eqref{eq:pauli-trace},
\[
 \Tr\left(
   P_{\pi_O(\bm{s})}^\dagger
   P_{\pi_O(\bm{u})}
 \right)=0.
\]
Using the tensor-product formula for the trace, we obtain
\[
 \begin{aligned}
 \Tr(U_{\bm{s}}^\dagger\eta)
 &=q^{-(n-|E|)}
   \sum_{\bm{u}\in S_{E^c}}
   \overline{\xi_{\bm{s}}}\xi_{\bm{u}}\\
 &\quad{}\times
   \Tr\left(
     P_{\pi_\Gamma(\bm{s})}^\dagger
     \mathcal R_{\Gamma,E}
       \bigl(P_{\pi_H(\bm{u})}\bigr)
   \right)\\
 &\quad{}\times
   \Tr\left(
     P_{\pi_O(\bm{s})}^\dagger
     P_{\pi_O(\bm{u})}
   \right)\\
 &=0.
 \end{aligned}
\]
This contradicts~\eqref{eq:stabilizer-expectation}.
Hence
$\pi_O(S)\subseteq\pi_O(S_{E^c})$.
The reverse inclusion follows from $S_{E^c}\subseteq S$.
Therefore,
\begin{equation}\label{eq:outside-stabilizer-projection}
 \pi_O(S)=\pi_O(S_{E^c}).
\end{equation}

Now let $\bm{s}\in S$.
By~\eqref{eq:outside-stabilizer-projection}, we may choose
$\bm{u}\in S_{E^c}$ such that
$\pi_O(\bm{u})=\pi_O(\bm{s})$.
Since $\bm{s},\bm{u}\in S$, their difference belongs to $S$.
Moreover,
\[
 \pi_O(\bm{s}-\bm{u})
 =\pi_O(\bm{s})-\pi_O(\bm{u})
 =\bm{0}.
\]
Thus $\bm{s}-\bm{u}$ is supported in $\Gamma$,
and hence $\bm{s}-\bm{u}\in S_\Gamma$.
Since $\bm{u}\in S_{E^c}$, we also have
$\pi_E(\bm{u})=\bm{0}$.
It follows that
\[
 \begin{aligned}
 \pi_E(\bm{s})
 =\pi_E(\bm{s})-\pi_E(\bm{u})
 =\pi_E(\bm{s}-\bm{u})
 \in\pi_E(S_\Gamma).
 \end{aligned}
\]
Since $\bm{s}\in S$ was arbitrary, this proves
$\pi_E(S)\subseteq\pi_E(S_\Gamma)$.
Combining the two inclusions, we obtain
$
 \pi_E(S_\Gamma)=\pi_E(S).
$
\end{proof}

We can now state the algebraic characterization of local erasure recovery for an SSC.

\begin{theorem}\label{thm:local}
Let $\mathcal Q$ be an SSC with additive gauge label space $G\le W$ and stabilizer label space $S=G\cap G^{\perpt}$.
Let $\emptyset\ne E\subseteq\Gamma\subseteq[n]$.
The following conditions are equivalent.
\begin{enumerate}
\item The erasure of $E$ is recoverable within $\Gamma$ in the sense of Definition~\ref{def:local}.
\item $W_E\cap S_\Gamma^{\perpt}=G_E$.
\item $\dim_p\pi_E(S_\Gamma)+\dim_pG_E=2m|E|$.
\item The erasure of $E$ is globally correctable and $\pi_E(S_\Gamma)=\pi_E(S)$.
\end{enumerate}
\end{theorem}

\begin{proof}
We first show that conditions 2) and 3) are equivalent.
By Lemma~\ref{lem:duality}, we have $$\dim_p G_E=\dim_p(W_E\cap S_\Gamma^{\perpt})=2m|E|-\dim_p \pi_E(S_\Gamma),$$
which proves that $2)\Rightarrow3)$.
By the definition of $S$, we have $S=G\cap G^{\perpt}$ and $S\subseteq G^{\perpt}$.
Then $G=(G^{\perpt})^{\perpt}\subseteq S^{\perpt}$.
By the definition of $G_E$,
$G_E=G\cap W_E\subseteq S^{\perpt}\cap W_E.$
By the definition of $S_\Gamma$,
$S_\Gamma=S\cap W_\Gamma\subseteq S$, which implies that $S^{\perpt}\subseteq S_\Gamma^{\perpt}$.
Then $W_E\cap S^{\perpt}\subseteq W_E\cap S_\Gamma^{\perpt}$.
Therefore, $G_E\subseteq W_E\cap S_\Gamma^{\perpt}$.
By condition 3) and Lemma~\ref{lem:duality}, we have
\[
 \begin{aligned}
\dim_p(W_E\cap S_\Gamma^{\perpt})&=2m|E|-\dim_p \pi_{E}(S_\Gamma)\\
                                 &=2m|E|-(2m|E|-\dim_p G_E)\\
                                 &=\dim_p G_E,
\end{aligned}
\]
which proves that $3)\Rightarrow2)$.

We next show that conditions 2) and 4) are equivalent.
On the one hand, $G_E\subseteq W_E\cap S^{\perpt}\subseteq W_E\cap S_\Gamma^{\perpt}$. By condition 2), we obtain
$G_E=W_E\cap S^{\perpt}$.
By Lemma~\ref{lem:global}, the erasure of $E$ is globally correctable.
On the other hand,
$$\dim_p G_E=\dim_p(W_E\cap S_\Gamma^{\perpt})=2m|E|-\dim_p\pi_E(S_\Gamma),$$
and
$$\dim_p G_E=\dim_p(W_E\cap S^{\perpt})=2m|E|-\dim_p \pi_E(S).$$
Then $\dim_p\pi_E(S_\Gamma)=\dim_p \pi_E(S)$.
Hence $\pi_E(S_\Gamma)=\pi_E(S)$, since $\pi_E(S_\Gamma)\subseteq\pi_E(S)$.
This proves $2)\Rightarrow4)$. Conversely, $4)\Rightarrow2)$ follows directly.

It remains to prove $1)\Rightarrow4)$.
Assume that condition 1) holds, and set
$H=\Gamma\setminus E$ and $O=\Gamma^c$.
By Definition~\ref{def:local}, there exist CPTP maps
\[
 \mathcal R_{\Gamma,E}:
 \BB(\HH_H)\longrightarrow\BB(\HH_\Gamma),~
 \Phi_{\Gamma,E}:
 \BB(\HH_B)\longrightarrow\BB(\HH_B)
\]
satisfying~\eqref{eq:recovery}.
Define
$\mathcal R_E:=\mathcal R_{\Gamma,E}\otimes\id_O$ and $\Phi_E:=\Phi_{\Gamma,E}$.
Since $E^c=H\sqcup O$ and $[n]=\Gamma\sqcup O$,
the map
$
 \mathcal R_E:
 \BB(\HH_{E^c})\rightarrow\BB(\HH_{[n]})
$
is CPTP.
Moreover, by~\eqref{eq:recovery}, we have
\[
 \begin{aligned}
 \mathcal R_E\circ\Tr_E\circ\cV
 &=(\mathcal R_{\Gamma,E}\otimes\id_O)
   \circ\Tr_E\circ\cV\\
 &=\cV\circ(\id_A\otimes\Phi_{\Gamma,E})\\
 &=\cV\circ(\id_A\otimes\Phi_E).
 \end{aligned}
\]
Thus $\mathcal R_E$ and $\Phi_E$ satisfy
the recovery identity in
Definition~\ref{def:global-recovery}, and the erasure of $E$ is globally correctable.
Finally, Lemma~\ref{le:1)to4)} implies that $\pi_E(S_\Gamma)=\pi_E(S)$. This proves $1)\Rightarrow4)$.

Combining the preceding implications with Proposition~\ref{lem:local-syndrome}, we obtain
$$1)\Rightarrow4)\Rightarrow3)\Rightarrow2)\Rightarrow1)$$
and complete the proof.
\end{proof}

\begin{remark}
Let $\chi$ be the syndrome map used in the proof of
Proposition~\ref{lem:local-syndrome}.
By condition 2) of Theorem~\ref{thm:local}, we have
$
 \ker\chi=W_E\cap S_\Gamma^{\perpt}=G_E.
$
By the first isomorphism theorem, we obtain an $\F_p$-linear isomorphism
\[
 W_E/G_E\longrightarrow\chi(W_E),~
 \bm{a}+G_E\longmapsto\chi(\bm{a}).
\]

By the definition of $S$,  we have $S=G\cap G^{\perpt}$ and $S\subseteq G$.
For $R=1$, we have $\dim_p G=\dim_p S$ by Lemma~\ref{lem:ssc-representation}~(i).
Hence, $G=S$ and $\mathcal Q$ is an ordinary stabilizer code with parameters $((n,K,d))_q$.
If $S$ is also $\F_q$-linear and $\emptyset\ne E\subsetneq\Gamma\subseteq[n]$, then $S^{\perpt}=S^{\perps}$ follows from the definitions of the symplectic and trace-symplectic forms.
By Lemma~\ref{lem:duality}, we obtain $\pi_\Gamma(S^{\perps})=\bigl(\operatorname{sh}_\Gamma(S)\bigr)^{\perps}$.
Since $\operatorname{sh}_\Gamma(S)=\pi_\Gamma(S_\Gamma)$, it follows from the definition of shortening that
\[
\begin{aligned}
\operatorname{sh}_E\bigl(\pi_\Gamma(S^{\perp_{\mathrm s}})\bigr)
&=
\left\{
\pi_E(\bm v):
\bm v\in W_E,\
\pi_\Gamma(\bm v)\in\pi_\Gamma(S^{\perp_{\mathrm s}})
\right\}\\
&=
\left\{
\pi_E(\bm v):
\bm v\in W_E,\
\bm v\in S_\Gamma^{\perp_{\mathrm s}}
\right\}\\
&=
\pi_E\bigl(W_E\cap S_\Gamma^{\perp_{\mathrm s}}\bigr).
\end{aligned}
\]
Since $G_E=S_E$ and $\operatorname{sh}_E(S)=\pi_E(S_E)$, condition 2) of Theorem~\ref{thm:local} is equivalent to
$
 \operatorname{sh}_E\bigl(\pi_\Gamma(S^{\perps})\bigr)
 =\operatorname{sh}_E(S),
$
which is the stabilizer local recovery criterion
in~\cite[Theorem~18]{galindo}.
\end{remark}

\begin{corollary}\label{cor:globaldistance}
An $(r,\delta)$-SSLRC with $K>1$ has $d\ge\delta$.
\end{corollary}
\begin{proof}
Suppose to the contrary that $d<\delta$.
By~\eqref{eq:distance}, choose $\bm{w}\in S^{\perpt}\setminus G$ with $\swt(\bm{w})=d$.
Let $i\in\supp(\bm{w})$, and let $\Gamma_i$ be a repair set for $i$.
Set $E=\supp(\bm{w})\cap\Gamma_i$, and let $\bm{v}$ be the zero extension of $\pi_{\Gamma_i}(\bm{w})$.
Then $\pi_{\Gamma_i}(\bm{v})=\pi_{\Gamma_i}(\bm{w}),\pi_{\Gamma_i^c}(\bm{v})=\bm{0}$.
Hence, $\supp(\bm{v})=\supp(\bm{w})\cap\Gamma_i=E,\bm{v}\in W_E$.
Since $i\in E$ and $E\subseteq\supp(\bm{w})$, we have $0<|E|\le|\supp(\bm{w})|=d<\delta$.
Thus, by Definition~\ref{def:lrc}, the erasure of $E$ is recoverable within $\Gamma_i$.

For every $\bm{z}\in S_{\Gamma_i}$, the support of $\bm{z}$ is contained in $\Gamma_i$.
Since $\pi_{\Gamma_i}(\bm{v})=\pi_{\Gamma_i}(\bm{w})$ and $\bm{w}\in S^{\perpt}$, we obtain
$
 [\bm{v},\bm{z}]_{\mathrm t}
 =[\bm{w},\bm{z}]_{\mathrm t}
 =0.
$
Therefore, $\bm{v}\in W_E\cap S_{\Gamma_i}^{\perpt}$.
By condition 2) of Theorem~\ref{thm:local}, $W_E\cap S_{\Gamma_i}^{\perpt}=G_E$, and hence $\bm{v}\in G_E$.
Since $G_E\subseteq G\subseteq S^{\perpt}$, we have $\bm{w}-\bm{v}\in S^{\perpt}$.
Moreover, $\bm{w}-\bm{v}\notin G$.
Indeed, if $\bm{w}-\bm{v}\in G$, then $\bm{w}=(\bm{w}-\bm{v})+\bm{v}\in G$, contrary to the choice of $\bm{w}$.
By the construction of $\bm{w}-\bm{v}$, we have
$
 \pi_{\Gamma_i}(\bm{w}-\bm{v})=\bm{0},
 \pi_{\Gamma_i^c}(\bm{w}-\bm{v})
 =\pi_{\Gamma_i^c}(\bm{w}).
$
Consequently,
$
 \supp(\bm{w}-\bm{v})
 =\supp(\bm{w})\setminus\Gamma_i
 =\supp(\bm{w})\setminus E.
$
Since $E\ne\emptyset$, it follows that
$
 \bm{w}-\bm{v}\in S^{\perpt}\setminus G,
 \swt(\bm{w}-\bm{v})=d-|E|<d.
$
This contradicts~\eqref{eq:distance}.
Therefore, $d\ge\delta$.
\end{proof}
\subsection{CSS and Hermitian Hull Constructions}

In this subsection, we apply Theorem~\ref{thm:local}
to the CSS and Hermitian hull constructions of
SSCs~\cite[Corollaries~4 and~5]{alylp}.
Let $C_X,C_Z\le\F_q^{[n]}$ be $\F_q$-linear codes
with full-row-rank generator matrices $M_X$ and $M_Z$.
Define
\[
 G=C_X\times C_Z,~
 H_X=C_X\cap C_Z^\perp,~
 H_Z=C_Z\cap C_X^\perp,
\]
and let $c_X=\dim_qC_X$ and $c_Z=\dim_qC_Z$.

We first determine the parameters of the CSS SSC.
\begin{lemma}\label{parameters:CSS}
Let $b=\rank_q(M_XM_Z^T)$. The gauge label space $G$ defines a CSS SSC $\mathcal Q$ with stabilizer label space $S=H_X\times H_Z$, logical dimension $K=q^{n-c_X-c_Z+b}$, and gauge dimension $R=q^b$.
If $K>1$, its minimum distance is $d=\min\{d_X,d_Z\}$, where
\begin{equation}\label{eq:25}
 \begin{aligned}
 d_X=\min_{\bm{a}\in H_Z^\perp\setminus C_X}
       \operatorname{wt}(\bm{a}),~
 d_Z=\min_{\bm{b}\in H_X^\perp\setminus C_Z}
       \operatorname{wt}(\bm{b}).
 \end{aligned}
\end{equation}
\end{lemma}

\begin{proof}
By~\cite[Corollary~4]{alylp}, $G=C_X\times C_Z$ defines a CSS SSC $\mathcal Q$ with stabilizer label space
\[
 S=(C_X\cap C_Z^\perp)\times(C_Z\cap C_X^\perp)=H_X\times H_Z.
\]
The gauge label space $G$ has the full-row-rank
generator matrix
\[
 M_G=
 \begin{pmatrix}
  M_X&0\\
  0&M_Z
 \end{pmatrix},
\]
whose two row blocks consist of $X$-type and
$Z$-type labels, respectively.
The parameter formulas for this construction imply that
\[
 k=n-\frac{c_X+c_Z+\dim_qS}{2},~
 g=\frac{c_X+c_Z-\dim_qS}{2},
\]
where $K=q^k$ and $R=q^g$ are the logical and
gauge dimensions, respectively.

Let $B=M_XM_Z^T$ and $\rank_qB=b$.
Since $M_X$ and $M_Z$ have full row rank, the codewords in $C_X$ and $C_Z$ can be written uniquely as $\bm{u}M_X$ and $\bm{v}M_Z$, respectively, where $\bm{u}\in\F_q^{c_X}$ and $\bm{v}\in\F_q^{c_Z}$.
By the definitions of $H_X$ and $H_Z$, we have
\[
 \begin{aligned}
 \bm{u}M_X\in H_X\Longleftrightarrow\bm{u}B=\bm{0},~
 \bm{v}M_Z\in H_Z\Longleftrightarrow\bm{v}^T=\bm{0}.
 \end{aligned}
\]
By~\eqref{eq:rank-nullity},
$$\dim_qH_X=c_X-b,\dim_qH_Z=c_Z-b.$$
Since $S=H_X\times H_Z$, it follows that
$
 \dim_qS
 =\dim_qH_X+\dim_qH_Z
 =c_X+c_Z-2b.
$
Hence $$k=n-c_X-c_Z+b,g=b,$$
which imply that $K=q^{n-c_X-c_Z+b}$ and $R=q^b$.

To determine the minimum distance of $\mathcal Q$, suppose that $K>1$, or equivalently, $k>0$.
By the above dimension formulas, we have
\[
 \begin{aligned}
 \dim_qH_Z^\perp-\dim_qC_X
 &=n-(c_Z-b)-c_X=k>0,\\
 \dim_qH_X^\perp-\dim_qC_Z
 &=n-(c_X-b)-c_Z=k>0.
 \end{aligned}
\]
Thus both $H_Z^\perp\setminus C_X$ and
$H_X^\perp\setminus C_Z$ are nonempty.
Since $S=H_X\times H_Z$ is $\F_q$-linear, $S^{\perpt}=S^{\perps}=H_Z^\perp\times H_X^\perp$.
By~\eqref{eq:distance},
\[
 d=
 \min\left\{
 \swt(\bm{a}\mid\bm{b}):
 (\bm{a}\mid\bm{b})
 \in(H_Z^\perp\times H_X^\perp)
   \setminus(C_X\times C_Z)
 \right\}.
\]
For every label in this set, either
$\bm{a}\in H_Z^\perp\setminus C_X$
or $\bm{b}\in H_X^\perp\setminus C_Z$.
Moreover,
\[
 \swt(\bm{a}\mid\bm{b})
 =|\supp(\bm{a})\cup\supp(\bm{b})|
 \ge
 \max\{\operatorname{wt}(\bm{a}),
       \operatorname{wt}(\bm{b})\}.
\]
It follows that $d\ge\min\{d_X,d_Z\}$.
Conversely, choose
$\bm{a}\in H_Z^\perp\setminus C_X$ and
$\bm{b}\in H_X^\perp\setminus C_Z$
such that
$
 \operatorname{wt}(\bm{a})=d_X,
 \operatorname{wt}(\bm{b})=d_Z.
$
Then
$
 (\bm{a}\mid\bm{0}),
 (\bm{0}\mid\bm{b})
 \in S^{\perpt}\setminus G.
$
Hence,
\[
 d\le\swt(\bm{a}\mid\bm{0})=d_X,~
 d\le\swt(\bm{0}\mid\bm{b})=d_Z,
\]
which, together with the preceding lower bound, imply that $d=\min\{d_X,d_Z\}$.
\end{proof}

We now characterize the $(r,\delta)$-locality of the CSS SSC
in Lemma~\ref{parameters:CSS}.

\begin{proposition}\label{CSS}
Let $\mathcal Q$ be the CSS SSC in
Lemma~\ref{parameters:CSS}.
Then $\mathcal Q$ is an $(r,\delta)$-SSLRC if and only if
it admits a repair cover satisfying~\eqref{eq:groupsize}
such that, for every repair set $\Gamma_i$ and every
nonempty $E\subseteq\Gamma_i$ with $|E|<\delta$,
\begin{align}
 \dim_q\pi_E((H_X)_{\Gamma_i})+\dim_q(C_Z)_E
 &=|E|,\label{eq:cssx}\\
 \dim_q\pi_E((H_Z)_{\Gamma_i})+\dim_q(C_X)_E
 &=|E|.\label{eq:cssz}
\end{align}
\end{proposition}

\begin{proof}
By Definition~\ref{def:lrc}, $\mathcal Q$ is an
$(r,\delta)$-SSLRC if and only if it admits a repair cover
satisfying~\eqref{eq:groupsize} such that every nonempty
$E\subseteq\Gamma_i$ with $|E|<\delta$ is recoverable
within $\Gamma_i$.
We first express the recovery criterion for a single erased set in terms of the dimensions of the component codes.

Fix $\emptyset\ne E\subseteq\Gamma\subseteq[n]$.
By condition 3) of Theorem~\ref{thm:local}, the erasure
of $E$ is recoverable within $\Gamma$ if and only if
\[
 \dim_p\pi_E(S_\Gamma)+\dim_pG_E=2m|E|.
\]
Since the spaces $\pi_E(S_\Gamma)$ and $G_E$ are $\F_q$-linear, the recovery condition is equivalent to
\[
 \dim_q\pi_E(S_\Gamma)+\dim_qG_E=2|E|.
\]

We now express these dimensions in terms of
$C_X$, $C_Z$, $H_X$, and $H_Z$.
By Lemma~\ref{parameters:CSS},
$G=C_X\times C_Z$ and $S=H_X\times H_Z.$
Hence, $S_\Gamma=(H_X)_\Gamma\times(H_Z)_\Gamma$ and $G_E=(C_X)_E\times(C_Z)_E$.
Since the projection $\pi_E$ acts on each component, we have
\[
 \pi_E(S_\Gamma)
 =
 \pi_E((H_X)_\Gamma)\times\pi_E((H_Z)_\Gamma).
\]
Therefore,
\[
 \begin{aligned}
 \dim_q\pi_E(S_\Gamma)
 &=\dim_q\pi_E((H_X)_\Gamma)
   +\dim_q\pi_E((H_Z)_\Gamma),\\
 \dim_qG_E
 &=\dim_q(C_X)_E+\dim_q(C_Z)_E.
 \end{aligned}
\]
It follows that the erasure of $E$ is recoverable
within $\Gamma$ if and only if
\[
 \begin{aligned}
 &\bigl(\dim_q\pi_E((H_X)_\Gamma)
        +\dim_q(C_Z)_E\bigr)\\
 &\quad+
 \bigl(\dim_q\pi_E((H_Z)_\Gamma)
        +\dim_q(C_X)_E\bigr)
 =2|E|.
 \end{aligned}
\]

We next prove that
$
 \dim_q\pi_E((H_X)_\Gamma)+\dim_q(C_Z)_E
 \le |E|.
$
To this end, we show that
$\pi_E((H_X)_\Gamma)$ and $\operatorname{sh}_E(C_Z)$
are orthogonal in $\F_q^E$.
Let $\bm{x}\in(H_X)_\Gamma$ and $\bm{z}\in(C_Z)_E$.
Since $\bm{z}$ is supported on $E$, we have $z_j=0$
for every $j\notin E$.
Hence
\[
 \begin{aligned}
 \pi_E(\bm{x})\cdot\pi_E(\bm{z})
 =\sum_{j\in E}x_jz_j
 =\sum_{j=1}^n x_jz_j
 =\bm{x}\cdot\bm{z}
 =0.
 \end{aligned}
\]
The last equality follows from
$\bm{x}\in H_X\subseteq C_Z^\perp$ and $\bm{z}\in C_Z$.
Since $\bm{x}$ and $\bm{z}$ are arbitrary, we obtain
$
 \pi_E((C_Z)_E)
 \subseteq
 \bigl(\pi_E((H_X)_\Gamma)\bigr)^\perp,
$
where the Euclidean orthogonal complement is taken
in $\F_q^E$.
By the definition of shortening,
$
 \operatorname{sh}_E(C_Z)=\pi_E((C_Z)_E).
$
Moreover, since $\pi_E$ is injective on $(C_Z)_E$, we have
$
 \dim_q(C_Z)_E=\dim_q\operatorname{sh}_E(C_Z).
$
The Euclidean inner product on $\F_q^E$ is nondegenerate.
Therefore,
\[
 \begin{aligned}
 \dim_q(C_Z)_E
 &=\dim_q\operatorname{sh}_E(C_Z)\\
 &\le\dim_q\bigl(\pi_E((H_X)_\Gamma)\bigr)^\perp\\
 &=|E|-\dim_q\pi_E((H_X)_\Gamma).
 \end{aligned}
\]
Thus
\[
 \dim_q\pi_E((H_X)_\Gamma)+\dim_q(C_Z)_E
 \le |E|.
\]
The same argument with $X$ and $Z$ interchanged shows that
\[
 \dim_q\pi_E((H_Z)_\Gamma)+\dim_q(C_X)_E
 \le |E|.
\]
Each left-hand side is at most $|E|$.
Their sum equals $2|E|$ if and only if both inequalities
are equalities.
By condition 3) of Theorem~\ref{thm:local} and the
preceding CSS dimension formulas, the erasure of $E$
is therefore recoverable within $\Gamma$ if and only if
\[
 \begin{aligned}
 \dim_q\pi_E((H_X)_\Gamma)+\dim_q(C_Z)_E
 &=|E|,\\
 \dim_q\pi_E((H_Z)_\Gamma)+\dim_q(C_X)_E
 &=|E|.
 \end{aligned}
\]

Suppose that $\mathcal Q$ is an $(r,\delta)$-SSLRC.
By Definition~\ref{def:lrc}, there is a repair cover
satisfying~\eqref{eq:groupsize} such that every nonempty
$E\subseteq\Gamma_i$ with $|E|<\delta$ is recoverable
within $\Gamma_i$.
The preceding equivalence with $\Gamma=\Gamma_i$
implies that~\eqref{eq:cssx} and~\eqref{eq:cssz}
hold for every such $E$.
Conversely, suppose that a repair cover satisfies
\eqref{eq:groupsize}.
Assume that~\eqref{eq:cssx} and~\eqref{eq:cssz}
hold for every repair set $\Gamma_i$ and every
nonempty $E\subseteq\Gamma_i$ with $|E|<\delta$.
By the same equivalence, each such $E$ is recoverable
within $\Gamma_i$.
Hence $\mathcal Q$ is an $(r,\delta)$-SSLRC
by Definition~\ref{def:lrc}.
\end{proof}

\begin{remark}\label{remarkIII.12}
If $C_X\subseteq C_Z^\perp$, then $H_X=C_X$, $H_Z=C_Z$, and $b=0$.
Thus $G=S=C_X\times C_Z$, and Proposition~\ref{CSS} reduces to the CSS stabilizer construction~\cite[Corollary~4]{alylp}.
\end{remark}

We next consider the Hermitian hull construction of SSCs, using the coordinate map $\psi$ in~\eqref{eq:psi} and Lemma~\ref{lem:hermitian-correspondence}.
Let $C\le\F_{q^2}^{[n]}$ be an $\F_{q^2}$-linear code with Hermitian hull
$
 H=\operatorname{Hull}_H(C)=C\cap C^{\perp_H}.
$
Write $c=\dim_{q^2}C$ and $h_H=\dim_{q^2}H$.

\begin{lemma}\label{parameters:Hermitian}
The gauge label space $G=\psi^{-1}(C)$ defines an SSC
$\mathcal Q$ with stabilizer label space $S=\psi^{-1}(H)$,
logical dimension $K=q^{n-c-h_H}$,
and gauge dimension $R=q^{c-h_H}$.
If $K>1$, or equivalently $c+h_H<n$, its minimum distance is
\[
 d=\min\{\wt(\bm{v}):
          \bm{v}\in H^{\perp_H}\setminus C\}.
\]
\end{lemma}

\begin{proof}
By the Hermitian subsystem construction
of~\cite[Corollary~5]{alylp}, the gauge label space
$G=\psi^{-1}(C)$ defines an SSC $\mathcal Q$.
Since $C$ is $\F_{q^2}$-linear,
Lemma~\ref{lem:hermitian-correspondence} implies
$
 G^{\perpt}=\psi^{-1}(C^{\perp_H}).
$
Hence the stabilizer label space is
\[
 \begin{aligned}
 S
 &=G\cap G^{\perpt}\\
 &=\psi^{-1}(C)\cap\psi^{-1}(C^{\perp_H})\\
 &=\psi^{-1}(C\cap C^{\perp_H})\\
 &=\psi^{-1}(H).
 \end{aligned}
\]

We next determine the logical and gauge dimensions.
The map $\psi$ is an $\F_q$-linear isomorphism,
and $\F_{q^2}$ has dimension two over $\F_q$.
Therefore, $\dim_qG=2c$ and $\dim_qS=2h_H$.
By~\eqref{eq:qdimensions},
$$
 k
 =n-\frac{\dim_qG+\dim_qS}{2}
 =n-\frac{2c+2h_H}{2}
 =n-c-h_H,
$$
and
$$
 g
 =\frac{\dim_qG-\dim_qS}{2}
 =\frac{2c-2h_H}{2}
 =c-h_H.
$$
Thus
$
 K=q^{n-c-h_H},R=q^{c-h_H}.
$

To determine the minimum distance of $\mathcal Q$, assume that $K>1$.
Then $k=n-c-h_H>0$.
Since $H\subseteq C^{\perp_H}$, Hermitian duality
implies $C\subseteq H^{\perp_H}$.
Moreover,
\[
 \dim_{q^2}H^{\perp_H}-\dim_{q^2}C
 =n-h_H-c
 =k>0.
\]
Thus $H^{\perp_H}\setminus C$ is nonempty.
The hull $H$ is $\F_{q^2}$-linear.
By Lemma~\ref{lem:hermitian-correspondence},
$
 S^{\perpt}=\psi^{-1}(H^{\perp_H}).
$
Consequently, $\psi$ restricts to a bijection
\[
 S^{\perpt}\setminus G
 \longrightarrow H^{\perp_H}\setminus C.
\]
Since the coordinate map $\psi$ preserves support, we have $\swt(\bm{w})=\wt(\psi(\bm{w}))$ for $\bm{w}\in W$.
By~\eqref{eq:distance},
\[
 \begin{aligned}
 d
 =\min_{\bm{w}\in S^{\perpt}\setminus G}
      \swt(\bm{w})
 =\min_{\bm{v}\in H^{\perp_H}\setminus C}
      \wt(\bm{v}).
 \end{aligned}
\]
\end{proof}

We now characterize the $(r,\delta)$-locality of the SSC
in Lemma~\ref{parameters:Hermitian}.

\begin{proposition}\label{prop:hermitian}
Let $\mathcal Q$ be the SSC in
Lemma~\ref{parameters:Hermitian}.
Then $\mathcal Q$ is an $(r,\delta)$-SSLRC if and only if
it admits a repair cover satisfying~\eqref{eq:groupsize}
such that, for every repair set $\Gamma_i$ and every
nonempty $E\subseteq\Gamma_i$ with $|E|<\delta$,
\begin{equation}\label{eq:hermitianlocal}
 \dim_{q^2}\pi_E(H_{\Gamma_i})
 +\dim_{q^2}C_E
 =|E|.
\end{equation}
\end{proposition}

\begin{proof}
We proceed as in the proof of Proposition~\ref{CSS}, using the corresponding dimension formulas.
By Lemma~\ref{parameters:Hermitian},
$\psi(G)=C$ and $\psi(S)=H$.
The coordinate map $\psi$ is an $\F_p$-linear isomorphism.
It preserves support and commutes with coordinate projections.
Since $C$ and $H$ are $\F_{q^2}$-linear and $q=p^m$,
we have
$$\dim_pG_E=2m\dim_{q^2}C_E,$$
$$\dim_p\pi_E(S_\Gamma)=2m\dim_{q^2}\pi_E(H_\Gamma)$$
for every $\emptyset\ne E\subseteq\Gamma\subseteq[n]$.
The assertion follows from condition 3) of
Theorem~\ref{thm:local} and Definition~\ref{def:lrc}.
\end{proof}

\begin{remark}
If $C\subseteq C^{\perp_H}$, then $H=C$ and $g=0$.
Thus Proposition~\ref{prop:hermitian} reduces to the Hermitian stabilizer construction~\cite[Corollary~5]{alylp}.
\end{remark}

\section{Bounds for $(r,\delta)$-SSLRCs Without Assuming Purity}\label{sec:bounds}

In this section, we derive three bounds for $(r,\delta)$-SSLRCs without assuming purity:
a redundancy bound for disjoint repair sets,
a supported-subspace bound from supported stabilizer and gauge subspaces,
and a mixed counting bound for known erasures within selected repair sets and errors at unknown positions outside their union.
Throughout this section, let $\mathcal Q$ be an $(r,\delta)$-SSLRC with parameters $((n,K,R,d))_q$, where $q=p^m$ and $K>1$.
The gauge label space $G$ is $\F_p$-linear.
Let $S=G\cap G^{\perpt}$ be the stabilizer label space, and set $s=\dim_pS$ and $h=\dim_pG$.
By~\eqref{eq:dimensions}, $\Lambda=\log_q(KR)=n-s/m$.
When $K=q^k$ and $R=q^g$ for nonnegative integers $k,g$, the parameters of $\mathcal Q$ are $[[n,k,g,d]]_q$ and $\Lambda=k+g$.
Every repair set satisfies $\delta\le|\Gamma|\le r+\delta-1$, and $d\ge\delta$ by Corollary~\ref{cor:globaldistance}.

\subsection{A Redundancy Bound for Disjoint Repair Sets}

We apply Theorem~\ref{thm:local} to pairwise disjoint repair sets to obtain the following bound on $\Lambda$.

\begin{proposition}\label{prop:redundancy}

Let $M\ge1$ be an integer, and let $\Gamma_1,\ldots,\Gamma_M$
be pairwise disjoint repair sets of $\mathcal Q$.
For each $j\in[M]$, let $E_j$ be a nonempty subset of
$\Gamma_j$ with $|E_j|<\delta$.
Then
\begin{equation}\label{eq:redundancy}
 \Lambda\le n-\sum_{j=1}^M
 \left(2|E_j|-\frac{\dim_pG_{E_j}}{m}\right).
\end{equation}
Moreover, for every $j\in[M]$,
\begin{equation}\label{eq:localfeasibility}
 r+\delta-1\ge|\Gamma_j|\ge2|E_j|-\frac{\dim_pG_{E_j}}{m}.
\end{equation}
\end{proposition}

\begin{proof}
Since the repair sets are pairwise disjoint, the stabilizer label subspaces $S_{\Gamma_j}$ supported in $\Gamma_j$ form a direct sum in $S$. Hence
\[
 \bigoplus_{j=1}^M S_{\Gamma_j}\subseteq S,~
 \sum_{j=1}^M\dim_pS_{\Gamma_j}\le s.
\]
Since $|E_j|<\delta$, condition 3) of Theorem~\ref{thm:local} implies
\begin{equation}\label{eq:30}
 2m|E_j|-\dim_pG_{E_j}
 =\dim_p\pi_{E_j}(S_{\Gamma_j})\le\dim_pS_{\Gamma_j}.
\end{equation}
Then
\[
 \begin{aligned}
 \sum_{j=1}^M\bigl(2m|E_j|-\dim_pG_{E_j}\bigr)
 &\le\sum_{j=1}^M\dim_pS_{\Gamma_j}\\
 &\le s=m(n-\Lambda),
 \end{aligned}
\]
which proves~\eqref{eq:redundancy}.

The space $W_{\Gamma_j}$ of Pauli labels supported in
$\Gamma_j$ is nondegenerate under the trace-symplectic form
and has $\F_p$-dimension $2m|\Gamma_j|$.
Since $S_{\Gamma_j}$ is isotropic, it follows from~\eqref{eq:orthogonal-complement} that
\[
 \dim_pS_{\Gamma_j}
 \le\dim_p(W_{\Gamma_j}\cap S_{\Gamma_j}^{\perpt})
 =2m|\Gamma_j|-\dim_pS_{\Gamma_j},
\]
which implies that $\dim_pS_{\Gamma_j}\le m|\Gamma_j|$.
Combining this inequality with \eqref{eq:30}, we obtain
\[
 2m|E_j|-\dim_pG_{E_j}\le\dim_pS_{\Gamma_j}\le m|\Gamma_j|.
\]
Dividing by $m$ and applying $|\Gamma_j|\le r+\delta-1$, we obtain~\eqref{eq:localfeasibility}, which completes the proof.
\end{proof}

\begin{remark}
Suppose that $|E_j|=\delta-1$ and $G_{E_j}=\{0\}$ for every
$j\in[M]$. By~\eqref{eq:redundancy}, we have
$
 \Lambda\le n-\sum_{j=1}^M2(\delta-1)=n-2M(\delta-1).
$
If $K=q^k$ and $R=q^g$ for nonnegative integers $k,g$, then
$\Lambda=k+g$, and we obtain
\begin{equation}\label{eq:pureredundancy}
 k+g\le n-2M(\delta-1).
\end{equation}
If the code is pure and $0\ne w\in G_{E_j}$, then
Definition~\ref{def:purity}, $\supp(w)\subseteq E_j$,
and Corollary~\ref{cor:globaldistance} imply
$
 d\le\swt(w)\le|E_j|\le\delta-1<d,
$
which is a contradiction. Hence $G_{E_j}=\{0\}$ for every $j$,
and~\eqref{eq:pureredundancy} applies to pure codes.
\end{remark}

\subsection{Bounds from Supported Stabilizer and Gauge Subspaces}
For two disjoint globally correctable erased sets, we derive bounds on $K$ and $R$ from their supported gauge spaces and the stabilizers on the remaining coordinates.
For $E\subseteq[n]$, let $S_E$ denote the subspace of stabilizer labels
supported in $E$. Since $S_E\subseteq G_E$,
define
\begin{equation}\label{eq:gaugeprofile}
 \beta(E)=\dim_pG_E-\dim_pS_E=\dim_p(G_E/S_E).
\end{equation}

\begin{theorem}\label{thm:profilebound}

Let $E_1,E_2\subseteq[n]$ be disjoint sets such that the erasure
of each $E_i$ is correctable with access to all surviving
coordinates in $E_i^c$.
Let $T=[n]\setminus(E_1\cup E_2)$. Then
\begin{equation}\label{eq:profilebound}
 \log_qK+\log_qR
 \le |T|-\frac{\dim_pS_T}{m}
       +\frac{\beta(E_1)+\beta(E_2)}{2m},
\end{equation}
where $S_T$ denotes the subspace of stabilizer labels
supported in $T$.
Moreover, $\beta(E_1)+\beta(E_2)\le2\log_pR$ and
\begin{equation}\label{eq:logicalprofile}
 \log_qK\le |T|-\frac{\dim_pS_T}{m}.
\end{equation}
\end{theorem}

\begin{proof}
Let $s_i=\dim_pS_{E_i}$ and $b_i=\dim_pG_{E_i}$ for
$i=1,2$, and let $s_T=\dim_pS_T$.
The map $\pi_{E_1}|_S$ is the coordinate projection onto
$E_1$ restricted to $S$. Since $E_1^c=E_2\cup T$,
its kernel consists of the stabilizer labels supported
in $E_2\cup T$. Thus
$
 \ker(\pi_{E_1}|_S)=S_{E_2\cup T}.
$
Since $S_{E_2},S_T\le S_{E_2\cup T}$ and
$S_{E_2}\cap S_T=S_{E_2\cap T}=\{0\}$, we have
$
 S_{E_2}\oplus S_T\subseteq S_{E_2\cup T}.
$
By~\eqref{eq:globalrank},
$
 \dim_p\pi_{E_1}(S)=2m|E_1|-b_1,
$
and by rank-nullity~\eqref{eq:rank-nullity},
$$
 s=\dim_p\pi_{E_1}(S)+\dim_pS_{E_2\cup T}\ge2m|E_1|-b_1+s_2+s_T.
$$
Similarly, for $E_2$, we obtain
$s\ge2m|E_2|-b_2+s_1+s_T.$
Then
\begin{align*}
 2s
 &\ge2m(|E_1|+|E_2|)-(b_1-s_1)-(b_2-s_2)+2s_T\\
 &=2m(n-|T|)-\beta(E_1)-\beta(E_2)+2s_T,
\end{align*}
which implies that
$$
n-\frac{s}{m}\leq|T|-\frac{\dim_pS_T}{m}+\frac{\beta(E_1)+\beta(E_2)}{2m}.
$$
By~\eqref{eq:dimensions}, we have
$\log_q(KR)=\log_{p^m}(p^{mn-s})=n-\frac{s}{m}$.
Hence~\eqref{eq:profilebound} holds.

Let $\vartheta:G\to G/S$ be the quotient map,
$\vartheta(g)=g+S$. For $i=1,2$, let
$M_i=(G_{E_i}+S)/S\le G/S$ be the image
$\vartheta(G_{E_i})$ of $G_{E_i}$ in $G/S$.
The restriction $\vartheta|_{G_{E_i}}$ has kernel
\[
 \ker(\vartheta|_{G_{E_i}})
 =G_{E_i}\cap S
 =(G\cap W_{E_i})\cap S
 =S\cap W_{E_i}
 =S_{E_i}.
\]
By
rank-nullity~\eqref{eq:rank-nullity},
$
 \dim_pM_i=b_i-s_i=\beta(E_i).
$
Since $E_1\cap E_2=\emptyset$, we have
$[G_{E_1},G_{E_2}]_{\mathrm t}=0$, and hence
$[M_1,M_2]_{\mathrm t}=0$.
Let $M_1^{\perpt}$ denote the orthogonal complement of $M_1$
in $G/S$ with respect to the induced trace-symplectic form.
Since $G/S$ is nondegenerate, it follows from the orthogonal-complement dimension formula in~\eqref{eq:orthogonal-complement} that
\begin{align*}
 \dim_pM_2&\le\dim_pM_1^{\perpt}
                   =\dim_p(G/S)-\dim_pM_1.\\
\end{align*}
Since $S\subseteq G$, we have
\[
\dim_p(G/S)=\dim_pG-\dim_pS=h-s.
\]
By~\eqref{eq:dimensions}, $\log_p R=\frac{h-s}{2}$.
Thus $\beta(E_1)+\beta(E_2)\le2\log_pR$.
Substituting this estimate into~\eqref{eq:profilebound},
we obtain
\[
 \log_qK+\log_qR
 \le |T|-\frac{s_T}{m}+\frac{2\log_pR}{2m}
 =|T|-\frac{s_T}{m}+\frac{m\log_{p^m}R}{m},
\]
which proves~\eqref{eq:logicalprofile}.
\end{proof}

\begin{remark}
The condition $K>1$ ensures that $2(d-1)\le n$.
Indeed, if $2(d-1)>n$, then $d-1\ge\lceil n/2\rceil$.
A partition of $[n]$ into sets of sizes $\lfloor n/2\rfloor$ and $\lceil n/2\rceil$ would then consist of two globally correctable sets.
Since $T=\emptyset$ and $S_\emptyset=\{0\}$, equation~\eqref{eq:logicalprofile} implies $\log_qK\le0$, a contradiction.
Therefore, we may choose disjoint sets $E_1,E_2\subseteq[n]$
with $|E_1|=|E_2|=d-1$.
Both are globally correctable by Lemma~\ref{lem:global}
and~\eqref{eq:distance}.
Let $T=[n]\setminus(E_1\cup E_2)$.
Since $|T|=n-2(d-1)$, it follows from Theorem~\ref{thm:profilebound} that

\[
 \Lambda+2(d-1)
 \le n-\frac{\dim_pS_T}{m}
       +\frac{\beta(E_1)+\beta(E_2)}{2m}.
\]
If the code is pure, then $G_{E_1}=G_{E_2}=\{0\}$
by Definition~\ref{def:purity}, and hence
$\beta(E_1)=\beta(E_2)=0$.
Since $\dim_pS_T\ge0$, we obtain
$\Lambda+2(d-1)\le n$, which is the pure
subsystem Singleton bound
$KR\le q^{n-2d+2}$ of~\cite[Theorem~11]{alylp}.
\end{remark}

As in Proposition~\ref{prop:redundancy}, we use pairwise disjoint repair sets in $T$ to obtain a lower bound on $\dim_pS_T$.
Substituting this bound into Theorem~\ref{thm:profilebound}, we obtain the following corollary.

\begin{corollary}\label{cor:profilelocal}
Let $E_1,E_2\subseteq[n]$ be disjoint sets of size $d-1$, and let $T=[n]\setminus(E_1\cup E_2)$.
Let $M\in\mathbb Z_{\ge0}$, and let
$\Gamma_1,\ldots,\Gamma_M\subseteq T$ be pairwise disjoint
repair sets of $\mathcal Q$.
For each $j\in[M]$, let $F_j\subseteq\Gamma_j$ with $|F_j|=\delta-1$.
Then
\begin{equation}\label{eq:profilelocal}
 \Lambda+2(d-1)
 +\sum_{j=1}^M
 \left(2(\delta-1)-\frac{\dim_pG_{F_j}}{m}\right)
 \le n+\frac{\beta(E_1)+\beta(E_2)}{2m},
\end{equation}
where $G_{F_j}$ denotes the subspace of gauge labels supported
in $F_j$.
\end{corollary}

\begin{proof}
Since $\Gamma_j\subseteq T$, the subspace of $S_T$ consisting
of labels supported in $\Gamma_j$ satisfies
\[
 \begin{aligned}
 (S_T)_{\Gamma_j}
 &=S_T\cap W_{\Gamma_j}\\
 &=S\cap W_T\cap W_{\Gamma_j}\\
 &=S\cap W_{\Gamma_j}\\
 &=S_{\Gamma_j}.
 \end{aligned}
\]
For $F_j\subseteq\Gamma_j$ with $0<|F_j|=\delta-1<\delta$, condition 3) of Theorem~\ref{thm:local} implies that
$
 \dim_p\pi_{F_j}(S_{\Gamma_j})
 =2m|F_j|-\dim_pG_{F_j}.
$
Since $\Gamma_i\cap\Gamma_j=\emptyset$ for $i\ne j$ and $\Gamma_j\subseteq T$ for every $j\in[M]$, we have
\[
 \bigoplus_{j=1}^M S_{\Gamma_j}\subseteq S_T.
\]
Hence
\[
 \begin{aligned}
 \dim_pS_T
 &\ge\sum_{j=1}^M\dim_pS_{\Gamma_j}\\
 &\ge\sum_{j=1}^M\dim_p\pi_{F_j}(S_{\Gamma_j})\\
 &=\sum_{j=1}^M
\bigl(2m(\delta-1)-\dim_pG_{F_j}\bigr),
 \end{aligned}
\]
which implies that
\[
 \frac{\dim_pS_T}{m}
 \ge\sum_{j=1}^M\left(2(\delta-1)-\frac{\dim_pG_{F_j}}{m}\right).
\]
Since $|T|=n-2(d-1)$, it follows from~\eqref{eq:profilebound} that
\begin{align*}
 &\log_qK+\log_qR\\
 \le& n-2(d-1)-\frac{\dim_pS_T}{m}
       +\frac{\beta(E_1)+\beta(E_2)}{2m}\\
 \le& n-2(d-1)
       -\sum_{j=1}^M\left(2(\delta-1)-\frac{\dim_pG_{F_j}}{m}\right)
       +\frac{\beta(E_1)+\beta(E_2)}{2m},
\end{align*}
which proves~\eqref{eq:profilelocal}.
\end{proof}
\begin{remark}
For $\F_q$-linear gauge spaces, the global subsystem Singleton bound
$k+g\le n-2(d-1)$ also applies~\cite[Theorem~3.6]{klappenecker}.
The bound in Corollary~\ref{cor:profilelocal} depends explicitly on the chosen pairwise disjoint repair sets and supported gauge spaces and requires only $\F_p$-linearity.
\end{remark}

\subsection{A Mixed Counting Bound via Gauge Equivalence Classes}

Local recovery corrects fewer than $\delta$ known erasures within each of several pairwise disjoint repair sets.
We consider these erasures together with errors at unknown positions outside the union of the repair sets, where the local recovery maps act trivially.

Let $b\in\mathbb Z_{\ge0}$, and let $\Gamma_1,\ldots,\Gamma_b$ be repair sets of $\mathcal Q$ satisfying $\Gamma_i\cap\Gamma_j=\emptyset$ for $1\le i<j\le b$.
For each $j\in[b]$, let $E_j\subseteq\Gamma_j$ be a known erased set with $e_j=|E_j|$ and $0\le e_j\le\delta-1$.
Let
\[
 T=\bigcup_{j=1}^b\Gamma_j,~
 U=T^c,~
 N=|U|.
\]
Since the repair sets are pairwise disjoint and
$|\Gamma_j|\le r+\delta-1$, we have
\[
 N=n-\sum_{j=1}^b|\Gamma_j|
 \ge n-b(r+\delta-1).
\]
Let $t\in\mathbb Z_{\ge0}$ satisfy $2t<d$.
Define the set of Pauli labels for errors affecting
at most $t$ qudits at unknown locations in $U$ by
\[
 \mathcal B_U(t)=\{u\in W_U:\swt(u)\le t\}.
\]
For each $j\in[b]$, we replace the erased qudits in $E_j$ with the maximally mixed state $I_{E_j}/q^{e_j}$.
By~\eqref{eq:twirl}, erasure followed by this replacement
is equivalent to applying a uniformly random Pauli operator
supported in $E_j$.
We therefore consider the error-label family
\[
 \mathcal E=\left\{\sum_{j=1}^b a_j+u:
           a_j\in W_{E_j},~u\in\mathcal B_U(t)\right\},
\]
where $a_j$ ranges over all Pauli labels supported in $E_j$.

Two Pauli labels are \emph{gauge-equivalent} if their difference
belongs to $G$.
For $D\in\{E_1,\ldots,E_b,U\}$, let $G_D=G\cap W_D$.
Then, for $w,w'\in W_D$,
\[
 w-w'\in G
 \Longleftrightarrow w-w'\in G_D
 \Longleftrightarrow w+G_D=w'+G_D.
\]
For each $j\in[b]$, let
\[
 \nu_j
 =2me_j-\dim_pG_{E_j}
 =\dim_p(W_{E_j}/G_{E_j}),~
 \nu=\sum_{j=1}^b\nu_j.
\]
Define the number of distinct gauge-equivalence classes
represented in $\mathcal B_U(t)$ by
\[
 M_U(t)=|\{u+G_U:u\in\mathcal B_U(t)\}|.
\]

The following bound compares the number of gauge-equivalence
classes represented in $\mathcal E$ with the number of available
stabilizer syndromes.

\begin{theorem}\label{thm:mixedpacking}
Under the assumptions and notation above,
\begin{equation}\label{eq:mixedpacking}
 p^\nu M_U(t)\le q^{n-\Lambda}.
\end{equation}
Equivalently,
\begin{equation}\label{eq:mixedpackinglog}
 \Lambda+\sum_{j=1}^b
 \left(2e_j-\frac{\dim_pG_{E_j}}{m}\right)
 +\log_qM_U(t)\le n.
\end{equation}
\end{theorem}

\begin{proof}
Let $s_1,\ldots,s_s$ be an $\F_p$-basis of $S$, with
$s=m(n-\Lambda)$.
For each Pauli label $v\in W$, define the syndrome map $\sigma:W\to\F_p^s$ by
\[
 \sigma(v)=([v,s_i]_{\mathrm t})_{i=1}^s.
\]
It follows from the definition of $S^{\perpt}$ and~\eqref{eq:orthogonal-complement} and~\eqref{eq:rank-nullity} that
\[
 \ker\sigma=S^{\perpt},~
 \rank_p\sigma=2mn-\dim_pS^{\perpt}=s.
\]

Define
\[
 \Theta:
 \left(\prod_{j=1}^b W_{E_j}/G_{E_j}\right)
 \times\{u+G_U:u\in\mathcal B_U(t)\}
 \longrightarrow\sigma(\mathcal E)
\]
by
\[
 \Theta(a_1+G_{E_1},\ldots,a_b+G_{E_b},u+G_U)
 =\sigma\left(\sum_{j=1}^b a_j+u\right),
\]
where the representatives satisfy $a_j\in W_{E_j}$ and $u\in\mathcal B_U(t)$.
Suppose that $a_j+G_{E_j}=a'_j+G_{E_j}$ for every $j$
and $u+G_U=u'+G_U$.
Then $a_j-a'_j\in G_{E_j}$ and $u-u'\in G_U$, so
\[
 \sum_{j=1}^b(a_j-a'_j)+(u-u')\in G\subseteq\ker\sigma.
\]
By linearity of $\sigma$,
\[
 \begin{aligned}
 &\sigma\left(\sum_{j=1}^b a_j+u\right)
 -\sigma\left(\sum_{j=1}^b a'_j+u'\right)\\
 \hspace{1em}
 =&\sigma\left(\sum_{j=1}^b(a_j-a'_j)+(u-u')\right)=0.
 \end{aligned}
\]
Thus $\Theta$ does not depend on the choice of representatives, and $\Theta$ is well defined.

To prove injectivity, let $x=\sum_j a_j+u$ and $x'=\sum_j a'_j+u'$ be elements of $\mathcal E$ with the same syndrome, and set $v=x-x'$.
By the linearity of $\sigma$, we have $\sigma(v)=\sigma(x-x')=\sigma(x)-\sigma(x')=0$, and hence $v\in\ker\sigma=S^{\perpt}$.
Fix $j\in[b]$. For every $z\in S_{\Gamma_j}$, we have
\[
 \supp(z)\subseteq\Gamma_j,~
 \supp(u-u')\subseteq U,~
 \supp(a_i-a'_i)\subseteq E_i\subseteq\Gamma_i.
\]
Since $\Gamma_j\cap U=\emptyset$ and
$\Gamma_j\cap\Gamma_i=\emptyset$ for $i\ne j$,
$[u-u',z]_{\mathrm t}=0$ and $[a_i-a'_i,z]_{\mathrm t}=0.$
Since $v\in S^{\perpt}$ and $z\in S_{\Gamma_j}\subseteq S$,
we obtain
\[
 \begin{aligned}
 0=[v,z]_{\mathrm t}
 &=\sum_{i=1}^b[a_i-a'_i,z]_{\mathrm t}
   +[u-u',z]_{\mathrm t}\\
 &=[a_j-a'_j,z]_{\mathrm t}.
 \end{aligned}
\]
Since this holds for every $z\in S_{\Gamma_j}$ and $a_j-a'_j\in W_{E_j}$, it follows from condition 2) of Theorem~\ref{thm:local} that
\[
 a_j-a'_j\in W_{E_j}\cap S_{\Gamma_j}^{\perpt}
 =G_{E_j},
\]
provided that $E_j\ne\emptyset$.
For $E_j=\emptyset$, we have $a_j-a'_j=0\in G_{E_j}=\{0\}$.
Since $v\in S^{\perpt}$ and $a_j-a'_j\in G_{E_j}\subseteq G\subseteq S^{\perpt}$ for every $j$, we have
\[
 u-u'=v-\sum_{j=1}^b(a_j-a'_j)\in S^{\perpt}.
\]
Moreover, since $u,u'\in\mathcal B_U(t)\subseteq W_U$, we have $u-u'\in W_U$ and
\[
 \swt(u-u')\le\swt(u)+\swt(u')\le2t<d.
\]
By~\eqref{eq:distance}, $u-u'\in G_U$.
Consequently, for $j\in[b]$,
\[
 a_j+G_{E_j}=a'_j+G_{E_j},~
 u+G_U=u'+G_U.
\]
Thus equal images under $\Theta$ imply equality of all
component cosets, and $\Theta$ is injective.

For any $y\in\sigma(\mathcal E)$, by the definition of $\mathcal E$, we can choose labels $a_j\in W_{E_j}$ and $u\in\mathcal B_U(t)$ such that
\[
 \begin{aligned}
 y
 &=\sigma\left(\sum_{j=1}^b a_j+u\right)\\
 &=\Theta(a_1+G_{E_1},\ldots,a_b+G_{E_b},u+G_U).
 \end{aligned}
\]
Thus $\Theta$ is surjective and therefore bijective.
By the definitions of $\nu_j$ and $M_U(t)$, we obtain
\begin{align*}
 |\sigma(\mathcal E)|
 &=\left(\prod_{j=1}^b|W_{E_j}/G_{E_j}|\right)M_U(t)\\
 &=\left(\prod_{j=1}^b p^{2me_j-\dim_pG_{E_j}}\right)M_U(t)\\
 &=p^\nu M_U(t)\le|\F_p^s|=p^s,
\end{align*}
where the inequality follows from
$\sigma(\mathcal E)\subseteq\F_p^s$.
Since $p^s=q^{n-\Lambda}$, this proves~\eqref{eq:mixedpacking}.
We take base-$q$ logarithms and use $\log_qp=1/m$
to obtain~\eqref{eq:mixedpackinglog}.
\end{proof}

\begin{remark}
The Pauli error family $\{P_x:x\in\mathcal E\}$ is jointly
correctable on the logical subsystem.
Let $P=VV^\dagger$ be the projector onto $\mathcal Q$.
For $x,x'\in\mathcal E$, distinct syndromes imply
$x'-x\notin S^{\perpt}$ and hence
\[
 P P_x^\dagger P_{x'}P=0.
\]
For equal syndromes, it follows from the proof above that $x'-x\in G$.
Equation~\eqref{eq:encoded-pauli} then shows that
$V^\dagger P_x^\dagger P_{x'}V$ acts only on the gauge
subsystem.
In both cases,
\[
 V^\dagger P_x^\dagger P_{x'}V=I_A\otimes M_{x,x'},~
 M_{x,x'}\in\BB(\HH_B),
\]
where $M_{x,x'}=0$ for distinct syndromes; for equal syndromes,
the scalar Pauli phase is absorbed into $M_{x,x'}$.
Thus the error family satisfies the subsystem Knill-Laflamme condition~\cite[Theorem~3.5]{kribs}.
Recovery of this combined error family may use global syndrome measurements.
The prescribed local recovery operations correct the erasures $E_j$ within their respective repair sets.
\end{remark}

\begin{remark}\label{remarkIV.9}
For $t=0$, we have $M_U(0)=1$. Thus~\eqref{eq:mixedpackinglog} reduces to the disjoint redundancy bound~\eqref{eq:redundancy}, with empty erased sets contributing zero.
For $b=0$, $U=[n]$ and $\nu=0$, so~\eqref{eq:mixedpacking} is a Hamming-type bound that counts gauge cosets represented by errors of weight at most $t$.

For a pure code, every nonzero gauge label has weight at least
$d$ by Definition~\ref{def:purity}.
Since $e_j<\delta\le d$ by
Corollary~\ref{cor:globaldistance}, we have $G_{E_j}=0$.
For distinct $u,u'\in\mathcal B_U(t)$,
$0<\swt(u-u')\le2t<d$.
Purity therefore implies $u-u'\notin G_U$, so the labels
represent distinct gauge cosets.
Let $V_Q(N,t)$ denote the volume of a Hamming ball of radius
$t$ in $N$ coordinates over an alphabet of size $Q$. Then
\[
 M_U(t)=V_{q^2}(N,t),~
 V_Q(N,t):=\sum_{i=0}^{\min(t,N)}\binom Ni(Q-1)^i.
\]
For $b=0$ and $t=\lfloor(d-1)/2\rfloor$, it follows from~\eqref{eq:mixedpacking} that
\[
 V_{q^2}(n,t)\le\frac{q^n}{KR},
\]
which is the pure subsystem Hamming bound
of~\cite[Lemma~13]{alylp}.
\end{remark}

\section{Bounds for Pure $(r,\delta)$-SSLRCs}\label{sec:pure}
In this section, we establish Singleton-type and alphabet-dependent
bounds for pure $(r,\delta)$-SSLRCs, together with explicit Griesmer,
Hamming, and Plotkin bounds. These bounds apply to overlapping repair sets.
Throughout this section, let $\mathcal Q$ be a pure
$(r,\delta)$-SSLRC with parameters $((n,K,R,d))_q$, where
$q=p^m$ and $K>1$.
Let $G$ be its $\F_p$-linear gauge label space and
$S=G\cap G^{\perpt}$ its stabilizer label space.
Set $s=\dim_pS$ and $\Lambda=\log_q(KR)$.
For parameters $[[n,k,g,d]]_q$, we have $\Lambda=k+g$.

\subsection{Reduction to Stabilizer Codes and Classical Comparison Codes}

By~\cite[Lemma~10]{alylp}, a pure $((n,K,R,d))_q$ SSC can be regarded as a pure $((n,KR,d))_q$ stabilizer code.

\begin{proposition}\label{prop:purereduction}
Let $\mathcal Q$ be a pure $(r,\delta)$-SSLRC with parameters
$((n,K,R,d))_q$, where $K>1$.
The common $+1$ eigenspace of its stabilizer group $\mathcal S$
is an $(r,\delta)$-locally recoverable stabilizer code with
parameters $((n,KR,d))_q$ and the same repair cover as $\mathcal Q$.
If $\mathcal Q$ has parameters $[[n,k,g,d]]_q$, then this
stabilizer code has parameters $[[n,k+g,d]]_q$.
\end{proposition}

\begin{proof}
Let $\mathcal Q_0$ denote the common $+1$ eigenspace of
$\mathcal S$, regarded as a stabilizer code.
The physical code space is unchanged, and the entire space
$\HH_A\otimes\HH_B$ is treated as its logical space.
By~\cite[Lemma~10]{alylp}, $\mathcal Q_0$ has dimension $KR$
and minimum distance $d$.
In the subsystem formalism, its gauge label space is
$G_0=S$, and its stabilizer label space remains $S$.

Let $\Gamma$ be a repair set in the prescribed repair cover
of $\mathcal Q$, and let
$\emptyset\ne E\subseteq\Gamma$ with $|E|<\delta$.
By Corollary~\ref{cor:globaldistance}, we have
$|E|<\delta\le d$.
Every label in $G_E$ has symplectic weight at most $|E|<d$,
whereas purity requires every nonzero gauge label to have
weight at least $d$.
Thus $G_E=\{0\}$ by Definition~\ref{def:purity}.
Since $S\subseteq G$, we also have $S_E=\{0\}$.
Since $E$ is recoverable within $\Gamma$ for $\mathcal Q$, condition 2) of Theorem~\ref{thm:local} implies that
\[
 W_E\cap S_\Gamma^{\perpt}
 =G_E=\{0\}=S_E=(G_0)_E.
\]
By the sufficient direction of the same criterion, $E$ is recoverable
within $\Gamma$ for $\mathcal Q_0$.
Since $\Gamma$ and $E$ were arbitrary and the repair-set
sizes are unchanged, $\mathcal Q_0$ has the same
$(r,\delta)$ repair cover as $\mathcal Q$.
\end{proof}

To bound $KR$, we use the additive classical code
$C=\psi(S^{\perpt})$, where $\psi$ is the coordinatewise map
defined in~\eqref{eq:psi}.
For an additive code $C\le\F_{q^2}^{[n]}$, we define its normalized dimension as $\log_{q^2}|C|$, which need not be an integer.
We call $C$ an \emph{additive $(r,\delta)$-cLRC} if it admits
a repair cover satisfying~\eqref{eq:groupsize} and
$d(C|_\Gamma)\ge\delta$ for every repair set $\Gamma$.

\begin{proposition}\label{prop:classicaltransfer}
Let $\mathcal Q$ be a pure $(r,\delta)$-SSLRC with parameters $((n,K,R,d))_q$, where $K>1$.
Then $C=\psi(S^{\perpt})$ is an additive $(r,\delta)$-cLRC with the same repair cover as $\mathcal Q$, minimum distance $d$, and normalized dimension
\begin{equation}\label{eq:classicalparameters}
 \kappa=\log_{q^2}|C|=n-\frac{s}{2m}
 =\frac{n+\Lambda}{2}.
\end{equation}
If $\mathcal Q$ has parameters $[[n,k,g,d]]_q$, then $\kappa=(n+k+g)/2$.
If $C$ is also $\F_{q^2}$-linear, then $\kappa$ is an integer and $C$ has parameters $[n,\kappa,d]_{q^2}$.
\end{proposition}

\begin{proof}
We first determine the normalized dimension of $C$.
Since $S^{\perpt}$ is $\F_p$-linear and $\psi$ is an $\F_p$-linear isomorphism, $C$ is additive.
By~\eqref{eq:orthogonal-complement}, we have
\[
 \begin{gathered}
 |C|=|S^{\perpt}|=p^{2mn-s},\\
 \kappa=\log_{q^2}|C|=\frac{2mn-s}{2m}=n-\frac{s}{2m}.
 \end{gathered}
\]
By~\eqref{eq:dimensions}, $\Lambda=n-s/m$, which proves~\eqref{eq:classicalparameters}.
If $\mathcal Q$ has parameters $[[n,k,g,d]]_q$, then $\Lambda=k+g$, so $\kappa=(n+k+g)/2$.
If $C$ is $\F_{q^2}$-linear, then $\kappa=\dim_{q^2}C$ is an integer.

We next prove that $d(C)=d$.
Let $G$ be the gauge label space of $\mathcal Q$.
By Definition~\ref{def:purity}, every nonzero label in $G$ has symplectic weight at least $d$.
By~\eqref{eq:distance}, every label in $S^{\perpt}\setminus G$ also has symplectic weight at least $d$.
Since $K>1$, the set $S^{\perpt}\setminus G$ is nonempty, and we may choose $\bm w_0\in S^{\perpt}\setminus G$ with $\swt(\bm w_0)=d$.
Since $\psi$ preserves weight, we obtain
\[
 d(C)=\min_{\bm0\ne\bm w\in S^{\perpt}}\swt(\bm w)=d.
\]

Finally, we prove that $C$ has the same repair cover as $\mathcal Q$.
Let $\Gamma$ be a prescribed repair set, and let $c\in C|_\Gamma$ satisfy $\wt(c)<\delta$.
Suppose that $c\ne0$, and choose $\bm v\in S^{\perpt}$ such that $\psi(\bm v)|_\Gamma=c$.
Let $\bm w$ be the zero extension of $\pi_\Gamma(\bm v)$, and set $E=\supp(\bm w)$.
Since $\psi$ acts coordinatewise and preserves support, we have
\[
 \varnothing\ne E\subseteq\Gamma,~
 |E|=\swt(\bm w)=\wt(c)<\delta.
\]
For every $\bm s\in S_\Gamma$, we have
\[
 [\bm w,\bm s]_{\mathrm t}
 =[\bm v,\bm s]_{\mathrm t}=0,
\]
since $\bm w$ and $\bm v$ agree on $\Gamma$ and $\bm s$ is supported in $\Gamma$.
Thus $\bm w\in W_E\cap S_\Gamma^{\perpt}$.
Since $E$ is recoverable within $\Gamma$ for $\mathcal Q$, it follows from condition~2) of Theorem~\ref{thm:local} that
\[
 \bm w\in W_E\cap S_\Gamma^{\perpt}=G_E\subseteq G.
\]
By Corollary~\ref{cor:globaldistance}, we have $\swt(\bm w)<\delta\le d$.
By Definition~\ref{def:purity}, we obtain $\bm w=0$, contrary to $c\ne0$.
Hence $d(C|_\Gamma)\ge\delta$ for every prescribed repair set.
Since the repair sets still cover $[n]$ and satisfy the same size bounds, $C$ is an additive $(r,\delta)$-cLRC with the same repair cover as $\mathcal Q$.
\end{proof}

\subsection{Singleton-Type and Alphabet-Dependent Bounds}

For an additive code $C\le\F_{q^2}^{[n]}$ and
$T\subseteq[n]$, let $\rho_C(T)$ denote the normalized
dimension of the punctured code $C|_T$, that is,
\[
 \rho_C(T)=\log_{q^2}|\pi_T(C)|
 =\frac{\dim_p\pi_T(C)}{2m}.
\]
We obtain the following lemma by adapting the arguments in~\cite[Lemma~1]{cadambe} and~\cite[proof of Corollary~2]{grezet} to additive codes.

\begin{lemma}\label{lem:rankaccumulation}
Let $C=\psi(S^{\perpt})$ be the comparison code in Proposition~\ref{prop:classicaltransfer} with normalized dimension
$\kappa$, and let $\ell=r+\delta-1$.
For every integer $a\ge0$ with $ar<\kappa$, there exists a
coordinate set $T\subseteq[n]$ such that
\[
 |T|\le a\ell,~
 \rho_C(T)\le ar,~
 |T|-\rho_C(T)\ge a(\delta-1).
\]
If $a\ell\le n$, the set $T$ can be enlarged to a set $T'$
satisfying $|T'|=a\ell$ and $\rho_C(T')\le ar$.
\end{lemma}

\begin{proof}
For $a=0$, take $T=T'=\emptyset$.
Assume $a\ge1$.
Expand each coordinate of $C$ in a fixed $\F_p$-basis of
$\F_{q^2}$, and let $M$ be a generator matrix of the resulting
$\F_p$-linear code.
Let
\[
 M=\bigl[M_1\mid M_2\mid\cdots\mid M_n\bigr],
\]
where $M_i$ consists of the $2m$ scalar columns corresponding
to coordinate $i\in[n]$.
For $T\subseteq[n]$, let $M_T$ be the submatrix formed by
the blocks indexed by $T$, and let $\mathcal C(T)$ be its
column space over $\F_p$.
The row space of $M_T$ is the coordinate expansion of
$\pi_T(C)$. Since row rank equals column rank, we have
\[
 \dim_p\mathcal C(T)
 =\rank_pM_T
 =\dim_p\pi_T(C)
 =2m\rho_C(T).
\]

Let $\Gamma$ be a repair set in the prescribed repair cover,
and let $D\subseteq\Gamma$ with $|D|\le\delta-1$.
The puncturing map
$C|_\Gamma\to C|_{\Gamma\setminus D}$ is surjective.
Any codeword in its kernel is supported in $D$ and has weight
at most $\delta-1$.
Since $d(C|_\Gamma)\ge\delta$ by
Proposition~\ref{prop:classicaltransfer}, the puncturing map
has trivial kernel.
By rank-nullity~\eqref{eq:rank-nullity}, the two punctured codes
have the same $\F_p$-dimension.
Consequently,
\[
 \mathcal C(\Gamma\setminus D)\subseteq\mathcal C(\Gamma),~
 \dim_p\mathcal C(\Gamma\setminus D)
 =\dim_p\mathcal C(\Gamma).
\]
Hence $\mathcal C(\Gamma\setminus D)=\mathcal C(\Gamma)$
for every such $D$.
Since $|\Gamma|\ge\delta$, we may take
$D_0\subseteq\Gamma$ with $|D_0|=\delta-1$.
By the equality of the column spaces, we have
$\rho_C(\Gamma)=\rho_C(\Gamma\setminus D_0).$
The matrix $M_{\Gamma\setminus D_0}$ contains
$|\Gamma|-(\delta-1)$ coordinate blocks, each consisting
of $2m$ scalar columns.
Since its rank is at most its number of columns,
\[
 \begin{aligned}
 \rho_C(\Gamma)
 &=\frac{\rank_pM_{\Gamma\setminus D_0}}{2m}\\
 &\le\frac{2m(|\Gamma|-(\delta-1))}{2m}\\
 &=|\Gamma|-(\delta-1)\le r,
 \end{aligned}
\]
where the last inequality follows from
$|\Gamma|\le r+\delta-1$.
Combining the column-space equality with this bound on $\rho_C(\Gamma)$,
we obtain, for every $D\subseteq\Gamma$ with $|D|\le\delta-1$,
\begin{equation}\label{eq:localcolumnspan}
 \begin{aligned}
 \mathcal C(\Gamma\setminus D)
 &=\mathcal C(\Gamma),\\
 \rho_C(\Gamma\setminus D)
 &=\rho_C(\Gamma)
 \le|\Gamma|-(\delta-1)\le r.
 \end{aligned}
\end{equation}

Now let $T\subseteq[n]$ satisfy $\rho_C(T)<\kappa$.
Since $\rho_C([n])=\kappa$, we have
\[
 \dim_p\mathcal C(T)
 =2m\rho_C(T)
 <2m\kappa
 =\dim_p\mathcal C([n]).
\]
Hence there is a coordinate $i\in[n]\setminus T$
whose block $M_i$ contains a column outside $\mathcal C(T)$,
so that $\mathcal C(\{i\})\not\subseteq\mathcal C(T)$.
Let $\Gamma$ be a repair set containing $i$ in the prescribed
repair cover. Then
\[
 \mathcal C(T)
 \subsetneq\mathcal C(T)+\mathcal C(\Gamma)
 =\mathcal C(T\cup\Gamma).
\]
Let $u$ denote the number of new coordinates and $b$ the increase in normalized dimension, that is,
\[
 \begin{aligned}
 u=|\Gamma\setminus T|=|T\cup\Gamma|-|T|,~
 b=\rho_C(T\cup\Gamma)-\rho_C(T)>0.
 \end{aligned}
\]
If $u\le\delta-1$, apply~\eqref{eq:localcolumnspan}
with $D=\Gamma\setminus T$.
Since $\Gamma\setminus D=\Gamma\cap T$, we obtain
$\mathcal C(\Gamma) =\mathcal C(\Gamma\cap T) \subseteq\mathcal C(T).$
It follows that
$\mathcal C(T\cup\Gamma) =\mathcal C(T)+\mathcal C(\Gamma) =\mathcal C(T).$
Thus $\rho_C(T\cup\Gamma)=\rho_C(T)$, and hence $b=0$, contrary to $b>0$.
Hence $u\ge\delta$.
Let $D\subseteq\Gamma\setminus T$ satisfy $|D|=\delta-1$.
By~\eqref{eq:localcolumnspan},
\begin{align*}
 \mathcal C(T\cup\Gamma)
 &=\mathcal C(T)+\mathcal C(\Gamma\setminus D)\\
 &=\mathcal C(T)
   +\mathcal C((\Gamma\setminus T)\setminus D),
\end{align*}
where the second equality uses
$\mathcal C(\Gamma\cap T)\subseteq\mathcal C(T)$.
The set $(\Gamma\setminus T)\setminus D$ has
$u-(\delta-1)$ coordinates, so its column space is generated
by $2m(u-(\delta-1))$ scalar columns.
Therefore,
\[
 \begin{aligned}
 2mb
 &=\dim_p\mathcal C(T\cup\Gamma)-\dim_p\mathcal C(T)\\
 &\le\dim_p\mathcal C((\Gamma\setminus T)\setminus D)\\
 &\le2m\bigl(u-(\delta-1)\bigr).
 \end{aligned}
\]
Since $u\le|\Gamma|\le\ell=r+\delta-1$, we obtain
\[
 u\le\ell,~
 0<b\le u-(\delta-1)\le|\Gamma|-(\delta-1)\le r.
\]
Moreover,
\[
 \begin{aligned}
 &\bigl(|T\cup\Gamma|-\rho_C(T\cup\Gamma)\bigr)
  -\bigl(|T|-\rho_C(T)\bigr)\\
 &\quad=\bigl((|T|+u)-(\rho_C(T)+b)\bigr)
  -\bigl(|T|-\rho_C(T)\bigr)\\
 &\quad=u-b\ge\delta-1.
 \end{aligned}
\]
Thus, for every $T\subseteq[n]$ with $\rho_C(T)<\kappa$,
there exists a repair set $\Gamma$ such that
\[
 \begin{aligned}
 |T\cup\Gamma|-|T|
 &\le\ell,\\
 0<\rho_C(T\cup\Gamma)-\rho_C(T)
 &\le r,\\
 |T\cup\Gamma|-\rho_C(T\cup\Gamma)
 &\ge|T|-\rho_C(T)+\delta-1.
 \end{aligned}
\]

Let $T_0=\emptyset$. We construct $T_1,\ldots,T_a$
inductively. For $0\le j<a$, suppose that $T_j$ has been
constructed.
Since $\rho_C(T_0)=0$ and each of the first $j$ steps
increases the normalized dimension by at most $r$, we have
\[
 \rho_C(T_j)
 \le\rho_C(T_0)+jr
 =jr
 \le ar
 <\kappa.
\]
There is therefore a repair set $\Gamma_{j+1}$ satisfying
the three inequalities above with $T=T_j$.
Set
$T_{j+1}=T_j\cup\Gamma_{j+1}.$
Then
\[
 \rho_C(T_{j+1})
 \le\rho_C(T_j)+r
 \le(j+1)r.
\]
Applying the three increment bounds successively for $a$ steps, we obtain
\[
 |T_a|\le a\ell,~
 \rho_C(T_a)\le ar,~
 |T_a|-\rho_C(T_a)\ge a(\delta-1).
\]
Hence $T=T_a$ satisfies the three required inequalities.

Finally, suppose $a\ell\le n$.
Since $|T_a|\le a\ell$, take $T'\supseteq T_a$ with
$|T'|=a\ell$.
Each of the $a\ell-|T_a|$ added coordinates contributes
$2m$ scalar columns and therefore increases the normalized
dimension by at most one.
Hence
\begin{align*}
 \rho_C(T')
 &\le\rho_C(T_a)+a\ell-|T_a|\\
 &=a\ell-\bigl(|T_a|-\rho_C(T_a)\bigr)\\
 &\le a\ell-a(\delta-1)=ar.
\end{align*}
\end{proof}

We now apply Lemma~\ref{lem:rankaccumulation} to derive a Singleton-type bound for $(r,\delta)$-SSLRCs.

\begin{theorem}\label{thm:singleton}
A pure $(r,\delta)$-SSLRC with parameters $[[n,k,g,d]]_q$
and $k>0$ satisfies
\begin{equation}\label{eq:lrcbound}
 k+g+2(d-1)+2\left(\left\lceil\frac{n+k+g}{2r}\right\rceil-1\right)
 (\delta-1)\le n.
\end{equation}
Moreover, a pure $(r,\delta)$-SSLRC with parameters
$((n,K,R,d))_q$ and $K>1$ satisfies
\begin{equation}\label{eq:classicalbound}
 \Lambda+2(d-1)+2\left(\left\lceil\frac{n+\Lambda}{2r}\right\rceil-1\right)
 (\delta-1)\le n,~\Lambda=\log_q(KR).
\end{equation}
\end{theorem}
\begin{proof}
Let $C=\psi(S^{\perpt})$ and $\kappa=(n+\Lambda)/2$, as in
Proposition~\ref{prop:classicaltransfer}.
Set $a=\lceil\kappa/r\rceil-1$.
Then $a\ge0$ and $ar<\kappa$.
Choose $T$ as in Lemma~\ref{lem:rankaccumulation}.
The kernel of $\pi_T|_C$ is $C_{T^c}$.
The shortened code $D=\operatorname{sh}_{T^c}(C)$ is obtained
from $C_{T^c}$ by deleting its zero coordinates in $T$.
By rank-nullity~\eqref{eq:rank-nullity} and the definition
of shortening, we have
\[
 \log_{q^2}|D|=\kappa-\rho_C(T)>0,~
 d(D)\ge d.
\]
Since the shortened code is nonzero, the cardinality form of the classical Singleton bound~\cite{macwilliams} implies that
\[
 |D|\le(q^2)^{n-|T|-d+1},~
 \kappa-\rho_C(T)\le n-|T|-d+1.
\]
It follows that
\begin{align*}
 d&\le n-\kappa+1-\bigl(|T|-\rho_C(T)\bigr)\\
  &\le n-\kappa+1-a(\delta-1).
\end{align*}
Since $2\kappa=n+\Lambda$, we obtain
\[
 \Lambda+2(d-1)+2a(\delta-1)\le n,~
 a=\left\lceil\frac{n+\Lambda}{2r}\right\rceil-1,
\]
which proves~\eqref{eq:classicalbound}.
For integer subsystem parameters, $\Lambda=k+g$.
Thus~\eqref{eq:lrcbound} follows from~\eqref{eq:classicalbound}.
\end{proof}

\begin{remark}
For $g=0$, \eqref{eq:lrcbound} reduces to the Singleton-type
bound for the pure stabilizer LRCs considered
in~\cite[Theorem~31 and Remark~34]{galindo}.
\end{remark}

\begin{corollary}
A pure $(r,\delta)$-SSLRC with parameters $((n,K,R,d))_q$
and $K>1$ satisfies $r\ge\delta$.
\end{corollary}

\begin{proof}
Let $\kappa=(n+\Lambda)/2$, $t=\delta-1\ge1$, and
$c=\lceil\kappa/r\rceil$.
By Corollary~\ref{cor:globaldistance} and Theorem~\ref{thm:singleton}, we have
\[
 t+1\le d\le n-\kappa+1-(c-1)t,~\kappa+ct\le n.
\]
Suppose $r\le t$. Since $c=\lceil\kappa/r\rceil\ge\kappa/r$,
$t/r\ge1$, and $\Lambda=\log_q(KR)>0$, we obtain
\[
 n\ge\kappa+ct\ge\kappa+\frac{\kappa t}{r}
 \ge2\kappa=n+\Lambda>n.
\]
This contradiction implies $r>\delta-1$, and hence $r\ge\delta$.
\end{proof}

For $Q=q^2$ and integers $N\ge0$ and $d\ge1$, let
$k_{\rm add}^{(Q)}(N,d)$ denote the maximum of $\log_Q|D|$
over $\F_p$-linear codes $D\le\F_Q^N$ in which every nonzero
codeword has Hamming weight at least $d$.
Let $k_{\rm opt}^{(Q)}(N,d)$ denote the maximum of $\dim_QD$
over $\F_Q$-linear codes satisfying the same weight condition.

\begin{theorem}\label{thm:alphabet}
Let $\mathcal Q$ be a pure $(r,\delta)$-SSLRC
with parameters $((n,K,R,d))_q$, where $q=p^m$ and $K>1$.
Let $\Lambda=\log_q(KR)$ and define
\[
 \mathcal A
 =\{a\in\mathbb Z_{\ge0}:
       2ar<n+\Lambda,~a(r+\delta-1)\le n\},
\]
where $a$ counts the repair sets added in Lemma~\ref{lem:rankaccumulation}.
Then
\begin{equation}\label{eq:alphabetbound}
 \Lambda\le\min_{a\in\mathcal A}
 \{2ar+2k_{\rm add}^{(q^2)}(n-a(r+\delta-1),d)-n\}.
\end{equation}
If $\mathcal Q$ has parameters $[[n,k,g,d]]_q$, then
\begin{equation}\label{eq:alphabetinteger}
 k+g\le\min_{\substack{a\in\mathbb Z_{\ge0},~2ar<n+k+g\\
                         a(r+\delta-1)\le n}}
 \{2ar+2k_{\rm add}^{(q^2)}(n-a(r+\delta-1),d)-n\}.
\end{equation}
Moreover, for $C=\psi(S^{\perpt})$
and every $T\subseteq[n]$, we have
\begin{equation}\label{eq:rankprofilebound}
 \Lambda\le2\rho_C(T)+2k_{\rm add}^{(q^2)}(n-|T|,d)-n.
\end{equation}
If $C$ is $\F_{q^2}$-linear, $k_{\rm add}^{(q^2)}$ can be
replaced by $k_{\rm opt}^{(q^2)}$ in all three inequalities.
\end{theorem}

\begin{proof}
Let $\kappa=\log_{q^2}|C|=(n+\Lambda)/2$, as in
Proposition~\ref{prop:classicaltransfer}.
For every $T\subseteq[n]$, we have the exact sequence
\[
 0\longrightarrow C_{T^c}\longrightarrow C
 \xrightarrow{\ \pi_T\ }C|_T\longrightarrow0.
\]
The shortened code $D_T=\operatorname{sh}_{T^c}(C)$ is obtained
from $C_{T^c}$ by deleting the zero coordinates in $T$,
which preserves its $\F_p$-dimension and Hamming weights.
By rank-nullity~\eqref{eq:rank-nullity}, we have
\[
 \log_{q^2}|D_T|=\kappa-\rho_C(T),~
 d(D_T)\ge d.
\]
By the definition of $k_{\rm add}^{(q^2)}$, we have
\begin{align*}
 \kappa-\rho_C(T)
 &\le k_{\rm add}^{(q^2)}(n-|T|,d),\\
 n+\Lambda=2\kappa
 &\le2\rho_C(T)+2k_{\rm add}^{(q^2)}(n-|T|,d).
\end{align*}
Thus~\eqref{eq:rankprofilebound} holds.

Let $\ell=r+\delta-1$ and $a\in\mathcal A$.
Then $ar<\kappa$ and $a\ell\le n$. By Lemma~\ref{lem:rankaccumulation}, we can choose a set $T'$ with
$|T'|=a\ell$ and $\rho_C(T')\le ar$. Applying~\eqref{eq:rankprofilebound} to $T'$, we obtain
\begin{align*}
 \Lambda
 &\le2\rho_C(T')+2k_{\rm add}^{(q^2)}(n-a\ell,d)-n\\
 &\le2ar+2k_{\rm add}^{(q^2)}(n-a(r+\delta-1),d)-n.
\end{align*}
Taking the minimum over $a\in\mathcal A$, we obtain~\eqref{eq:alphabetbound}.
For integer subsystem parameters, $\Lambda=k+g$,
which implies~\eqref{eq:alphabetinteger}.

If $C$ is $\F_{q^2}$-linear, every shortened code $D_T$
is also $\F_{q^2}$-linear.
In this case,
\[
 \log_{q^2}|D_T|
 =\dim_{q^2}D_T
 \le k_{\rm opt}^{(q^2)}(n-|T|,d).
\]
The same argument establishes all three inequalities with $k_{\rm opt}^{(q^2)}$ in place of $k_{\rm add}^{(q^2)}$.
\end{proof}
\begin{remark}
For $g=0$ and an $\F_{q^2}$-linear comparison code
$C=\psi(S^{\perpt})$, the linear version
of~\eqref{eq:alphabetbound} reduces to the classical
alphabet-dependent bound in~\cite[Corollary~2]{grezet}
under the substitution $\kappa=(n+k)/2$, with the same
admissible range of $a$.
\end{remark}

\begin{remark}
The alphabet-dependent bound~\eqref{eq:alphabetbound}
is at least as strong as the Singleton-type
bound~\eqref{eq:classicalbound}.
By the classical Singleton bound~\cite{macwilliams}, we have
\[
 k_{\rm add}^{(q^2)}(N,d)\le\max\{0,N-d+1\}.
\]
Let $a_*=\lceil\kappa/r\rceil-1$.
Then $a_*\ge0$ and $a_*r<\kappa$.
By Theorem~\ref{thm:singleton} and
Corollary~\ref{cor:globaldistance}, we have
\[
 \begin{aligned}
 n
 &\ge\kappa+(a_*+1)(\delta-1)\\
 &>a_*r+a_*(\delta-1)\\
 &=a_*(r+\delta-1).
 \end{aligned}
\]
Thus $a_*\in\mathcal A$.
For $a=a_*$, let $T'$ be the coordinate set used in the proof
of Theorem~\ref{thm:alphabet}.
The shortened code $D_{T'}$ has length
$N_*=n-a_*(r+\delta-1)$, and
\[
 \log_{q^2}|D_{T'}|
 =\kappa-\rho_C(T')
 \ge\kappa-a_*r
 >0.
\]
Hence $D_{T'}$ is nonzero.
Since $d(D_{T'})\ge d$, its length satisfies $N_*\ge d$.
By the Singleton estimate, we therefore have
$k_{\rm add}^{(q^2)}(N_*,d)\le N_*-d+1$.
At $a=a_*$, \eqref{eq:alphabetbound} yields
\[
 \begin{aligned}
 \Lambda
 &\le2a_*r+2k_{\rm add}^{(q^2)}(N_*,d)-n\\
 &\le2a_*r+2(N_*-d+1)-n\\
 &=n-2(d-1)-2a_*(\delta-1).
 \end{aligned}
\]
This is equivalent to~\eqref{eq:classicalbound}.
If $k_{\rm add}^{(q^2)}(N_*,d)<N_*-d+1$, the bound
at $a=a_*$ is strictly stronger.
\end{remark}

\subsection{Explicit Griesmer, Hamming and Plotkin Bounds}

Let $C=\psi(S^{\perpt})\le\F_{q^2}^{[n]}$ be the additive
classical code in Proposition~\ref{prop:classicaltransfer}.
Recall that $\kappa=(n+\Lambda)/2$ and $\ell=r+\delta-1$,
and define
\begin{align*}
 \mathcal A_+
 &=\{a\in\mathbb Z_{\ge0}:ar<\kappa,~a\ell<n\},\\
 N_a&=n-a\ell,~
 b_a=2m(\kappa-ar),~
 M_a=p^{b_a}=q^{n+\Lambda-2ar}.
\end{align*}
For each $a\in\mathcal A_+$, by Lemma~\ref{lem:rankaccumulation}, we can choose a set $T'$ with $|T'|=a\ell$ and $\rho_C(T')\le ar$.
We select all codewords of $C$ that vanish on $T'$ and delete these coordinates from each selected codeword. The resulting shortened code
$D_a=\operatorname{sh}_{(T')^c}(C)$ has length
$n-|T'|=n-a\ell=N_a$.
By rank-nullity~\eqref{eq:rank-nullity}, we have
\[
 \dim_pD_a
 =2m\bigl(\kappa-\rho_C(T')\bigr)
 \ge2m(\kappa-ar)=b_a.
\]
Since $|D_a|=p^{\dim_pD_a}$ and shortening does not
decrease the minimum distance, we obtain
\begin{equation}\label{eq:explicitshortening}
 \dim_pD_a\ge b_a,~
 |D_a|\ge M_a,~
 d(D_a)\ge d.
\end{equation}

\begin{theorem}\label{cor:griesmer}
Let $\mathcal Q$ be a pure $(r,\delta)$-SSLRC with parameters
$((n,K,R,d))_q$, where $q=p^m$ and $K>1$.
Set $L_p=(p^{2m}-1)/(p-1)$ and $\omega_p=p^{2m-1}$.
For every $a\in\mathcal A_+$,
\begin{equation}\label{eq:additivegriesmer}
 n\ge a(r+\delta-1)+
 \left\lceil\frac{1}{L_p}\sum_{i=0}^{b_a-1}
 \left\lceil\frac{\omega_p d}{p^i}\right\rceil\right\rceil.
\end{equation}
If $b_a\ge2m$, \eqref{eq:additivegriesmer} is equivalent to
\begin{equation}\label{eq:additivegriesmersimplified}
 n\ge a(r+\delta-1)+d+
 \left\lceil\frac{p-1}{p^{2m}-1}
 \sum_{j=1}^{b_a-2m}\left\lceil\frac{d}{p^j}\right\rceil\right\rceil.
\end{equation}
If $C=\psi(S^{\perpt})$ is $\F_{q^2}$-linear, then
$h_a=\kappa-ar$ is a positive integer and
\begin{equation}\label{eq:lineargriesmer}
 n\ge a(r+\delta-1)+\sum_{i=0}^{h_a-1}
 \left\lceil\frac{d}{q^{2i}}\right\rceil.
\end{equation}
\end{theorem}

\begin{proof}
Fix $a\in\mathcal A_+$.
By~\eqref{eq:explicitshortening}, $D_a$ contains an
$\F_p$-linear subcode $D'_a$ of dimension $b_a$.
Its minimum distance is at least $d$.
Choose an $\F_p$-basis of $\F_{q^2}$ and identify each symbol
with its coordinate vector in $\F_p^{2m}$.
We concatenate $D'_a$ with a $p$-ary simplex code with
parameters $[L_p,2m,\omega_p]_p$, as in the proof
of~\cite[Theorem~12]{ball}.
The simplex encoder is $\F_p$-linear and injective, and every
nonzero input vector is mapped to a codeword of weight $\omega_p$.
The concatenated code therefore has parameters
\[
 [L_pN_a,b_a,d'_a]_p,~d'_a\ge\omega_p d.
\]
By the classical Griesmer bound~\cite{macwilliams},
\[
 L_pN_a
 \ge\sum_{i=0}^{b_a-1}\left\lceil\frac{d'_a}{p^i}\right\rceil
 \ge\sum_{i=0}^{b_a-1}
 \left\lceil\frac{\omega_p d}{p^i}\right\rceil.
\]
We divide by $L_p$ and take the ceiling, since $N_a$ is an integer. Using $n=a\ell+N_a$, we obtain~\eqref{eq:additivegriesmer}.

Suppose $b_a\ge2m$.
Since $\omega_p=p^{2m-1}$, the first $2m$ terms and
the remaining terms satisfy
\begin{align*}
 \sum_{i=0}^{2m-1}
 \left\lceil\frac{p^{2m-1}d}{p^i}\right\rceil
 &=d\sum_{j=0}^{2m-1}p^j=L_pd,\\
 \sum_{i=2m}^{b_a-1}
 \left\lceil\frac{p^{2m-1}d}{p^i}\right\rceil
 &=\sum_{j=1}^{b_a-2m}\left\lceil\frac{d}{p^j}\right\rceil.
\end{align*}
The second sum is zero when $b_a=2m$.
Since $d$ is an integer, we have
\[
 \left\lceil\frac{1}{L_p}\sum_{i=0}^{b_a-1}
 \left\lceil\frac{\omega_p d}{p^i}\right\rceil\right\rceil
 =d+\left\lceil\frac{1}{L_p}\sum_{j=1}^{b_a-2m}
 \left\lceil\frac{d}{p^j}\right\rceil\right\rceil.
\]
Thus~\eqref{eq:additivegriesmer} takes the form~\eqref{eq:additivegriesmersimplified}, since $1/L_p=(p-1)/(p^{2m}-1)$.

If $C$ is $\F_{q^2}$-linear, so is $D_a$, and
$\dim_{q^2}D_a\ge\kappa-ar=h_a$.
Choose an $h_a$-dimensional $\F_{q^2}$-linear subcode of $D_a$.
Since its minimum distance is at least $d$, the classical
Griesmer bound over $\F_{q^2}$ implies
\[
 N_a\ge\sum_{i=0}^{h_a-1}
 \left\lceil\frac{d}{q^{2i}}\right\rceil.
\]
Since $N_a=n-a(r+\delta-1)$, we
obtain~\eqref{eq:lineargriesmer}.
\end{proof}

\begin{remark}
For $1\le d\le q^2$, the $\F_{q^2}$-linear
bound~\eqref{eq:lineargriesmer} reduces to the Singleton
inequality $N_a\ge d+h_a-1$.
\end{remark}

\begin{theorem}\label{cor:hammingplotkin}
Let $\mathcal Q$ be a pure $(r,\delta)$-SSLRC with parameters
$((n,K,R,d))_q$, where $q=p^m$ and $K>1$.
Let $t=\lfloor(d-1)/2\rfloor$.
For every $a\in\mathcal A_+$,
\begin{align}
 \Lambda+2a(\delta-1)+\log_q V_{q^2}(N_a,t)
 &\le n,\nonumber\\
 d&\le
 \left\lfloor
 \frac{(q^2-1)N_aM_a}{q^2(M_a-1)}
 \right\rfloor,
 \label{eq:pureplotkin}
\end{align}
where $V_{q^2}(N_a,t)$ is the Hamming-ball volume defined in
Remark~\ref{remarkIV.9}.
\end{theorem}

\begin{proof}
Fix $a\in\mathcal A_+$, and let $Q=q^2$ and $M=|D_a|$.
By~\eqref{eq:explicitshortening}, $M\ge M_a>1$ and
$d(D_a)\ge d$.
Since $2t<d$, the Hamming balls of radius $t$ centered at
distinct codewords of $D_a$ are disjoint.
By the classical Hamming bound~\cite{macwilliams}, we have
\[
 M_aV_Q(N_a,t)\le MV_Q(N_a,t)\le Q^{N_a}.
\]
We take base-$q$ logarithms and use
$M_a=q^{n+\Lambda-2ar}$ and $N_a=n-a\ell$ to obtain
\begin{align*}
 n+\Lambda-2ar+\log_qV_Q(N_a,t)
 &\le2(n-a\ell),\\
 \Lambda+2a(\ell-r)+\log_qV_Q(N_a,t)
 &\le n.
\end{align*}
Since $\ell-r=\delta-1$, we obtain the stated Hamming bound.

By the distance form of the classical Plotkin bound~\cite{macwilliams}, we have
\[
 d\le d(D_a)\le\frac{(Q-1)N_aM}{Q(M-1)}.
\]
The inequality $M\ge M_a>1$ implies
\[
 \frac{M}{M-1}-\frac{M_a}{M_a-1}
 =\frac{M_a-M}{(M-1)(M_a-1)}\le0.
\]
Consequently,
\[
 d\le\frac{(Q-1)N_aM_a}{Q(M_a-1)}.
\]
Since $d$ is an integer, we take the floor of the right-hand side and substitute $Q=q^2$ to obtain~\eqref{eq:pureplotkin}.
\end{proof}

\section{Some Constructions of Optimal $\F_q$-Linear Pure $(r,\delta)$-SSLRCs}\label{sec:construction}
In this section, we construct nine classes of $\F_q$-linear pure $(r,\delta)$-SSLRCs from MDS stabilizer codes, RS codes, and linear codes with MDS Hermitian hulls. The explicit families have $d=\delta$ and simultaneously attain the Singleton-type bound~\eqref{eq:lrcbound}, the alphabet-dependent bound~\eqref{eq:alphabetbound}, the $\F_{q^2}$-linear Griesmer bound~\eqref{eq:lineargriesmer}, and the disjoint redundancy bound~\eqref{eq:pureredundancy}. These optimal constructions are summarized in Table~\ref{tab:constructions}, where $q$ is a prime power, $B\ge1$, and all remaining parameters are integers.

\begin{table*}[!htbp]
\centering
\caption{Optimal pure $\F_q$-linear $(r,\delta)$-SSLRCs attaining the bounds~\eqref{eq:lrcbound},~\eqref{eq:alphabetbound},~\eqref{eq:lineargriesmer} and~\eqref{eq:pureredundancy}.}
\label{tab:constructions}
\small
\renewcommand{\arraystretch}{1.2}
\setlength{\tabcolsep}{4pt}
\resizebox{\textwidth}{!}{%
\begin{tabular}{@{}llll@{}}
\toprule
Parameters & $(r,\delta)$ & Conditions & References\\
\midrule
$[[BN,B(N-2d_0+2)-g,g,d_0]]_q$
& $(N-d_0+1,d_0)$
& \begin{tabular}[t]{@{}l@{}}
$3\le N\le q$,\\
$2\le d_0\le\lfloor(N+1)/2\rfloor$,\\
$0\le g\le B(N-2d_0+2)-1$
\end{tabular}
& Cor.~\ref{cor:mdsfamilies}(i)\\
\addlinespace[3pt]
$[[B(\nu+1)q,B(\nu+1)(q-2)-g,g,\nu+2]]_q$
& $((\nu+1)(q-1),\nu+2)$
& \begin{tabular}[t]{@{}l@{}}
$q\ge3$, $0\le\nu\le q-2$,\\
$0\le g\le B(\nu+1)(q-2)-1$
\end{tabular}
& Cor.~\ref{cor:mdsfamilies}(ii)\\
\addlinespace[3pt]
$[[B(q-1),B(q-2s-1)-g,g,s+1]]_q$
& $(q-s-1,s+1)$
& \begin{tabular}[t]{@{}l@{}}
$1\le s<(q-1)/2$,\\
$0\le g\le B(q-2s-1)-1$
\end{tabular}
& Cor.~\ref{cor:mdsfamilies}(iii)\\
\addlinespace[3pt]
$[[Bq,B(q-2s-2)-g,g,s+2]]_q$
& $(q-s-1,s+2)$
& \begin{tabular}[t]{@{}l@{}}
$0\le s\le\lfloor(q-3)/2\rfloor$,\\
$0\le g\le B(q-2s-2)-1$
\end{tabular}
& Cor.~\ref{cor:mdsfamilies}(iv)\\
\addlinespace[3pt]
$[[B(q^2-1),B(q^2-2s-1)-g,g,s+1]]_q$
& $(q^2-s-1,s+1)$
& \begin{tabular}[t]{@{}l@{}}
$1\le s<q-1$,\\
$0\le g\le B(q^2-2s-1)-1$
\end{tabular}
& Cor.~\ref{cor:mdsfamilies}(v)\\
\addlinespace[3pt]
$[[Bq^2,B(q^2-2s-2)-g,g,s+2]]_q$
& $(q^2-s-1,s+2)$
& \begin{tabular}[t]{@{}l@{}}
$0\le s<q-1$,\\
$0\le g\le B(q^2-2s-2)-1$
\end{tabular}
& Cor.~\ref{cor:mdsfamilies}(vi)\\
\addlinespace[3pt]
$[[Bq,B(q-2t)-g,g,t+1]]_q$
& $(q-t,t+1)$
& \begin{tabular}[t]{@{}l@{}}
$1\le t<q/2$,\\
$0\le g\le B(q-2t)-1$
\end{tabular}
& Thm.~\ref{thm:rsfamily}\\
\addlinespace[3pt]
$[[BN,B(N-2h_H)-g,g,h_H+1]]_q$
& $(N-h_H,h_H+1)$
& \begin{tabular}[t]{@{}l@{}}
$1\le h_H<c$,\\
$c+h_H<N\le q-1$,\\
$0\le g\le B(N-2h_H)-1$
\end{tabular}
& Thm.~\ref{thm:explicitgrs}\\
\addlinespace[3pt]
$[[Bq^2,B(q^2-2q+2)-g,g,q]]_q$
& $(q^2-q+1,q)$
& $0\le g\le B(q^2-2q+2)-1$
& Prop.~\ref{prop:fullfield}\\
\bottomrule
\end{tabular}%
}
\end{table*}

\subsection{Tensor Products and Dimension Trading}
In this subsection, we determine the parameters of tensor products of SSCs and give conditions under which their coordinate blocks form repair sets.
We then show how to trade the logical and gauge dimensions of pure $(r,\delta)$-SSLRCs while preserving the minimum distance and the repair cover.

\begin{proposition}\label{prop:blocks}
Let $B\ge1$ be an integer.
For each $j\in[B]$, let $\mathcal Q_j$ be an SSC over $\F_q$ with parameters $[[\ell_j,k_j,g_j,d_j]]_q$, where $k_j>0$.
Set $n=\sum_{j=1}^{B}\ell_j$ and $\mathcal Q=\bigotimes_{j=1}^{B}\mathcal Q_j$.
Let $\Gamma_j$ denote the coordinate block of $\mathcal Q_j$ in $\mathcal Q$ so that $[n]=\bigsqcup_{j=1}^{B}\Gamma_j$ and $|\Gamma_j|=\ell_j$.
Then the following statements hold.
\begin{itemize}
  \item [(i)] $\mathcal Q$ is an SSC with parameters
\[
 \left[\left[
 n,\ \sum_{j=1}^{B}k_j,\ \sum_{j=1}^{B}g_j,\
 \min_{j\in[B]}d_j
 \right]\right]_q.
\]
  \item [(ii)] If each $\mathcal Q_j$ is pure, then $\mathcal Q$ is pure.
  \item [(iii)] If $r\ge1$ and $\delta\ge2$ are integers satisfying $\delta\le\min_{j\in[B]}d_j$ and $\delta\le\ell_j\le r+\delta-1$ for every $j\in[B]$,
then $\mathcal Q$ is an $(r,\delta)$-SSLRC with $\Gamma_1,\ldots,\Gamma_B$ as repair sets.
\end{itemize}
\end{proposition}

\begin{proof}
For each $j\in[B]$, let $G_j\le\F_q^{\Gamma_j}\times\F_q^{\Gamma_j}$ be the gauge label space of $\mathcal Q_j$, and set $S_j=G_j\cap G_j^{\perpt}$, where $G_j^{\perpt}$ and $S_j^{\perpt}$ denote the orthogonal complements taken in $\F_q^{\Gamma_j}\times\F_q^{\Gamma_j}$.
We extend $G_j$, $S_j$, and their orthogonal complements by zero outside $\Gamma_j$.
The tensor product $\mathcal Q$ has gauge label space $G=\bigoplus_{j=1}^{B}G_j$.
Since labels supported on distinct blocks are orthogonal under the trace-symplectic form, we obtain
\[
 \begin{aligned}
 G^{\perpt}
 &=\bigoplus_{j=1}^{B}G_j^{\perpt},\\
 S=G\cap G^{\perpt}
 &=\bigoplus_{j=1}^{B}(G_j\cap G_j^{\perpt})
  =\bigoplus_{j=1}^{B}S_j,\\
 S^{\perpt}
 &=\bigoplus_{j=1}^{B}S_j^{\perpt}.
 \end{aligned}
\]
By~\eqref{eq:dimensions} and the additivity of dimensions under direct sums,
\[
 K=\prod_{j=1}^{B}q^{k_j}
  =q^{\sum_{j=1}^{B}k_j},~
 R=\prod_{j=1}^{B}q^{g_j}
  =q^{\sum_{j=1}^{B}g_j}.
\]
Let $\bm{w}=\sum_{j=1}^{B}\bm{w}_j\in S^{\perpt}\setminus G$, where $\bm{w}_j\in S_j^{\perpt}$.
Then $\bm{w}_j\notin G_j$ for some $j\in[B]$.
By~\eqref{eq:distance},
\[
 \swt(\bm{w})
 =\sum_{i=1}^{B}\swt(\bm{w}_i)
 \ge\swt(\bm{w}_j)
 \ge d_j
 \ge\min_{i\in[B]}d_i,
\]
which implies that $d(\mathcal Q)\geq\min_{i\in[B]}d_i$.
Conversely, choose $j_0\in[B]$ with $d_{j_0}=\min_{j\in[B]}d_j$.
Since $k_{j_0}>0$, we may choose $\bm{w}_0\in S_{j_0}^{\perpt}\setminus G_{j_0}$ of weight $d_{j_0}$ by~\eqref{eq:distance}.
Its zero extension belongs to $S^{\perpt}\setminus G$ and has the same weight.
Hence $d(\mathcal Q)=\min_{j\in[B]}d_j$.
This proves (i).

Suppose that every $\mathcal Q_j$ is pure.
For any nonzero $\bm{v}=\sum_{j=1}^{B}\bm{v}_j\in G$, choose $j$ with $\bm{v}_j\ne\bm{0}$.
By Definition~\ref{def:purity},
\[
 \swt(\bm{v})
 =\sum_{i=1}^{B}\swt(\bm{v}_i)
 \ge\swt(\bm{v}_j)
 \ge d_j
 \ge d(\mathcal Q).
\]
Thus $\mathcal Q$ is pure, proving (ii).

To prove (iii), fix $j\in[B]$ and $\emptyset\ne E\subseteq\Gamma_j$ with $|E|<\delta$.
Since $|E|<\delta\le d_j$, no label in $S_j^{\perpt}\setminus G_j$ is supported in $E$ by~\eqref{eq:distance}.
Together with $G_j\subseteq S_j^{\perpt}$, this implies $W_E\cap S_j^{\perpt}=(G_j)_E$.
Since $S_{\Gamma_j}=S_j$, the orthogonal complement of $S_{\Gamma_j}$ in $W$ satisfies
\[
 S_{\Gamma_j}^{\perpt}
 =S_j^{\perpt}\oplus W_{\Gamma_j^c}.
\]
Since $E\subseteq\Gamma_j$ and $G=\bigoplus_{i=1}^{B}G_i$, we have
\[
 \begin{aligned}
 W_E\cap S_{\Gamma_j}^{\perpt}
 =W_E\cap S_j^{\perpt}
 =(G_j)_E
  =G_E.
 \end{aligned}
\]
By condition~2) of Theorem~\ref{thm:local}, $E$ is recoverable within $\Gamma_j$.
The blocks partition $[n]$ and satisfy $\delta\le|\Gamma_j|\le r+\delta-1$.
Therefore, $\mathcal Q$ is an $(r,\delta)$-SSLRC with the stated repair sets.
This completes the proof.
\end{proof}

By~\cite[Theorem~7]{aly}, an $\F_q$-linear MDS stabilizer code with parameters $[[n,k_0,d]]_q$ can be converted into a pure $\F_q$-linear SSC with parameters $[[n,k_0-\gamma,\gamma,d]]_q$ for $0\le\gamma<k_0$.
We next apply the symplectic construction underlying~\cite[Theorems~2 and~3]{aly} to trade dimensions in pure $\F_q$-linear $(r,\delta)$-SSLRCs with a prescribed repair cover.

\begin{proposition}\label{prop:local-dimension-trading}
Let $\mathcal Q$ be a pure $\F_q$-linear $(r,\delta)$-SSLRC with parameters $[[n,k,g,d]]_q$, where $k>0$.
Let $S$ be its stabilizer label space, and fix a repair cover $\mathscr R$ of $\mathcal Q$.
For every integer $0\le\gamma\le k+g-1$, there exists a pure $\F_q$-linear $(r,\delta)$-SSLRC $\mathcal Q_\gamma$ with parameters $[[n,k+g-\gamma,\gamma,d]]_q$, stabilizer label space $S$, and repair cover $\mathscr R$.
\end{proposition}

\begin{proof}
We first construct $\F_q$-linear spaces $G_\gamma$ with $G_\gamma\cap G_\gamma^{\perpt}=S$.
Let $G$ be the gauge label space of $\mathcal Q$.
Since $S=G\cap G^{\perpt}$, we have $S\subseteq G$.
For any $\bm v\in G$ and $\bm s\in S\subseteq G^{\perpt}$, we have $[\bm v,\bm s]_{\mathrm t}=-[\bm s,\bm v]_{\mathrm t}=0$.
Hence $G\subseteq S^{\perpt}$.
By Definition~\ref{def:purity}, every nonzero label in $G$ has weight at least $d$.
By~\eqref{eq:distance}, every label in $S^{\perpt}\setminus G$ also has weight at least $d$.
Therefore,
\begin{equation}\label{eq:trading-centralizer-weight}
 \swt(\bm v)\ge d
 ~
 \text{for every }\bm0\ne\bm v\in S^{\perpt}.
\end{equation}
Since $k>0$, choose $\bm w_*\in S^{\perpt}\setminus G$ with $\swt(\bm w_*)=d$ by~\eqref{eq:distance}.
Since $S\subseteq G$, we have $\bm w_*\notin S$.
By Corollary~\ref{cor:globaldistance}, we have $d\ge\delta\ge2$.
Hence $S\ne\{\bm0\}$; otherwise, $S^{\perpt}=W$ would contain a label of weight one, contrary to~\eqref{eq:trading-centralizer-weight}.
Since $S$ is $\F_q$-linear, we have $S^{\perpt}=S^{\perps}$.
By~\eqref{eq:qdimensions} and~\eqref{eq:orthogonal-complement}, we obtain
\[
 \dim_qS=n-k-g,
 ~
 \dim_qS^{\perps}-\dim_qS=2(k+g).
\]
Moreover, $S^{\perps}\cap(S^{\perps})^{\perps}=S.$ Set $\bm x_1=\bm w_*$.
Since $(S^{\perps})^{\perps}=S$ and $\bm x_1\notin S$, choose $\bm z_1\in S^{\perps}$ with $[\bm x_1,\bm z_1]_{\mathrm s}=1$.
By symplectic basis extension, choose $\bm x_i,\bm z_i\in S^{\perps}$ for $2\le i\le k+g$ such that
\[
 S^{\perps}
 =
 S\oplus
 \operatorname{span}_{\F_q}
 \{\bm x_i,\bm z_i:1\le i\le k+g\}
\]
and for $1\le i,j\le k+g$,
\[
 \begin{aligned}
 [\bm x_i,\bm z_j]_{\mathrm s}=\delta_{ij},
 [\bm x_i,\bm x_j]_{\mathrm s}
 =[\bm z_i,\bm z_j]_{\mathrm s}=0.
 \end{aligned}
\]
For $0\le\gamma\le k+g-1$, define
\begin{equation}\label{eq:51}
\begin{aligned}
 G_\gamma
 =
 S\oplus
 \operatorname{span}_{\F_q}
 \{\bm x_i,\bm z_i:2\le i\le\gamma+1\},
\end{aligned}
\end{equation}
where the span of the empty set is $\{\bm0\}$.
Since $S\subseteq G_\gamma\subseteq S^{\perps}$, we have $S\subseteq G_\gamma\cap G_\gamma^{\perps}.$ Conversely, let $\bm v\in G_\gamma\cap G_\gamma^{\perps}$ and write
\[
 \bm v
 =
 \bm s+\sum_{i=2}^{\gamma+1}
 (a_i\bm x_i+b_i\bm z_i),
\]
where $\bm s\in S$ and $a_i,b_i\in\F_q$. For every $2\le j\le\gamma+1$, we have
\[
 a_j=[\bm v,\bm z_j]_{\mathrm s}=0,
 ~
 -b_j=[\bm v,\bm x_j]_{\mathrm s}=0.
\]
These equalities follow from the symplectic basis relations and $\bm v\in G_\gamma^{\perps}$.

Hence $\bm v=\bm s\in S$.
Since $G_\gamma$ is $\F_q$-linear, we conclude that
\[
 G_\gamma\cap G_\gamma^{\perpt}
 =
 G_\gamma\cap G_\gamma^{\perps}
 =
 S.
\]

We next determine the logical and gauge dimensions.
By~\cite[Theorem~1]{aly}, we obtain an $\F_q$-linear SSC $\mathcal Q_\gamma$ with gauge label space $G_\gamma$ and stabilizer label space $S$.
By~\eqref{eq:51}, we have
\[
 \dim_qG_\gamma
 =\dim_qS+2\gamma
 =n-k-g+2\gamma.
\]
By~\eqref{eq:qdimensions}, we obtain
\[
 \begin{aligned}
 k_\gamma
 &=n-\frac{\dim_qG_\gamma+\dim_qS}{2}
   =k+g-\gamma>0,\\
 g_\gamma
 &=\frac{\dim_qG_\gamma-\dim_qS}{2}
   =\gamma.
 \end{aligned}
\]

We next prove that $d(\mathcal Q_\gamma)=d$ and that $\mathcal Q_\gamma$ is pure.
By the direct sum decomposition of $S^{\perps}$, we obtain $\bm w_*=\bm x_1\notin G_\gamma$.
Thus $\bm w_*\in S^{\perpt}\setminus G_\gamma$.
By~\eqref{eq:distance} and~\eqref{eq:trading-centralizer-weight}, we obtain $d\le d(\mathcal Q_\gamma)\le\swt(\bm w_*)=d.$ Hence $d(\mathcal Q_\gamma)=d$.
For every $\bm0\ne\bm v\in G_\gamma\subseteq S^{\perpt}$, we also have $\swt(\bm v)\ge d$ by~\eqref{eq:trading-centralizer-weight}.
Thus $\mathcal Q_\gamma$ is pure by Definition~\ref{def:purity}.

Finally, we prove that $\mathscr R$ is a repair cover of $\mathcal Q_\gamma$.
Fix $\Gamma\in\mathscr R$ and $\emptyset\ne E\subseteq\Gamma$ with $|E|<\delta$.
For every $\bm v\in W_E$, we have $\swt(\bm v)\le|E|<\delta\le d$.
By~\eqref{eq:trading-centralizer-weight}, we obtain $W_E\cap S^{\perpt}=\{\bm0\}.$ Since $G,G_\gamma\subseteq S^{\perpt}$, we have $G_E=(G_\gamma)_E=\{\bm0\}$.
By condition~2) of Theorem~\ref{thm:local}, applied to $\mathcal Q$, we obtain $W_E\cap S_\Gamma^{\perpt} =G_E =\{\bm0\} =(G_\gamma)_E.$ Since $\mathcal Q_\gamma$ has the same stabilizer label space $S$, we conclude by the same theorem that $E$ is recoverable within $\Gamma$ for $\mathcal Q_\gamma$.
Since $\Gamma$ and $E$ were arbitrary, $\mathscr R$ is also a repair cover of $\mathcal Q_\gamma$.
Therefore, for every $0\le\gamma\le k+g-1$, there exists a pure $\F_q$-linear $(r,\delta)$-SSLRC with parameters $[[n,k+g-\gamma,\gamma,d]]_q$, stabilizer label space $S$, and repair cover $\mathscr R$.
\end{proof}

Combining Propositions~\ref{prop:blocks} and~\ref{prop:local-dimension-trading}, we obtain the following corollary.

\begin{corollary}\label{cor:block-dimension-trading}
With the notation of Proposition~\ref{prop:blocks}, suppose that every $\mathcal Q_j$ is pure and $\F_q$-linear.
Let $r\ge1$ and $\delta\ge2$ be integers such that $\delta\le\min_{j\in[B]}d_j$ and $\delta\le\ell_j\le r+\delta-1$ for every $j\in[B]$.
For every integer $0\le\gamma\le \sum_{j=1}^{B}(k_j+g_j)-1$, there exists a pure $\F_q$-linear $(r,\delta)$-SSLRC with parameters
\[
 \left[\left[n,\sum_{j=1}^{B}(k_j+g_j)-\gamma,\gamma,\min_{j\in[B]}d_j\right]\right]_q
\]
and repair sets $\Gamma_1,\ldots,\Gamma_B$.
Its stabilizer label space is $\bigoplus_{j=1}^{B}S_j$, where $S_j$ is the stabilizer label space of $\mathcal Q_j$.
\end{corollary}

{
\subsection{Optimal Pure $(r,\delta)$-SSLRCs from MDS Stabilizer Codes}
}
{ Applying Corollary~\ref{cor:block-dimension-trading} to MDS stabilizer codes, we obtain the following construction.

\begin{theorem}\label{thm:mdsblocks}
Let $\mathcal Q_0$ be an $\F_q$-linear MDS stabilizer code with parameters $[[N,k_0,d_0]]_q$, where $k_0>0$ and $d_0\ge2$.
Let $S_0$ be its stabilizer label space.
For every pair of integers $B\ge1$ and $0\le g\le Bk_0-1$, there exists a pure $\F_q$-linear $(r,\delta)$-SSLRC $\mathcal Q$ with parameters
\begin{equation}\label{eq:mdsblocks}
 [[BN,\ Bk_0-g,\ g,\ d_0]]_q,
 ~ (r,\delta)=(N-d_0+1,d_0).
\end{equation}
Its stabilizer label space is $S=S_0^{\oplus B}$, and its $B$ coordinate blocks of length $N$ form a pairwise disjoint repair cover.
The code attains the Singleton-type bound~\eqref{eq:lrcbound}, the alphabet-dependent bound~\eqref{eq:alphabetbound}, and the redundancy bound~\eqref{eq:pureredundancy} for these blocks.
If $\psi(S_0^{\perpt})$ is $\F_{q^2}$-linear, then $\mathcal Q$ also attains the $\F_{q^2}$-linear Griesmer bound~\eqref{eq:lineargriesmer}.
\end{theorem}

\begin{proof}
We first construct $\mathcal Q$.
Since $\mathcal Q_0$ is MDS, it is pure and $k_0=N-2d_0+2$; see the proof of~\cite[Theorem~7]{aly}.
Let $r=N-d_0+1$ and $\delta=d_0$.
Since $k_0>0$ and $\delta\le N=r+\delta-1$, we may apply Corollary~\ref{cor:block-dimension-trading} to $B$ copies of $\mathcal Q_0$.
For every $0\le g\le Bk_0-1$, we obtain a pure $\F_q$-linear $(r,\delta)$-SSLRC $\mathcal Q$ with the parameters in~\eqref{eq:mdsblocks} and stabilizer label space $S=S_0^{\oplus B}$.
Its $B$ coordinate blocks of length $N$ form a pairwise disjoint repair cover.

We next show that equality holds in the Singleton-type bound~\eqref{eq:lrcbound} and the disjoint redundancy bound~\eqref{eq:pureredundancy}.
Let $n=BN$, $k=Bk_0-g$, and $d=d_0$.
Since $N+k_0=2(N-d_0+1)=2r$, we have
\[
 \left\lceil\frac{n+k+g}{2r}\right\rceil
 =\left\lceil\frac{B(N+k_0)}{2r}\right\rceil
 =B.
\]
Consequently,
\begin{align*}
 &k+g+2(d-1)+2(B-1)(\delta-1)\\
 =&Bk_0+2(d_0-1)+2(B-1)(d_0-1)\\
 =&B(k_0+2d_0-2)=BN=n.
\end{align*}
Thus $\mathcal Q$ attains the Singleton-type bound~\eqref{eq:lrcbound}.
Since $\mathcal Q$ is pure, the redundancy bound~\eqref{eq:pureredundancy} applies with $M=B$.
Moreover,
\[
 n-k-g=B(N-k_0)=2B(d_0-1)=2B(\delta-1).
\]
Hence the redundancy bound~\eqref{eq:pureredundancy} is also attained.

We next prove that the alphabet-dependent bound~\eqref{eq:alphabetbound} is attained.
Set $\kappa=(n+k+g)/2=Br$ and choose $a=B-1$.
This value is admissible because
\[
 ar=(B-1)r<\kappa,
 ~ a(r+\delta-1)=(B-1)N<n.
\]
For this choice, the remaining length is $n-a(r+\delta-1)=N$.
By purity and~\eqref{eq:distance}, every nonzero label in $S_0^{\perpt}$ has weight at least $d_0$, and some label in $S_0^{\perpt}\setminus S_0$ has weight $d_0$.
Since $\psi$ preserves weight, $C_0:=\psi(S_0^{\perpt})$ has minimum distance $d_0$.
By~\eqref{eq:qdimensions} and~\eqref{eq:orthogonal-complement}, we have
\[
 |C_0|=q^{\dim_q S_0^{\perpt}}
      =q^{N+k_0}=(q^2)^r.
\]
Combining this equality with the cardinality form of the classical Singleton bound~\cite{macwilliams}, we obtain
\[
 r\le k_{\rm add}^{(q^2)}(N,d_0)\le N-d_0+1=r.
\]
Therefore, $k_{\rm add}^{(q^2)}(N,d_0)=r$.
Substituting $a=B-1$ into the right-hand side of~\eqref{eq:alphabetbound}, we obtain
\[
 \begin{aligned}
 k+g
 &\le\min_{a\in\mathcal A}
 \bigl\{2ar+2k_{\rm add}^{(q^2)}(n-a(r+\delta-1),d)-n\bigr\}\\
 &\le2(B-1)r+2r-BN\\
 &=B(2r-N)=Bk_0=k+g.
 \end{aligned}
\]
Hence the alphabet-dependent bound~\eqref{eq:alphabetbound} is attained.

Finally, suppose that $C_0$ is $\F_{q^2}$-linear.
Since $S=S_0^{\oplus B}$, we have
\[
 C:=\psi(S^{\perpt})=C_0^{\oplus B}.
\]
Thus $C$ is $\F_{q^2}$-linear. By Proposition~\ref{prop:classicaltransfer},
\[
 \dim_{q^2}C=\frac{n+k+g}{2}=\kappa=Br.
\]
Therefore, $\mathcal Q$ satisfies the $\F_{q^2}$-linear Griesmer bound~\eqref{eq:lineargriesmer}.
For the same admissible choice $a=B-1$, we have $N_a=N$ and $h_a=\kappa-ar=r$.
Since $\lceil d_0/q^{2i}\rceil\ge1$ for $i\ge1$, we obtain
\[
 \begin{aligned}
 BN
 &\ge(B-1)N+\sum_{i=0}^{r-1}
       \left\lceil\frac{d_0}{q^{2i}}\right\rceil\\
 &\ge(B-1)N+d_0+r-1\\
 &=BN.
 \end{aligned}
\]
Thus $\mathcal Q$ also attains the $\F_{q^2}$-linear Griesmer bound~\eqref{eq:lineargriesmer}.
\end{proof}
}
For an integer $1\le t<q/2$, let $H_t=\{(f(a))_{a\in\F_q}:f\in\F_q[x],\ \deg f<t\}.$ The code $H_t$ is an $[q,t,q-t+1]_q$ Reed-Solomon code with generator matrix $A_t=(a^j)_{0\le j<t,\ a\in\F_q}$ by~\cite[Ch.~10, Sec.~8]{macwilliams}. { Since $H_t=\operatorname{GRS}_t(\boldsymbol a,\boldsymbol1)$, we have $H_t^\perp=H_{q-t}$ by the GRS duality theorem~\cite[Ch.~10, Theorem~4]{macwilliams}.
}

\begin{lemma}\label{lemmaVI.2}
For $1\le t<q/2$, $H_t\subseteq H_t^\perp$, and $H_t^\perp$ is an $[q,q-t,t+1]_q$ MDS code.
The isotropic space $S_0=H_t\times H_t$ defines a pure $\F_q$-linear MDS stabilizer code with parameters $[[q,q-2t,t+1]]_q$.
\end{lemma}
\begin{proof}
Since $t<q-t$, we have $H_t\subseteq H_{q-t}=H_t^\perp$.
By Lemma~\ref{le:GRS}~(ii), $H_t^\perp$ has parameters $[q,q-t,t+1]_q$.
Apply Lemma~\ref{parameters:CSS} with $C_X=C_Z=H_t$ and $M_X=M_Z=A_t$.
Since $A_tA_t^T=0$, we have $b=0$ and
\[
 G_0=S_0=H_t\times H_t,~
 K=q^{q-2t}>1,~
 R=1.
\]
Thus $S_0=G_0\cap G_0^{\perpt}\subseteq S_0^{\perpt}$, and the resulting code $\mathcal Q_0$ is an $\F_q$-linear stabilizer code.
By~\eqref{eq:25} in Lemma~\ref{parameters:CSS},
\[
 d(\mathcal Q_0)
 =\min\{\wt(\bm{u}):\bm{u}\in H_t^\perp\setminus H_t\}
 =t+1,
\]
where the last equality follows from $d(H_t)=q-t+1>t+1=d(H_t^\perp)$.
For every nonzero $(\bm{x},\bm{z})\in G_0$, at least one component is a nonzero codeword of $H_t$, and hence
\[
 \swt(\bm{x},\bm{z})
 =|\supp(\bm{x})\cup\supp(\bm{z})|
 \ge q-t+1>t+1.
\]
Therefore, $\mathcal Q_0$ is pure by Definition~\ref{def:purity}.
Finally, $(q-2t)+2\bigl(d(\mathcal Q_0)-1\bigr)=q,$ so $\mathcal Q_0$ attains the quantum Singleton bound~\cite[Theorem~11]{alylp}.
\end{proof}

Applying Corollary~\ref{cor:block-dimension-trading} to copies of the stabilizer code in Lemma~\ref{lemmaVI.2}, we obtain the following family of pure $(r,\delta)$-SSLRCs.

\begin{theorem}\label{thm:rsfamily}
Let $q$ be a prime power, and let $t$, $B$, and $g$ be integers such that $1\le t<q/2,B\ge1$ and $0\le g\le B(q-2t)-1$.
Then there exists a pure $\F_q$-linear $(r,\delta)$-SSLRC $\mathcal Q$ with parameters
\begin{equation}\label{eq:rsfamily}
 [[Bq,\ B(q-2t)-g,\ g,\ t+1]]_q,~(r,\delta)=(q-t,t+1).
\end{equation}
Moreover, $\mathcal Q$ attains the Singleton-type bound~\eqref{eq:lrcbound}, the alphabet-dependent bound~\eqref{eq:alphabetbound}, and the $\F_{q^2}$-linear Griesmer bound~\eqref{eq:lineargriesmer}.
It also attains the redundancy bound~\eqref{eq:pureredundancy} for its $B$ pairwise disjoint coordinate blocks.
\end{theorem}
\begin{proof}
We first construct $\mathcal Q$.
By Lemma~\ref{lemmaVI.2}, there exists a pure $\F_q$-linear stabilizer code $\mathcal Q_0$ with parameters $[[q,q-2t,t+1]]_q$ and stabilizer label space $S_0=H_t\times H_t$.
Let $r=q-t$ and $\delta=t+1$.
Since $q-2t\ge1$, $d(\mathcal Q_0)=\delta$, and $\delta\le q=r+\delta-1$, we may apply Corollary~\ref{cor:block-dimension-trading} to $B$ copies of $\mathcal Q_0$.
For every $0\le g\le B(q-2t)-1$, we obtain a pure $\F_q$-linear $(r,\delta)$-SSLRC $\mathcal Q$ with the parameters in~\eqref{eq:rsfamily} and stabilizer label space $S=S_0^{\oplus B}$.
Its $B$ coordinate blocks of length $q$ form a pairwise disjoint repair cover.

{ Finally, we verify the $\F_{q^2}$-linearity required in Theorem~\ref{thm:mdsblocks}. Since $S_0=H_t\times H_t$ is $\F_q$-linear, its trace-symplectic and symplectic orthogonal complements coincide, and hence
\[
 S_0^{\perpt}=S_0^{\perps}=H_t^\perp\times H_t^\perp.
\]
By the coordinatewise definition of $\psi$ in~\eqref{eq:psi}, we obtain
\[
 \psi(S_0^{\perpt})
 =H_t^\perp+\alpha H_t^\perp
 =\operatorname{span}_{\F_{q^2}}H_t^\perp,
\]
where the last equality follows from the $\F_q$-linearity of $H_t^\perp$ and $\F_{q^2}=\F_q+\alpha\F_q$.
Thus $\psi(S_0^{\perpt})$ is $\F_{q^2}$-linear.
By Theorem~\ref{thm:mdsblocks}, $\mathcal Q$ attains the four stated bounds.
}
\end{proof}

{ We use the following Lagrange interpolation identity~\cite{macwilliams} to choose the multipliers of a GRS code.
}

\begin{lemma}\label{lem:lagrange}\cite{macwilliams}
Let $N\ge1$, let $a_1,\ldots,a_N$ be distinct elements of a field, and set $\ell_i=\prod_{j\ne i}(a_i-a_j)$.
For every integer $0\le u\le N-1$,
\[
 \sum_{i=1}^N\frac{a_i^u}{\ell_i}
 =
 \begin{cases}
  0,&0\le u\le N-2,\\
  1,&u=N-1.
 \end{cases}
\]
\end{lemma}
Using the parameter families in~\cite[Lemma~8]{aly}, we obtain the following optimal $(r,\delta)$-SSLRCs. We use $s$ to denote the distance parameter in parts~(iii)--(vi) of that lemma.

\begin{corollary}\label{cor:mdsfamilies}
Let $q$ be a prime power, and let $B\ge1$ be an integer. Then the following statements hold.
\begin{enumerate}
\item[(i)] A family of $\F_q$-linear pure $(N-d_0+1,d_0)$-SSLRCs with parameters $[[BN,B(N-2d_0+2)-g,g,d_0]]_q$  exists for integers $3\le N\le q$, $2\le d_0\le\lfloor(N+1)/2\rfloor$, and $0\le g\le B(N-2d_0+2)-1$.
\item[(ii)] A family of $\F_q$-linear pure  $((\nu+1)(q-1),\nu+2)$-SSLRCs with parameters $[[B(\nu+1)q,B(\nu+1)(q-2)-g,g,\nu+2]]_q$ exists for $q\ge3$ and integers $0\le\nu\le q-2$ and $0\le g\le B(\nu+1)(q-2)-1$.
\item[(iii)] A family of $\F_q$-linear pure $(q-s-1,s+1)$-SSLRCs with parameters $[[B(q-1),B(q-2s-1)-g,g,s+1]]_q$ exists for integers $1\le s<(q-1)/2$ and $0\le g\le B(q-2s-1)-1$.
\item[(iv)] A family of $\F_q$-linear pure $(q-s-1,s+2)$-SSLRCs with parameters $[[Bq,B(q-2s-2)-g,g,s+2]]_q$ exists for integers $0\le s\le\lfloor(q-3)/2\rfloor$ and $0\le g\le B(q-2s-2)-1$.
\item[(v)] A family of $\F_q$-linear pure $(q^2-s-1,s+1)$-SSLRCs with parameters $[[B(q^2-1),B(q^2-2s-1)-g,g,s+1]]_q$ exists for integers $1\le s<q-1$ and $0\le g\le B(q^2-2s-1)-1$.
\item[(vi)] A family of $\F_q$-linear pure $(q^2-s-1,s+2)$-SSLRCs with parameters $[[Bq^2,B(q^2-2s-2)-g,g,s+2]]_q$ exists for integers $0\le s<q-1$ and $0\le g\le B(q^2-2s-2)-1$.
\end{enumerate}
Moreover, these families of pure SSLRCs attain the four bounds stated in Theorem~\ref{thm:mdsblocks}. Their $B$ coordinate blocks of length $n/B$ form a pairwise disjoint repair cover, where $n$ denotes the code length.
\end{corollary}

\begin{proof}
We first verify that the MDS stabilizer codes can be chosen with $\F_{q^2}$-linear comparison codes.
For each case, let $N$ be the block length and let $d_0=\delta$.
Set $k_0=N-2d_0+2$.
For parts~(i), (iii), and~(iv), we have $N\le q$ and $k_0>0$.
Choose distinct $a_1,\ldots,a_N\in\F_q$, set $\ell_i=\prod_{j\ne i}(a_i-a_j)$, and choose $v_i\in\F_{q^2}^{\times}$ with $v_i^{q+1}=\ell_i^{-1}$.
Since $\ell_i^{-1}\in\F_q^{\times}$ and the norm map onto $\F_q^{\times}$ is surjective, such multipliers exist.
Let $X=\operatorname{GRS}_{d_0-1}(\boldsymbol a,\boldsymbol v)$ over $\F_{q^2}$.
For $0\le j,j'\le d_0-2$, we have
\[
 \sum_{i=1}^N(v_i a_i^j)(v_i a_i^{j'})^q
 =\sum_{i=1}^N\frac{a_i^{j+j'}}{\ell_i}=0
\]
by Lemma~\ref{lem:lagrange}, since $a_i^q=a_i$ and $j+j'\le2d_0-4\le N-2$.
Hence $X\subseteq X^{\perp_H}$.

For part~(ii), let $A$ be the length-$q^2$ RS code obtained by evaluating polynomials of degree at most $\nu$ over $\F_{q^2}$.
By~\cite[Lemma~8 and the proof of Theorem~5]{sarvepalli-rm}, there exists a vector $\boldsymbol\lambda\in\F_q^{q^2}$ of weight $N=(\nu+1)q$ such that
\[
 \sum_{i=1}^{q^2}\lambda_i a_i b_i^q=0
 ~\text{for all }\boldsymbol a,\boldsymbol b\in A.
\]
Let $I=\supp(\boldsymbol\lambda)$, and choose $v_i\in\F_{q^2}^{\times}$ with $v_i^{q+1}=\lambda_i$ for $i\in I$.
Then
\[
 X=\{(v_i a_i)_{i\in I}:\boldsymbol a\in A\}
\]
is an $\F_{q^2}$-linear $[N,\nu+1,N-\nu]_{q^2}$ GRS code.
Moreover, for all $\boldsymbol a,\boldsymbol b\in A$,
\[
 \sum_{i\in I}(v_i a_i)(v_i b_i)^q
 =\sum_{i\in I}\lambda_i a_i b_i^q=0.
\]
Thus $X\subseteq X^{\perp_H}$.
For parts~(v) and~(vi), we may choose a Hermitian self-orthogonal $[N,d_0-1,N-d_0+2]_{q^2}$ MDS code $X$ by the constructions in the proof of~\cite[Theorem~10]{grassl-optimal}.

We next determine the parameters of the stabilizer codes.
In every case, $X$ is an $[N,d_0-1,N-d_0+2]_{q^2}$ MDS code and $X\subseteq X^{\perp_H}$.
By MDS duality~\cite{macwilliams}, $X^{\perp_H}$ has parameters $[N,N-d_0+1,d_0]_{q^2}$.
Since $k_0=N-2d_0+2>0$, we have $d(X)=N-d_0+2>d_0$.
Hence
\[
 \min\{\wt(\bm w):\bm w\in X^{\perp_H}\setminus X\}=d_0.
\]
By Lemma~\ref{parameters:Hermitian}, $S_0=\psi^{-1}(X)$ defines a pure $\F_q$-linear MDS stabilizer code with parameters $[[N,k_0,d_0]]_q$ and
\[
 \psi(S_0^{\perpt})=X^{\perp_H}.
\]
Finally, $X^{\perp_H}$ is $\F_{q^2}$-linear, so the result follows from Theorem~\ref{thm:mdsblocks}.
\end{proof}

\begin{example}\label{ex:mdsfamilies}
Theorem~\ref{thm:rsfamily} and Corollary~\ref{cor:mdsfamilies} provide the pure $(r,\delta)$-SSLRCs listed in Table~\ref{tab:mdsexamples}.
For example, take $q=3$, $\nu=1$, and $B=2$ in Corollary~\ref{cor:mdsfamilies}(ii).
Then $N=6$, $k_0=2$, and $d_0=3$.
For every $0\le g\le3$, we obtain
\[
 [[12,4-g,g,3]]_3,~(r,\delta)=(4,3).
\]
For $g=1$, the parameters are $[[12,3,1,3]]_3$.
The two coordinate blocks of length $6$ form a pairwise disjoint repair cover.
Every code in Table~\ref{tab:mdsexamples} attains the four bounds in Theorem~\ref{thm:mdsblocks}.

\begin{table}[!hbp]
\centering
\caption{Optimal Pure $(r,\delta)$-SSLRCs from Theorem~\ref{thm:rsfamily} and Corollary~\ref{cor:mdsfamilies}.}
\label{tab:mdsexamples}
\small
\renewcommand{\arraystretch}{1.0}
\setlength{\tabcolsep}{5pt}
\begin{tabular}{@{}ccccccc@{}}
\toprule
$q$ & $\mathcal Q_0$ & $B$ & $g$ & $\mathcal Q$ & $(r,\delta)$ & References\\
\midrule
$5$ & $[[5,3,2]]_{5}$ & $2$ & $1$ & $[[10,5,1,2]]_{5}$ & $(4,2)$ & Thm.~\ref{thm:rsfamily}\\
$7$ & $[[7,3,3]]_{7}$ & $3$ & $2$ & $[[21,7,2,3]]_{7}$ & $(5,3)$ & Thm.~\ref{thm:rsfamily}\\
$5$ & $[[4,2,2]]_{5}$ & $2$ & $1$ & $[[8,3,1,2]]_{5}$ & $(3,2)$ & Cor.~\ref{cor:mdsfamilies}(i)\\
$7$ & $[[5,1,3]]_{7}$ & $3$ & $2$ & $[[15,1,2,3]]_{7}$ & $(3,3)$ & Cor.~\ref{cor:mdsfamilies}(i)\\
$3$ & $[[6,2,3]]_{3}$ & $2$ & $1$ & $[[12,3,1,3]]_{3}$ & $(4,3)$ & Cor.~\ref{cor:mdsfamilies}(ii)\\
$4$ & $[[12,6,4]]_{4}$ & $2$ & $5$ & $[[24,7,5,4]]_{4}$ & $(9,4)$ & Cor.~\ref{cor:mdsfamilies}(ii)\\
$7$ & $[[6,2,3]]_{7}$ & $2$ & $3$ & $[[12,1,3,3]]_{7}$ & $(4,3)$ & Cor.~\ref{cor:mdsfamilies}(iii)\\
$9$ & $[[8,4,3]]_{9}$ & $2$ & $3$ & $[[16,5,3,3]]_{9}$ & $(6,3)$ & Cor.~\ref{cor:mdsfamilies}(iii)\\
$5$ & $[[5,1,3]]_{5}$ & $2$ & $1$ & $[[10,1,1,3]]_{5}$ & $(3,3)$ & Cor.~\ref{cor:mdsfamilies}(iv)\\
$7$ & $[[7,3,3]]_{7}$ & $2$ & $3$ & $[[14,3,3,3]]_{7}$ & $(5,3)$ & Cor.~\ref{cor:mdsfamilies}(iv)\\
$3$ & $[[8,6,2]]_{3}$ & $2$ & $5$ & $[[16,7,5,2]]_{3}$ & $(7,2)$ & Cor.~\ref{cor:mdsfamilies}(v)\\
$4$ & $[[15,11,3]]_{4}$ & $2$ & $7$ & $[[30,15,7,3]]_{4}$ & $(13,3)$ & Cor.~\ref{cor:mdsfamilies}(v)\\
$3$ & $[[9,5,3]]_{3}$ & $2$ & $1$ & $[[18,9,1,3]]_{3}$ & $(7,3)$ & Cor.~\ref{cor:mdsfamilies}(vi)\\
$4$ & $[[16,10,4]]_{4}$ & $2$ & $5$ & $[[32,15,5,4]]_{4}$ & $(13,4)$ & Cor.~\ref{cor:mdsfamilies}(vi)\\
\bottomrule
\end{tabular}
\end{table}
\end{example}

\subsection{\texorpdfstring{Optimal Pure $(r,\delta)$-SSLRCs from MDS Hermitian Hulls}{Optimal Pure (r,delta)-SSLRCs from MDS Hermitian Hulls}}
In this subsection, we construct pure $\F_q$-linear $(r,\delta)$-SSLRCs from Hermitian self-orthogonal MDS codes.
We then apply the construction to explicit MDS Hermitian hulls.
The resulting codes attain the four bounds in Theorem~\ref{thm:mdsblocks}.

\begin{theorem}\label{thm:hullblocks}
Let $q$ be a prime power, and let $H$ be an $[N,h_H,N-h_H+1]_{q^2}$ MDS code.
Suppose that $H\subseteq H^{\perp_H}$ and $1\le h_H<N/2$.
For every pair of integers $B\ge1$ and $0\le g\le B(N-2h_H)-1$, there exists a pure $\F_q$-linear $(r,\delta)$-SSLRC with parameters
\[
 [[BN,\ B(N-2h_H)-g,\ g,\ h_H+1]]_q,
 ~ (r,\delta)=(N-h_H,h_H+1).
\]
Its stabilizer label space is $S=(\psi^{-1}(H))^{\oplus B}$.
The code attains the Singleton-type bound~\eqref{eq:lrcbound}, the alphabet-dependent bound~\eqref{eq:alphabetbound}, and the $\F_{q^2}$-linear Griesmer bound~\eqref{eq:lineargriesmer}.
It also attains the redundancy bound~\eqref{eq:pureredundancy} for its $B$ pairwise disjoint coordinate blocks of length $N$.
\end{theorem}

\begin{proof}
We first construct a pure MDS stabilizer code.
By MDS duality~\cite{macwilliams}, $H^{\perp_H}$ is an $[N,N-h_H,h_H+1]_{q^2}$ MDS code.
Since $2h_H<N$, we have
\[
 d(H)=N-h_H+1>h_H+1=d(H^{\perp_H}).
\]
Let $\bm w_0\in H^{\perp_H}$ have weight $h_H+1$.
Since $d(H)>h_H+1$, we have $\bm w_0\notin H$.
Thus
\[
 \min_{\bm w\in H^{\perp_H}\setminus H}\wt(\bm w)
 =d(H^{\perp_H})=h_H+1.
\]
By Lemma~\ref{parameters:Hermitian}, $S_0=\psi^{-1}(H)$ defines an $\F_q$-linear stabilizer code $\mathcal Q_0$ with parameters
\[
 [[N,\ N-2h_H,\ h_H+1]]_q.
\]
Since $\psi$ preserves weight, every nonzero label in $S_0$ has symplectic weight at least $d(H)>h_H+1$.
Thus $\mathcal Q_0$ is pure.
Moreover, $(N-2h_H)+2((h_H+1)-1)=N$, so $\mathcal Q_0$ is MDS.

Finally, we apply Theorem~\ref{thm:mdsblocks}.
By Lemma~\ref{lem:hermitian-correspondence}, we have $\psi(S_0^{\perpt})=H^{\perp_H}$, which is $\F_{q^2}$-linear.
Since $N-2h_H>0$, Theorem~\ref{thm:mdsblocks} applies with $k_0=N-2h_H$ and $d_0=h_H+1$.
For every $0\le g\le B(N-2h_H)-1$, we obtain a code with the stated parameters, stabilizer label space, and repair cover. This code attains all four bounds.
\end{proof}

For any $\F_{q^2}$-linear code $C$ with Hermitian hull $H=C\cap C^{\perp_H}$, we have $H\subseteq C$ and $H\subseteq C^{\perp_H}\subseteq H^{\perp_H}$.
Thus Theorem~\ref{thm:hullblocks} applies whenever $H$ is MDS and $1\le\dim_{q^2}H<N/2$.
If $C$ is an $[N,c,N-c+1]_{q^2}$ MDS code and $c+h_H<N$, then choosing $g=B(c-h_H)$ recovers the parameters $[[BN,B(N-c-h_H),B(c-h_H),h_H+1]]_q$.

We next apply Lemma~\ref{lem:lagrange} to determine the Hermitian hull of an explicit GRS code.
}

\begin{theorem}\label{thm:explicitgrs}
Let $q$ be a prime power, and let $N,c,h_H$ be integers satisfying $1\le h_H<c$ and $c+h_H<N\le q-1.$ Let $a_1,\ldots,a_N\in\F_q^\times$ be distinct, set $\ell_i=\prod_{j\ne i}(a_i-a_j)$, and choose $v_i\in\F_{q^2}^\times$ such that
\begin{equation}\label{eq:grsnorms}
 \mu=N-c-h_H,
 ~
 v_i^{q+1}=\frac{a_i^\mu}{\ell_i}~(1\le i\le N).
\end{equation}
For $\boldsymbol a=(a_1,\ldots,a_N)$ and $\boldsymbol v=(v_1,\ldots,v_N)$, the code $C=\operatorname{GRS}_c(\boldsymbol a,\boldsymbol v)$ has Hermitian hull $C\cap C^{\perp_H} =\operatorname{GRS}_{h_H}(\boldsymbol a,\boldsymbol v).$ { Consequently, for every pair of integers $B\ge1$ and $0\le g\le B(N-2h_H)-1$, there exists a pure $\F_q$-linear $(r,\delta)$-SSLRC with parameters
\[
 [[BN,\ B(N-2h_H)-g,\ g,\ h_H+1]]_q,
 ~ (r,\delta)=(N-h_H,h_H+1).
\]
Its stabilizer label space is $S=(\psi^{-1}(\operatorname{GRS}_{h_H}(\boldsymbol a,\boldsymbol v)))^{\oplus B}$.
}
The code attains all the bounds stated in Theorem~\ref{thm:hullblocks}.
\end{theorem}

\begin{proof}
We first show that the required multipliers exist.
Since $a_1,\ldots,a_N\in\F_q^\times$ are distinct, $a_i^\mu/\ell_i\in\F_q^\times$ for $1\leq i\leq N$.
The norm map $v\mapsto v^{q+1}$ from $\F_{q^2}^\times$ to $\F_q^\times$ is surjective.
Thus, for each $i$, there exists $v_i\in\F_{q^2}^\times$ satisfying~\eqref{eq:grsnorms}.
We next determine the rank of the Hermitian Gram matrix.
Let $M=(v_i a_i^j)_{0\le j<c,\,1\le i\le N}$ be the generator matrix of $C$, and set $\Omega=M(M^{(q)})^T$.
Since $a_i^q=a_i$, the $(j,j')$-entry of $\Omega$ satisfies
\[
 \begin{aligned}
 \Omega_{jj'}
 =\sum_{i=1}^N(v_i a_i^j)(v_i a_i^{j'})^q
 =\sum_{i=1}^N v_i^{q+1}a_i^{j+j'}
 =\sum_{i=1}^N\frac{a_i^{\mu+j+j'}}{\ell_i},
 \end{aligned}
\]
with $0\le j,j'<c$.
If $j<h_H$ or $j'<h_H$, then
\[
 \begin{aligned}
 0\le\mu+j+j'
 &\le(N-c-h_H)+(h_H-1)+(c-1)\\
 &=N-2.
 \end{aligned}
\]
By Lemma~\ref{lem:lagrange}, we have $\Omega_{jj'}=0$ whenever $j<h_H$ or $j'<h_H$.
Let $g_0=c-h_H$, and let $\Delta=(\Delta_{ab})_{0\le a,b<g_0}$ be the lower-right block of $\Omega$.

For $0\le a,b<g_0$, set $j=h_H+a$ and $j'=h_H+b$.
Since $\mu=N-c-h_H$ and $g_0=c-h_H$, we have
\[
 \begin{aligned}
 \Delta_{ab}
 =\Omega_{h_H+a,h_H+b}
 =\sum_{i=1}^N
   \frac{a_i^{\mu+2h_H+a+b}}{\ell_i}
 =\sum_{i=1}^N
   \frac{a_i^{N-g_0+a+b}}{\ell_i}.
 \end{aligned}
\]
By Lemma~\ref{lem:lagrange},
$$
\Delta_{ab}=
\left\{
  \begin{array}{ll}
    0, & \hbox{if $a+b\leq g_0-2$;} \\
    1, & \hbox{if $a+b=g_0-1$.}
  \end{array}
\right.
$$
Thus
\[
 \Omega=
 \begin{pmatrix}
  0&0\\
  0&\Delta
 \end{pmatrix}.
\]
Let $J_{g_0}$ be the $g_0\times g_0$ matrix with ones on the anti-diagonal and zeros elsewhere.
Then $\Delta J_{g_0}$ is obtained from $\Delta$ by reversing the order of its columns. { Since $(\Delta J_{g_0})_{ab}=\Delta_{a,g_0-1-b}$, we have $(\Delta J_{g_0})_{ab}=0$ for $a<b$ and $(\Delta J_{g_0})_{aa}=1$. Thus $\Delta J_{g_0}$ is lower triangular with unit diagonal, and
\[
 \det\Delta\cdot\det J_{g_0}=\det(\Delta J_{g_0})=1.
\]
Hence $\det\Delta\ne0$ and $\rank_{q^2}\Omega=g_0$.
}

{ We then determine the Hermitian hull of $C$. Let $x=(x_0,x_1,\ldots,x_{c-1})$ and $y=(y_0,y_1,\ldots,y_{c-1})$ be coefficient rows in $\F_{q^2}^c$. For every $1\le i\le N$, we have
}
\[
(xM)_i=\sum_{j=0}^{c-1}x_jv_i a_i^j,~
(yM)_i=\sum_{j'=0}^{c-1}y_{j'}v_i a_i^{j'}.
\]
Then
\[
\begin{aligned}
\langle xM,yM\rangle_H
&=\sum_{i=1}^{N}(xM)_i\bigl((yM)_i\bigr)^q\\
&=\sum_{i=1}^{N}
  \left(\sum_{j=0}^{c-1}x_jv_i a_i^j\right)
  \left(\sum_{j'=0}^{c-1}y_{j'}v_i a_i^{j'}\right)^q\\
&=\sum_{i=1}^{N}\sum_{j=0}^{c-1}\sum_{j'=0}^{c-1}
  x_jy_{j'}^q(v_i a_i^j)(v_i a_i^{j'})^q\\
&=\sum_{j=0}^{c-1}\sum_{j'=0}^{c-1}
  x_j\left(\sum_{i=1}^{N}
  (v_i a_i^j)(v_i a_i^{j'})^q\right)y_{j'}^q\\
&=\sum_{j=0}^{c-1}\sum_{j'=0}^{c-1}x_j\Omega_{jj'}y_{j'}^q\\
&=x\Omega(y^{(q)})^T.
\end{aligned}
\]
Since $C=\{yM:y\in\F_{q^2}^c\}$ and $y\mapsto y^{(q)}$ is a bijection on $\F_{q^2}^c$, we have
\[
 \begin{aligned}
 xM\in C\cap C^{\perp_H}
 &\Longleftrightarrow
 x\Omega(y^{(q)})^T=0\quad\text{for every }y\in\F_{q^2}^c\\
 &\Longleftrightarrow x\Omega=0.
 \end{aligned}
\]
By the block form of $\Omega$ and the nonsingularity of $\Delta$,
\[
 \begin{aligned}
 x\Omega=0
 &\Longleftrightarrow
 (x_{h_H},\ldots,x_{c-1})\Delta=0\\
 &\Longleftrightarrow
 x_j=0\quad\text{for every }h_H\le j<c.
 \end{aligned}
\]

Thus the hull is generated by the first $h_H$ rows of $M$, and hence $C\cap C^{\perp_H} =\operatorname{GRS}_{h_H}(\boldsymbol a,\boldsymbol v).$ The codes $C$ and its hull are MDS, with parameters $[N,c,N-c+1]_{q^2}$ and $[N,h_H,N-h_H+1]_{q^2}$, respectively. { Finally, since $H=C\cap C^{\perp_H}$ is Hermitian self-orthogonal and $2h_H<c+h_H<N$, we may apply Theorem~\ref{thm:hullblocks}. For every $0\le g\le B(N-2h_H)-1$, we obtain a pure code with the stated quantum parameters, stabilizer label space, and locality. This code attains all four bounds.
}
\end{proof}

{ Combining the hull identity in~\cite[Theorem~10]{luomds} with Theorem~\ref{thm:hullblocks}, we obtain the following family.
}

\begin{proposition}\label{prop:fullfield}
Let $q$ be a prime power, let $\boldsymbol a$ list all elements of $\F_{q^2}$, and let $\boldsymbol1$ be the all-one vector of length $q^2$.
Then $\operatorname{Hull}_H \bigl(\operatorname{GRS}_q(\boldsymbol a,\boldsymbol1)\bigr) = \operatorname{GRS}_{q-1}(\boldsymbol a,\boldsymbol1).$ { For every pair of integers $B\ge1$ and $0\le g\le B(q^2-2q+2)-1$, there exists a pure $\F_q$-linear $(r,\delta)$-SSLRC with parameters
\[
 [[Bq^2,\ B(q^2-2q+2)-g,\ g,\ q]]_q,
 ~ (r,\delta)=(q^2-q+1,q).
\]
Its stabilizer label space is $S=(\psi^{-1}(\operatorname{GRS}_{q-1}(\boldsymbol a,\boldsymbol1)))^{\oplus B}$.
}
The code attains all the bounds stated in Theorem~\ref{thm:hullblocks}.
\end{proposition}

{
\begin{proof}
By~\cite[Theorem~10]{luomds}, we have
\[
 H:=\operatorname{Hull}_H(\operatorname{GRS}_q(\boldsymbol a,\boldsymbol1))
   =\operatorname{GRS}_{q-1}(\boldsymbol a,\boldsymbol1).
\]
Thus $H$ is an $[q^2,q-1,q^2-q+2]_{q^2}$ MDS code and $H\subseteq H^{\perp_H}$.
Since $q^2-2(q-1)=(q-1)^2+1>0$, we may apply Theorem~\ref{thm:hullblocks} with $N=q^2$ and $h_H=q-1$.
Hence the resulting codes have the stated parameters and attain all four bounds.
\end{proof}
}

\begin{example}

For Table~\ref{tab:explicitgrs}, let $\zeta$ be a primitive element of $\F_q$, set $a_i=\zeta^{i-1}$ for $1\le i\le N$, and choose the multipliers as in~\eqref{eq:grsnorms}.
For Table~\ref{tab:fullfield}, let $\boldsymbol a$ list all elements of $\F_{q^2}$ and set $\boldsymbol v=\boldsymbol1$. { By Theorem~\ref{thm:explicitgrs} and Proposition~\ref{prop:fullfield}, respectively, we obtain the pure $\F_q$-linear $(r,\delta)$-SSLRCs in the two tables. Each code has $d=\delta$, and its $B$ coordinate blocks form pairwise disjoint repair sets. The choices $g=B(c-h_H)$ and $g=B$ recover the original examples in Tables~\ref{tab:explicitgrs} and~\ref{tab:fullfield}, respectively. The remaining rows correspond to other admissible gauge dimensions.
}
All listed codes attain the Singleton-type bound~\eqref{eq:lrcbound}, the alphabet-dependent bound~\eqref{eq:alphabetbound}, the $\F_{q^2}$-linear Griesmer bound~\eqref{eq:lineargriesmer}, and the disjoint redundancy bound~\eqref{eq:pureredundancy}.

\begin{table}
\centering
\caption{Optimal Pure $(r,\delta)$-SSLRCs from
Theorem~\ref{thm:explicitgrs}.}
\label{tab:explicitgrs}
\small
\renewcommand{\arraystretch}{1.15}
\setlength{\tabcolsep}{5pt}
\begin{tabular}{@{}cccccc@{}}
\toprule
$q$ & $C$ & $\operatorname{Hull}_H(C)$ & $B$
    & $\mathcal Q$ & $(r,\delta)$\\
\midrule
$5$ & $[4,2,3]_{5^2}$ & $[4,1,4]_{5^2}$ & $1$
    & $[[4,1,1,2]]_{5}$ & $(3,2)$\\
\addlinespace[1pt]
$7$ & $[6,3,4]_{7^2}$ & $[6,2,5]_{7^2}$ & $2$
    & $[[12,2,2,3]]_{7}$ & $(4,3)$\\
\addlinespace[1pt]
$8$ & $[7,3,5]_{8^2}$ & $[7,2,6]_{8^2}$ & $2$
    & $[[14,4,2,3]]_{8}$ & $(5,3)$\\
$8$ & $[7,4,4]_{8^2}$ & $[7,2,6]_{8^2}$ & $3$
    & $[[21,3,6,3]]_{8}$ & $(5,3)$\\
\addlinespace[1pt]
$9$ & $[8,4,5]_{9^2}$ & $[8,3,6]_{9^2}$ & $3$
    & $[[24,3,3,4]]_{9}$ & $(5,4)$\\
\addlinespace[1pt]
$11$ & $[8,3,6]_{11^2}$ & $[8,2,7]_{11^2}$ & $2$
     & $[[16,6,2,3]]_{11}$ & $(6,3)$\\
$11$ & $[8,4,5]_{11^2}$ & $[8,2,7]_{11^2}$ & $2$
     & $[[16,4,4,3]]_{11}$ & $(6,3)$\\
$11$ & $[8,5,4]_{11^2}$ & $[8,2,7]_{11^2}$ & $2$
     & $[[16,2,6,3]]_{11}$ & $(6,3)$\\
$11$ & $[10,5,6]_{11^2}$ & $[10,4,7]_{11^2}$ & $2$
     & $[[20,2,2,5]]_{11}$ & $(6,5)$\\
\addlinespace[1pt]
$13$ & $[10,5,6]_{13^2}$ & $[10,3,8]_{13^2}$ & $2$
     & $[[20,4,4,4]]_{13}$ & $(7,4)$\\
\addlinespace[2pt]
$7$ & $[6,3,4]_{7^2}$ & $[6,2,5]_{7^2}$ & $2$
 & $[[12,3,1,3]]_7$ & $(4,3)$\\
$11$ & $[8,3,6]_{11^2}$ & $[8,2,7]_{11^2}$ & $2$
 & $[[16,5,3,3]]_{11}$ & $(6,3)$\\
\bottomrule
\end{tabular}
\end{table}
\begin{table}
\centering
\caption{Optimal Pure $(r,\delta)$-SSLRCs from
Proposition~\ref{prop:fullfield}.}
\label{tab:fullfield}
\small
\renewcommand{\arraystretch}{1.15}
\setlength{\tabcolsep}{5pt}
\begin{tabular}{@{}cccccc@{}}
\toprule
$q$ & $C$ & $\operatorname{Hull}_H(C)$ & $B$
    & $\mathcal Q$ & $(r,\delta)$\\
\midrule
$2$ & $[4,2,3]_{2^2}$ & $[4,1,4]_{2^2}$ & $1$
    & $[[4,1,1,2]]_{2}$ & $(3,2)$\\
$2$ & $[4,2,3]_{2^2}$ & $[4,1,4]_{2^2}$ & $2$
    & $[[8,2,2,2]]_{2}$ & $(3,2)$\\
\addlinespace[1pt]
$3$ & $[9,3,7]_{3^2}$ & $[9,2,8]_{3^2}$ & $1$
    & $[[9,4,1,3]]_{3}$ & $(7,3)$\\
$3$ & $[9,3,7]_{3^2}$ & $[9,2,8]_{3^2}$ & $2$
    & $[[18,8,2,3]]_{3}$ & $(7,3)$\\
\addlinespace[1pt]
$4$ & $[16,4,13]_{4^2}$ & $[16,3,14]_{4^2}$ & $1$
    & $[[16,9,1,4]]_{4}$ & $(13,4)$\\
$4$ & $[16,4,13]_{4^2}$ & $[16,3,14]_{4^2}$ & $2$
    & $[[32,18,2,4]]_{4}$ & $(13,4)$\\
\addlinespace[1pt]
$5$ & $[25,5,21]_{5^2}$ & $[25,4,22]_{5^2}$ & $1$
    & $[[25,16,1,5]]_{5}$ & $(21,5)$\\
$5$ & $[25,5,21]_{5^2}$ & $[25,4,22]_{5^2}$ & $2$
    & $[[50,32,2,5]]_{5}$ & $(21,5)$\\
\addlinespace[1pt]
$7$ & $[49,7,43]_{7^2}$ & $[49,6,44]_{7^2}$ & $1$
    & $[[49,36,1,7]]_{7}$ & $(43,7)$\\
$7$ & $[49,7,43]_{7^2}$ & $[49,6,44]_{7^2}$ & $2$
    & $[[98,72,2,7]]_{7}$ & $(43,7)$\\
\addlinespace[2pt]
$3$ & $[9,3,7]_{3^2}$ & $[9,2,8]_{3^2}$ & $2$
 & $[[18,9,1,3]]_3$ & $(7,3)$\\
$3$ & $[9,3,7]_{3^2}$ & $[9,2,8]_{3^2}$ & $2$
 & $[[18,1,9,3]]_3$ & $(7,3)$\\
$4$ & $[16,4,13]_{4^2}$ & $[16,3,14]_{4^2}$ & $2$
 & $[[32,15,5,4]]_4$ & $(13,4)$\\
\bottomrule
\end{tabular}
\end{table}
\end{example}

\section{Conclusion and Outlook}\label{con-out}
In this paper, we characterized exact local erasure recovery for $(r,\delta)$-SSLRCs algebraically without assuming purity.
Using this criterion, we derived redundancy, supported-subspace, and mixed counting bounds.
For pure codes, we used associated additive classical codes to obtain Singleton-type and alphabet-dependent bounds, together with explicit Griesmer, Hamming, and Plotkin inequalities.
We also proved that logical and gauge dimensions can be traded while preserving the stabilizer label space, minimum distance, and repair cover.
Using MDS stabilizer codes, RS codes, and MDS Hermitian hulls, we constructed pure families with $d=\delta$ that simultaneously attain the Singleton-type, alphabet-dependent, $\F_{q^2}$-linear Griesmer, and disjoint redundancy bounds.

The constructions in Section~\ref{sec:construction} do not address equality in every bound proved here.
One problem for future work is to construct pure families attaining the mixed counting bound~\eqref{eq:mixedpacking} in the presence of both local erasures and errors outside the selected repair sets.
Another problem is to construct pure $(r,\delta)$-SSLRCs with $d>\delta$ that attain the bounds in Section~\ref{sec:pure} while preserving the prescribed repair cover.
The supported-subspace bounds also suggest a separate study of optimal impure families and their gauge-equivalence classes.

\end{document}